\PassOptionsToPackage{colorlinks=true,
                     urlcolor=BrightLinks,
                     linkcolor=Blue,
                     citecolor=Mulberry}{hyperref}

\documentclass[12pt]{thesis}  

\usepackage{titlesec}
   \titleformat{\chapter}
      {\normalfont\large}{Chapter \thechapter:}{1em}{}
\usepackage{graphicx}
\usepackage{cite}
\usepackage{amsmath}
\usepackage{amssymb}
\usepackage{mathtools}
\usepackage{tensor}
\usepackage{amsthm}
\usepackage{mathrsfs}
\usepackage[greek, english]{babel} 
\usepackage{teubner}  
\usepackage{enumitem} 
\usepackage{ragged2e} 
\usepackage{bbm} 

\usepackage[usenames, dvipsnames]{color} 
\definecolor{darkblue}{rgb}{0.2, 0, 0.8}
\definecolor{darkgreen}{rgb}{0.2, 0.71, 0}  
\definecolor{LinkGreen}{cmyk}{1,0.05,.95,0.5}
\definecolor{BrightLinks}{cmyk}{.9,0,1,0.0}

\renewcommand{\baselinestretch}{2}
\renewcommand\thesubsection{\thesection.\alph{subsection}} 
\renewcommand\thesubsubsection{\thesubsection.\roman{subsubsection}}

\newcommand{\dd}{\mathrm{d}}
\newcommand{\lie}{\pounds}
\newcommand{\op}{\mathcal{O}}
\newcommand{\vep}{\varepsilon}
\newcommand{\tr}{\operatorname{Tr}}
\newcommand{\dt}{\frac{d}{2}}
\renewcommand{\D}{\Delta}
\newcommand{\pd}{\mathcal{D}}

\newcommand{\eul}{\gamma_E}
\newcommand{\vev}[1]{\langle #1 \rangle}
\newcommand{\fvev}[1]{\left\langle #1 \right \rangle}
\newcommand{\mvev}[2]{#1\langle #2 #1\rangle}
\newcommand{\logp}[1]{\log\left( #1 \right)}

\newcommand{\beq}{\begin{equation}}
\newcommand{\eeq}{\end{equation}}

\newtheorem*{hypothesis*}{Hypothesis}
 
\newcommand{\req}[1]{(\ref{#1})} 

\newcommand{\Hi}{\mathcal{H}}
\newcommand{\SSS}{\mathcal{S}}
\newcommand{\PP}{\mathcal{P}}
\newcommand{\Ph}{\Phi}
\newcommand{\del}{\delta}
\newcommand{\al}{\alpha}
\newcommand{\dar}{\rchi}
    \DeclareRobustCommand{\rchi}{{\mathpalette\irchi\relax}}
    \newcommand{\irchi}[2]{\raisebox{\depth}{$#1\chi$}} 
\newcommand{\dy}{{\dar_Y}}
\newcommand{\hdy}{{\hat\dar_Y}}
\newcommand{\Om}{\Omega}
\newcommand{\dx}{{\dar}}
\newcommand{\hdx}{{\hat\dar}}
\newcommand{\Th}{\Theta}
\newcommand{\ep}{\epsilon}
\newcommand{\qo}{\text{\textqoppa}}
\newcommand{\ddiff}{{\text{Dir}_{\partial\Sigma} } } 

\begin{document}
\pagestyle{empty}

\hbox{\ }

\renewcommand{\baselinestretch}{1}
\small \normalsize

\begin{center}
\large{{ABSTRACT}}

\vspace{3em}

\end{center}
\hspace{-.15in}
\begin{tabular}{ll}
Title of dissertation:    & {\large  INVESTIGATIONS ON ENTANGLEMENT}\\
&                     {\large      ENTROPY IN GRAVITY } \\
\ \\
&                          {\large Antony John Speranza} \\
&                           {\large Doctor of Philosophy, 2018} \\
\ \\
Dissertation directed by: & {\large  Professor Theodore Jacobson} \\
&               {\large  Department of Physics } \\
\end{tabular}

\vspace{3em}

\renewcommand{\baselinestretch}{2}
\large \normalsize

Entanglement entropy first arose from attempts to understand 
the entropy of black holes, and is believed to play a crucial role in
a complete  description of quantum gravity.  
This thesis explores
some proposed connections between entanglement entropy and the geometry of spacetime.
One such connection is the ability to derive  gravitational field equations 
from entanglement  identities.  I will discuss a specific derivation of the Einstein
equation from an equilibrium condition satisfied by entanglement entropy, and explore 
a subtlety in the construction when the matter fields are not conformally invariant.  As a further
generalization, I extend the argument to include higher curvature theories of gravity, whose 
consideration is necessitated by the presence of subleading divergences in the entanglement
entropy beyond the area law.  

A deeper issue in this construction, as well as in more general considerations identifying
black hole entropy with entanglement entropy, is that the  entropy is ambiguous 
for gauge fields and gravitons.  The ambiguity
stems from how one handles edge modes at the entangling surface, which parameterize 
the gauge transformations that are broken by the presence of the boundary.  The final part
of this thesis is devoted to identifying the edge modes in arbitrary diffeomorphism-invariant
theories.  
Edge modes are conjectured to provide a statistical description of the black hole entropy,
and this work takes some initial 
steps toward checking this conjecture in  higher curvature theories.


\thispagestyle{empty}
\hbox{\ }
\vspace{1in}
\renewcommand{\baselinestretch}{1}
\small\normalsize
\begin{center}

\large{{INVESTIGATIONS ON ENTANGLEMENT  \\ ENTROPY IN GRAVITY}}\\
\ \\
\ \\
\large{by} \\
\ \\
\large{Antony John Speranza}
\ \\
\ \\
\ \\
\ \\
\normalsize
Dissertation submitted to the Faculty of the Graduate School of the \\
University of Maryland, College Park in partial fulfillment \\
of the requirements for the degree of \\
Doctor of Philosophy \\
2018
\end{center}

\vspace{7.5em}

\noindent Advisory Committee: \\
Professor Theodore Jacobson, Chair/Advisor \\
Professor Raman Sundrum \\
Professor Bei-Lok Hu \\
Professor Brian Swingle \\
Professor Jonathan Rosenberg

\thispagestyle{empty}
\hbox{\ }

\vfill
\renewcommand{\baselinestretch}{1}
\small\normalsize

\vspace{-.65in}

\begin{center}
\large{\copyright \hbox{ }Copyright by\\
Antony John Speranza  
\\
2018}
\end{center}

\vfill

\newpage

\hbox{\ }
\newpage

\pagestyle{plain} \pagenumbering{roman} \setcounter{page}{2}
\addcontentsline{toc}{chapter}{Acknowledgements}

\renewcommand{\baselinestretch}{2}
\small\normalsize
\hbox{\ }
 
\vspace{-.65in}

\begin{center}
\large{Acknowledgments} 
\end{center} 

\vspace{1ex}

I would like to thank Anton de la Fuente, 
Will Donnelly, Tom Faulkner, Laurent Freidel, Dami\'{a}n Galante, Sungwoo Hong, 
Arif Mohd, Rob Myers, Vladimir Rosenhaus, 
Jos\'{e} Senovilla, Raman Sundrum, and Aron Wall for helpful discussions related 
to various aspects of this thesis.
I'm also grateful to Pablo Bueno, Vincent Min, and Manus Visser
with whom I collaborated on
the work that comprises chapter \ref{ch:EEhigher}.
I acknowledge support from Monroe H. Martin Graduate Research Fellowship.  

I thank Steve Ragole, Gina Quan, and Kevin Palm for their friendship during my 
time in College Park.  

I am incredibly grateful to my advisor, Ted Jacobson,
for his support, encouragement, advice, and countless hours of discussions about physics.
I am truly lucky to have had the opportunity to work with him over the past several years.  

Finally, 
a huge debt of gratitude is owed to my parents, 
Dominick and Ellen, and family members, Rosa, Emily, Dominick, Rebecca,
and Sebastian, for all they have done for me.  


\renewcommand{\baselinestretch}{1}
\small\normalsize
\tableofcontents 
\newpage

\listoffigures 
\newpage

\newpage
\setlength{\parskip}{0em}
\renewcommand{\baselinestretch}{2}
\small\normalsize

\setcounter{page}{1}
\pagenumbering{arabic}

\chapter{Entanglement entropy in quantum gravity}

\section{Gravity as a regulator}
\label{sec:grreg}
The notion of quantizing the gravitational field is nearly as old as general relativity itself.  In Einstein's
1916 paper on gravitational waves, he remarked that 
electrons would be able to radiate gravitationally as well as electromagnetically, and
inferred from this that 
the arguments for quantizing the electromagnetic 
field 
applied
equally well to gravity \cite{Einstein1916}.
The following century saw a set of ideas emerge for how a 
theory of quantum gravity might look, applying lessons from 
the rapidly developing fields of quantum mechanics, quantum field theory, and 
classical general relativity (see \cite{Stachel1999, Carlip2015, Rovelli2000, DeWitt1967b} for
historical reviews).  By the late 1960's, DeWitt had formulated the perturbative theory in terms
of interacting gravitons \cite{DeWitt1964a, DeWitt1967c, DeWitt1967d}, and in particular
had shown that the theory was not renormalizable in the power-counting sense of Dyson
\cite{Dyson1949}.  The divergent structure of pure general relativity proved to have 
better ultraviolet behavior than naive power-counting would suggest, being 
one-loop finite in four spacetime dimensions \cite{tHooft1974}; even so, it was 
eventually shown to diverge at two loops  \cite{Goroff1985}, dashing
any prospects for a perturbatively renomalizable theory of quantum gravity
(although there remains some hope that its maximal supersymmetric extension in four
dimensions may yet turn out to be perturbatively finite \cite{Bern2007}).

From one perspective, nonrenormalizability seems to doom the perturbative theory
as lacking any predictive power; however, this is overly pessimistic.  
The modern interpretation \cite{Weinberg1979}  
treats perturbative quantum gravity as an effective field theory, valid
at energies small compared to some high energy scale \cite{Gomis1996a, Burgess2004a, 
Donoghue2012a}.  The cutoff for the effective theory could be taken to be the Planck scale, at which
gravity becomes strongly coupled, or it may be a lower scale where additional 
physical degrees of freedom
become important.  The effective description allows one to be agnostic about the precise value
of the cutoff or the details of the UV completion, and becomes predictive after a finite number 
of renormalized couplings are fixed experimentally.  
This approach leads to some unambiguous results in quantum gravity, such as corrections
to the Newtonian potential \cite{Donoghue1994, Bjerrum-Bohr2003}, 
and can also be usefully applied to 
classical post-Newtonian calculations
\cite{Goldberger2006a}.

Beyond the realm of effective field theory, there has long been a hope that 
nonrenormalizability is only relic of perturbation theory, and that when the full
nonlinear structure of general relativity is taken into account, the quantum theory is 
UV finite.  
This expectation extends to theories of matter coupled to 
gravity, so that if it is true, gravity takes on the 
privileged role of a universal regulator for the 
divergences of quantum field theory. 
Such a radical  statement of UV finiteness
might only be considered possible in the presence of  
a powerful symmetry principle.  Fortunately, in 
gravitational theories, a candidate symmetry is available: invariance under the diffeomorphism
group of a manifold.  

It was noted early on by Bergmann that this symmetry imbues the theory with certain 
holographic 
properties,\footnote{\renewcommand{\baselinestretch}{1}\footnotesize Although, the 
term ``holographic''  would not be applied to gravity until
much later
\cite{tHooft:1993dmi, Susskind1995a}. } 
namely that the energy-momentum contained within a subregion 
can be represented in terms of a boundary surface integral  \cite{Bergmann1949,
Bergmann1949a}.  This property
had previously been applied by Einstein, Infeld, and Hoffmann to show that the classical
 gravitational vacuum field equations  fully determine the motion of point particle
singularities \cite{Einstein1938a},\footnote{\renewcommand{\baselinestretch}{1}\footnotesize In
reality, such singularities are not pointlike, since they actually represent black holes with 
finite area in the classical theory.  They are handled in the Einstein, Infeld, Hoffman
work by cutting off the solution at a finite radius larger than the horizon area, and evaluating the 
energy and momentum of the particle through an integral over the cutoff surface.  The 
Einstein equations then constrain the evolution of the energy and momentum associated
with the surface integral.}
completely avoiding the difficulties encountered in classical
electromagnetism in which point particles require an infinite mass subtraction to compensate
the 
divergent self-energy \cite{Dirac, Infeld1940}.  
Bergmann believed that this holographic property of gravity would persist in the quantum
theory, and suspected that it might help alleviate the infinities encountered in the 
renormalization of quantum field theories \cite{Bergmann1956}.  
The idea he seemed to have in mind was that one could regulate the short distance interactions
leading to the divergence, and then try to argue that gravity nonperturbatively determines the 
behavior of the correlation function in the UV, analogous to how it determined the point particle 
motion in the classical theory.  The hope was that in the limit that the regulator is taken to zero,
the final answer would be finite, rather than divergent, and exhibit an effective cutoff near
the Planck scale.  While this program was never fully brought to fruition, various aspects of 
this proposal have appeared in several investigations of classical and quantum gravity
\cite{DeWitt1979}.

A particularly lucid example due to Arnowitt, Deser, and 
Misner serves to illustrate the general features of such a gravitational regularization,
albeit for the classical theory
\cite{Arnowitt1960a, Arnowitt1960, Arnowitt1960b}.  They consider a charged 
shell of radius $R$, total charge $e$,  and bare mass $m_0$.
In the Newtonian limit, the total mass 
is given by the sum of the bare mass, the  energy stored in the 
Coulomb field, and the energy in the Newtonian gravitational field, which is negative on account of 
gravity's universal attractiveness,
\beq \label{eqn:mtotNewt}
m_\text{tot} = m_0 + \frac{e^2/4\pi}{2R} -\frac{G m_0^2}{2R}.
\eeq
Absent a precise tuning between the charge and bare mass, 
this clearly diverges in the point particle limit $R\rightarrow 0$.  However, in general relativity,
the electric and gravitational fields are themselves sources of gravity, which suggests 
the total mass $m_\text{tot}$ should appear in the term involving the gravitational energy,
\beq
m_\text{tot} = m_0 + \frac{e^2/4\pi}{2R} - \frac{G m_\text{tot}^2}{2R}.
\eeq
Solving for the total mass gives
\beq \label{eqn:mtot}
m_\text{tot} = \frac{R}{G} \left(-1+\sqrt{1+\frac{2G}{R}\left(m_0+\frac{e^2/4\pi}{2R}\right)} \;\right)
\overset{R\rightarrow 0}{\longrightarrow} \frac{\sqrt{e^2/4\pi}}{\sqrt{G}}.
\eeq
Although this argument was heuristic, it can be made rigorous by solving the Einstein-Maxwell
equations exactly and computing the ADM mass \cite{Arnowitt1960a, Arnowitt1960}, and the 
result coincides with (\ref{eqn:mtot}).  One can recognize the $R\rightarrow 0$ 
mass as that of an extremal Reissner-Nordstr\"om black hole with charge $e$. 

This result is quite remarkable.  The renormalized mass is finite, 
and diverges with weakening strength of 
the gravitational interaction, $G\rightarrow 0$, verifying that gravity is responsible for taming
the divergent self-energy of the charge.  Furthermore, the nonlinearity of gravitational
interactions plays an essential role, since the linear Newtonian result (\ref{eqn:mtotNewt}) 
does not produce a finite renormalized mass, except in the case of a precisely tuned bare 
mass.\footnote{Amusingly,  the required tuning is that
the bare mass be equal to the $R\rightarrow 0$ limit of (\ref{eqn:mtot}). 
Gravity has naturally provided the necessary ``counterterm.'' 
\renewcommand{\baselinestretch}{1}\footnotesize  } It is also worth noting the that since the $R\rightarrow 0$ 
mass is
proportional
to $G^{-1/2}$, an attempt to compute it perturbatively in integer powers of $G$ would lead
to a divergent result at any finite order in perturbation theory.  
The finiteness exhibited in the point particle mass can be related to the holographic 
nature of gravity.  The mass is determined by an integral of the fields well-separated from the 
point particle, which is finite since the point particle defines a regular solution to the field equations.
Hence, although the electromagnetic fields tend to give divergent energy density near the point
particle,\footnote{\renewcommand{\baselinestretch}{1}\footnotesize Since the solution is 
extremal Reissner-Nordstr\"om, the point particle is replaced by a black hole throat with
nonzero radius, at which the electric field remains finite.  However, the throat becomes 
infinitely long in the extremal limit, and the integral of the electromagnetic energy density
up to the horizon is still divergent due to this infinite volume.} 
the gravitational field is required to provide compensating negative energy density to 
keep the total energy finite; this is simply the negative energy density in the Newtonian potential.
A more detailed analysis of this example is given in \cite{Ashtekar1991a, Woodard2009a},
and especially \cite{Woodard1984}.

The arguments so far have focused on the classical regulating effects of gravity,
but there exist various cases where these improvements occur in  quantum theories
as well.  One set of results performs partial resummations of the graviton loop expansion, which
lead
to nondivergent expressions \cite{DeWitt1964, Isham1971, Isham1972, Woodard2002, Mora2012},
although this may not be special to gravitational theories, since similar resummations
have been carried out in other nonrenormalizable theories \cite{Lee1962, Feinberg1963,
FRANK1971}.
Other results 
involve the idea that graviton fluctuations smear out the lightcone, and 
hence soften divergences along lightlike directions in the propagators for quantum fields
\cite{Klein1956, Pauli1956, Deser1957, DeWitt1981, Ford1995}.  
In fact, at distances short compared to the Planck length, large fluctuations in geometry and 
topology are unsuppressed, suggesting the smooth manifold picture of spacetime degenerates 
into a sort of  topological foam \cite{Wheeler1964, Hawking1978}.
This would imply a complete breakdown of the usual notion of a continuum quantum field
theory, which was essential to producing divergences in the first place.  

\section{Black hole entropy}
Perhaps the most compelling evidence for gravity's UV finiteness comes from
the physics of black
holes.  
Based on thought experiments in which the entropy of the universe is decreased by 
sending packets of thermal matter 
into a black hole, Bekenstein conjectured that black holes must possess an
intrinsic entropy in order to preserve the second law of thermodynamics \cite{Bekenstein1972,
Bekenstein1973a}.  He further reasoned that the black hole entropy should be proportional to 
the area $A$ of its event horizon, in light of the findings that, assuming the null energy
condition, no process can decrease this area
\cite{Penrose1971, Christodoulou1970, Christodoulou1971, Hawking1971, Hawking1972}. 
This led to his formula for black hole entropy,
\beq \label{eqn:Sbh}
S_{\text{b.h.}} = \eta \frac{A}{G},
\eeq
where $\eta$ is a dimensionless constant of order unity, and the factor of $1/G$ is fixed on 
dimensional grounds ($\hbar=c=1$ unless  otherwise stated).  

Determining the precise value of $\eta$ would seem to require a complete knowledge of the 
quantum gravity theory, including an accounting of all the black hole microstates.  
Surprisingly, no such detailed description is required, and one can 
determine $\eta = 1/4$ by combining two important results.  
One is the first law of black hole mechanics,
which states that small changes in the area and angular momentum
$J$  of a stationary black hole are 
related to the change in its mass through the equation
\beq \label{eqn:flbhm}
\delta M = \frac{\kappa}{8\pi G} \delta A + \Omega_H \delta J,
\eeq
where $\kappa$ is the surface gravity of the black hole horizon and $\Omega_H$ its angular velocity
\cite{Smarr1973, Bekenstein1973a, Bardeen1973}.  
The relation bears an obvious resemblance to the first law of thermodynamics,
by identifying $M$ with the internal energy, and $\frac{\kappa}{8 \pi \eta}$ with the temperature,
given Bekenstein's entropy formula (\ref{eqn:Sbh}).  The term $\Omega_H \delta J$ 
is analogous to 
a
chemical potential $\mu dN$ term from thermodynamics \cite{Gibbons1977a}.
The other key result is Hawking's stunning discovery that quantum fields propagating in a 
black hole spacetime radiate thermally, at a temperature $T = \kappa/2\pi$ \cite{Hawking1974,
Hawking1975}.   Equating this 
temperature with the one obtained from the first law (\ref{eqn:flbhm}) 
fixes $\eta = 1/4$, and gives the Bekenstein-Hawking formula for black hole entropy,
\beq \label{eqn:SBH}
S_\text{BH} = \frac{A}{4 G}.
\eeq

One puzzling aspect of this result is that it seems almost independent of any quantum
aspects of gravity.  Hawking's calculation is quantum mechanical, but involves quantum field
theory on a nondynamical background spacetime.  There is no mention of wildly  fluctuating
geometry at Planckian length scales, smeared lightcones, or other quantum gravitational
phenomena, and the resulting entropy is highly robust 
and derivable using a variety of disparate methods \cite{Carlip2014}.  
It does, however, incorporate a crucial, nonperturbative gravitational effect in applying the
first law of black hole mechanics, (\ref{eqn:flbhm}).  This formula is a statement of 
gravity's holographic nature, since it relates the mass and angular momentum, which are 
 boundary integrals at infinity, to the horizon area, which is a property of a
surface in the interior.  Furthermore, it is a direct consequence of diffeomorphism invariance,
and analogous relations can be derived for any diffeomorphism-invariant theory \cite{Wald1993a,
Iyer1994a}.  One therefore might view the 
Bekenstein-Hawking  formula (\ref{eqn:SBH}), as well as Wald's generalization \cite{Wald1993a},
 as the only entropies
consistent with Hawking's calculation that also incorporate diffeomorphism invariance 
in the gravitational theory.

Returning to Bekenstein's original motivation, the resolution of 
the entropy loss conundrum
is that the second law of thermodynamics applies to the {\it total} generalized entropy
of the universe, which consists of both black hole entropy and the entropy of matter outside the 
horizon,
\beq\label{eqn:Sgenintro}
S_\text{gen} = \frac{A}{4G} + S_\text{out}.
\eeq
While offering a means to salvage the second law from the entropy-reducing machinations
of black holes, 
the Bekenstein-Hawking
and generalized entropies (\ref{eqn:SBH}), (\ref{eqn:Sgenintro})
introduce a number of new puzzles.  
The first concerns the statistical interpretation of $S_\text{BH}$.  Being
proportional to the area in Planck units, it suggests a picture of quantum gravitational degrees of 
freedom confined to a membrane at the horizon, holographically accounting for the physics in
the black hole interior \cite{tHooft:1993dmi}.  
A second puzzle is how to give a precise definition to the 
outside matter entropy, $S_\text{out}$.  
Given Hawking's semiclassical analysis, one might
expect this entropy to be related to the von Neumann entropy of the quantum 
fields restricted to the black hole exterior
that participate in Hawking radiation. 

\section{Generalized entropy as entanglement entropy} \label{sec:SgenSEE}

Underlying both of these puzzles is a deeper question: why are these entropies 
finite? 
The hypothetical membrane theory on the black hole horizon must not be a continuum 
field theory, with an infinitude of states, but rather should be discrete at Planckian length scales to 
give the correct value for $S_\text{BH}$.  The field theory definition for $S_\text{out}$ also 
presents a problem.  Continuum quantum fields propagating on the black hole spacetime
are highly entangled between spatial regions in low energy states.  A sharp restriction of the 
quantum state to the black hole exterior produces a divergent von Neumann entropy, due to 
infinitely many degrees of freedom entangled at arbitrarily short distances across the black 
hole horizon.  
While this  threatens to deprive the generalized entropy of any useful
meaning, a more detailed analysis of the divergence reveals possible resolutions to many of the 
above issues.  

The process of tracing out degrees of freedom in a spatial subregion $\bar\Sigma$ 
produces a mixed reduced
density matrix $\rho_\Sigma$, and its von Neumann entropy is known as the entanglement entropy,
\beq \label{eqn:trrlogr}
S_\text{EE} = -\tr \rho_\Sigma\log\rho_\Sigma.
\eeq  
This construction was originally introduced in order to understand aspects of black hole entropy
\cite{Sorkin:2014kta, Bombelli1986, tHooft1985, Srednicki1993a, Frolov1993}, 
although it has since found important applications in a variety of other areas of physics 
\cite{Amico2008, Eisert2010, Horodecki2009, VanRaamsdonk2017, Calabrese2009, Casini2017}.
In quantum field theories, this entropy is UV divergent, but upon regularization, it takes the 
form
\beq\label{eqn:SEEintro}
S_\text{EE} = c_0 \frac{A}{\ep^{d-2}} +\{ \text{subleading divergences} \} + S_\text{finite},
\eeq
where $\epsilon$ is a short-distance cutoff, $c_0$ is some dimensionless parameter
that in general depends on the regularization scheme, and $d$ is the spacetime dimension.  
The similarity of (\ref{eqn:SEEintro}) to 
the generalized entropy (\ref{eqn:Sgenintro}) is immediately apparent.  Identifying $c_0/\ep^{d-2}$
with $1/4G$ allows the generalized entropy to be attributed entirely to the entanglement
entropy.
The justification of this  invokes the universal regulating properties of gravity: 
it cuts off the infinitely many short distance degrees of freedom of the quantum 
fields at the Planck scale, producing a finite entanglement entropy whose leading term
matches the Bekenstein-Hawking entropy.  

One issue with this identification is that the coefficient $c_0$ appearing in the area term
of (\ref{eqn:SEEintro})
is not universal, and depends on the choice of regularization scheme \cite{Kabat:1994vj, 
Callan1994a, Susskind1994}.  However, this difficulty has a rather clever resolution.  The 
quantum fields responsible for the entanglement entropy divergence also 
produce divergences that renormalize $G$, and these divergences conspire to
ensure that the generalized entropy is independent of the choice of 
regulator \cite{Susskind1994}.  More explicitly, if $S_{\text{EE}}$ in (\ref{eqn:SEEintro})
is split in a regulator-dependent way into an area term $c_0 A/\ep^{d-2}$ and a finite 
piece $S_\text{finite}^{(\ep)}$ (ignoring the subleading divergences  for the moment), 
and if the same regularization scheme is used in the matter field loops that change
the bare Newton's constant $G_0$ to its renormalized value $G_\text{ren}^{(\ep)}$,
the following relationship holds
\begin{align} \label{eqn:Sgencancel}
S_\text{gen} = \frac{A}{4G_0} + c_0 \frac{A}{\ep^{d-2}} + S^{(\ep)}_\text{finite} = 
\frac{A}{4 G^{(\ep)}_\text{ren}} + S_\text{finite}^{(\ep)}.
\end{align}
This suggests that $S_\text{gen}$ is invariant under renormalization group flow.  
If, as has been argued above, gravity becomes strongly coupled near the Planck scale, 
it would make sense for the bare Newton constant to diverge there,
$G_0\rightarrow\infty$. 
This would lead to the conclusion  that  \cite{Jacobson1994a}
\beq \label{eqn:SgenSEE}
S_\text{gen} = S_\text{EE},
\eeq
with the entanglement entropy being rendered finite by the strong quantum gravitational effects
at the Planck scale.

The best way to demonstrate this miraculous cancellation of divergences  is through
 a technique for computing entanglement entropy known as 
the replica trick \cite{Susskind:1993ws, Susskind1994, Callan1994a, Holzhey:1994we}
(reviewed in \cite{Solodukhin2011a, Bousso2016}).  Using the path integral 
representation of the density matrix (see section \ref{sec:EEballs}), one can
show that (\ref{eqn:trrlogr}) is equivalent to an expression in terms of the gravitational
effective action $W(n) = -\log Z(n)$, ($Z(n)$ is the partition function), given by
\beq\label{eqn:Sreplica}
S_\text{EE} = (n \partial_n -1) W(n)\big|_{n=1}.
\eeq
The effective action is evaluated on a manifold with a conical singularity at the entangling surface, 
with an excess angle of $2\pi(n-1)$.  Some 
terms in $W(n)$ will take the form of local, diffeomorphism-invariant integrals over the manifold.
These are extracted
from the path integral in a saddle point approximation, and, crucially, include all UV-divergent
counterterms for the quantum fields.  They appear alongside the local terms coming from the 
saddle point approximation of the classical gravitational action, and hence have the effect of 
simply renormalizing the gravitational couplings.  In particular, one counterterm for the quantum
fields involves the Ricci scalar, and its divergent coefficient renormalizes $G_0$.  
When (\ref{eqn:Sreplica}) is evaluated for these local terms, the only contribution comes from
the entangling surface, and is given by the Wald entropy for the corresponding integrand
\cite{Iyer1995b, Nelson1994} (which, for the Ricci scalar, gives the area).  
From this perspective, the divergences
in the entaglement entropy and the counterterms for the gravitational couplings have a common
origin in the gravitational effective action, demystifying the precise cancellation observed in 
equation (\ref{eqn:Sgencancel}). 

The above construction has the added bonus of providing an interpretation for the subleading
entanglement entropy divergences that appear in (\ref{eqn:SEEintro}).  These simply 
arise from the higher curvature counterterms that can appear in $W(n)$.  Such higher curvature
corrections arise generically in quantum gravity theories 
\cite{Burgess2004a}, in which case $S_\text{BH}$ is replaced by the Wald entropy
\cite{Wald1993a, Iyer1994a},
\beq
S_\text{gen} = S_\text{Wald} + S_\text{out}.
\eeq
The subleading divergences are then seen to simply correspond to the renormalization of the 
higher curvature gravitational couplings appearing in $S_\text{Wald}$, by the same argument
as before \cite{Solodukhin1995, Fursaev1996, Cooperman:2013iqr}.  

It is worth clarifying that the finiteness of $S_\text{gen}$ is the key nonperturbative
effect in this discussion.  The dominant contribution comes from $S_\text{BH}$, which,
being proportional to $1/G$, is similarly nonperturbative.
This dependence on $G$ is calculated using nonperturbative techniques, namely 
the replica trick and the saddle point approximation to the effective action.  
Note that similar to the ADM example of section \ref{sec:grreg}, 
turning off gravity by sending
$G\rightarrow 0$ 
causes the generalized entropy to diverge.  In light of equation (\ref{eqn:SgenSEE}), 
this divergence is just the familiar fact that entanglement entropy is infinite for continuum 
(non-gravitational) quantum field theories.  This makes apparent 
an important relationship between entanglement and gravity, namely that larger entanglement
is associated with weaker gravitational interactions, i.e.\ entanglement screens Newton's constant.  

Note  one perturbative aspect of 
the above discussion is that the divergences that renormalize $G$ can be computed
perturbatively in a loop expansion.  
This is justified from the effective field theory point of view \cite{Burgess2004a},
and hence assumes a cutoff that is well-separated from the Planck scale.  
The renormalization-group-invariance of $S_\text{gen}$ has therefore been demonstrated by 
the above arguments only within the regime of validity of the effective theory, and 
extending invariance and finiteness to the Planck scale involves some amount of extrapolation.  
One consequence of this effective field theory viewpoint is that the splitting of the $S_\text{gen}$
into $S_\text{BH}$ and $S_\text{out}$ as in (\ref{eqn:Sgencancel}) depends on the cutoff 
for the effective description, and changing the cutoff causes entropy to shift between the 
two terms \cite{Kabat1995a, Jacobson:2012ek}.  
Furthermore, this leads to the interesting 
viewpoint that many theorems of classical general relativity can be extended to the semiclassical
regime simply by replacing areas of surfaces with the RG-invariant generalization, $S_\text{gen}$.  
Bekenstein's generalized second law \cite{Bekenstein1973a} (proved in \cite{Wall2012}) 
is thus interpreted as a semiclassical
improvement of Hawking's area theorem \cite{Hawking1972}, and 
similar generalizations include \cite{Wall2013, Engelhardt2015, Bousso2016}.
Pushing these results to their ultimate conclusion suggests 
that spacetime geometry may be viewed as fundamentally 
reflecting the entanglement structure of the underlying theory
\cite{VanRaamsdonk2010}.  

\section{Examples where $S_\text{gen}=S_\text{EE}$}
The identification of black hole entropy with entanglement entropy may seem like a radical
proposal at first, but luckily it can be checked in situations 
where a UV completion for gravity is known in some detail.  
One such example comes from the AdS/CFT correspondence
\cite{Maldacena1999a, Aharony2000a},
in which a quantum gravity theory in anti-de Sitter (AdS) space has a dual description 
in terms of a non-gravitational conformal field theory residing at the conformal 
boundary.  
The bulk theory admits spherically symmetric AdS-Schwarzschild black hole solutions, 
whose entropy can be understood from the perspective of the CFT \cite{Witten1998a, 
Maldacena2003a, Harlow2016a}.  
A thermal state at temperature $\beta^{-1}$
in the CFT can be represented as a pure state on two copies of 
the CFT, $L$ and $R$, with a specific entanglement structure, 
\beq \label{eqn:PsiTFD}
|\Psi_\text{TFD}\rangle = \frac1{Z(\beta)} \sum_n e^{-\beta E_n/2} |n\rangle_L 
\otimes |n\rangle_R.
\eeq
This state can be prepared using a Euclidean path integral, and this maps 
via the holographic dictionary \cite{Witten:1998qj, Gubser1998} to 
a Hartle-Hawking path integral
in the bulk, which prepares an AdS-Schwarzschild black hole with a Hawking 
temperature matching the CFT 
\cite{Barvinsky1995, Jacobson1994c}.  
Tracing out the left CFT in (\ref{eqn:PsiTFD}) 
produces a mixed thermal state on the right CFT, whose entropy is given by the 
Bekenstein-Hawking entropy of the dual black hole.  This leads to the conclusion that 
$S_\text{BH}$ is precisely the entanglement entropy between the left and right CFTs.

The identification of CFT entanglement entropy with areas of bulk surfaces 
occurs in much more general contexts in AdS/CFT.  This is due to the 
Ryu-Takayanagi (RT) formula \cite{Ryu:2006bv, Ryu:2006ef}, which states that the entanglement
entropy of a subregion in the CFT is equal to the Bekenstein-Hawking formula, applied 
to a minimal surface in the bulk which asymptotes to the boundary subregion,
\beq
S_\text{EE} = \frac{A_\text{min}}{4 G}.
\eeq
The application of this formula to the AdS-Schwarzschild example above immediately
reproduces the black hole entropy, since the horizon is the minimal area surface in the throat
of the wormhole separating the two asymptotic boundaries.  The RT formula can be used to 
demonstrate the equality of black hole entropy and entanglement entropy in other contexts
as well, such as 
Randall-Sundrum models \cite{Randall1999} of induced gravity \cite{Emparan2006}.  
Additional examples demonstrating
the equality are reviewed in \cite{Solodukhin2011a}.

\section{Gravitational dynamics from entanglement} \label{sec:enteq}

When viewed as entanglement entropy, 
it is clear that a generalized entropy can be assigned to surfaces other than black hole
horizon cross sections 
\cite{Jacobson1999, Jacobson2003a, 
Bianchi2012a, Bousso2016}.  
This is borne out explicitly in AdS/CFT, where the quantum-corrected RT formula 
\cite{Faulkner2013a}
maps the generalized entropy of minimal-area surfaces to the entanglement entropy in the CFT.
Even without assuming  holographic duality, 
the arguments of section \ref{sec:SgenSEE} strongly suggest that 
generalized entropy gives a  UV finite quantity that is naturally associated with both
entanglement entropy and the geometry of surfaces, providing a vital
link between
the two.  
When supplemented with thermodynamic 
information, this link can in fact reproduce the dynamical equations for gravity.  
The first demonstration of this was Jacobson's derivation of the Einstein equation as an 
equation of state for local causal horizons possessing an entropy proportional to their area 
\cite{Jacobson1995a}.  Subsequent work using entropic arguments 
\cite{Verlinde2010,Verlinde2016} and holographic entanglement entropy 
\cite{Lashkari2013, Faulkner:2013ica,Swingle2014, Faulkner2017, Haehl2017, Lewkowycz2018}
  confirmed that entanglement thermodynamics is connected to 
gravitational dynamics.  A review of some of these approaches is given in
section \ref{sec:comparison}.

Chapters \ref{ch:EEinCPT} and \ref{ch:EEhigher} of this thesis are devoted to studying a particular 
 approach to deriving geometry from entanglement, which is 
Jacobson's {\it entanglement equilibrium} argument \cite{Jacobson2015a}.  
This proposal begins with  a geometrical identity similar to the first law of black hole mechanics
(\ref{eqn:flbhm})
but applicable to spherical ball-shaped regions in maximally symmetric
spaces (MSS), as opposed to black hole horizons.  This {\it first law of causal diamond
mechanics} reads 
\beq\label{eqn:flcd}
\frac{\kappa}{8\pi G} {\delta A}\big|_V+ \delta H_\text{matter} = 0,
\eeq
where $H_\text{matter}$ is the matter energy associated with translation along a conformal
Killing vector that preserves the causal diamond, and $\kappa$ is the 
surface gravity of this conformal Killing vector \cite{Jacobson1993}.  
The radius of the ball must be adjusted 
when taking the variation in such a way that the total volume of the ball is held fixed,
which is indicated by $\delta A\big|_V$ in this equation.  This relation holds when the 
Einstein equation is satisfied; when working off-shell, the right hand side of (\ref{eqn:flcd})
is proportional to the constraint equation of general relativity, integrated over the interior
of the ball.  The argument then proceeds by interpreting the terms on the left hand side 
of (\ref{eqn:flcd}) in terms of a variation of 
the generalized entropy of the state restricted to the ball
interior.  The area term is associated with the Bekenstein-Hawking entropy of the surface, and 
$\delta H_\text{matter}$ can be associated with a variation of the (renormalized)
entanglement entropy of the matter fields within the ball using the {\it first law of 
entanglment entropy} \cite{Bhattacharya:2012mi, Blanco2013a}.  Then, applying the equality of 
generalized entropy and entanglement entropy, (\ref{eqn:flcd}) states that 
\beq\label{eqn:dSEEtotintro}
\frac{\kappa}{2\pi} \delta S_\text{EE}^\text{total}\big|_V = \int_\Sigma \delta C_\zeta  =0,
\eeq
where $S_\text{EE}^\text{total}$ is the total entanglement entropy, including the area law divergence, 
and the integral is over the ball $\Sigma$ of a component of the linearized Einstein equation,
see equation (\ref{eqn:constreq}).

This equation  states that maximizing\footnote{Or more precisely, extremizing.
Showing maximality would require consideration of second order perturbations. 
\renewcommand{\baselinestretch}{1}\footnotesize  }
the entanglement entropy of a fixed-volume subregion is equivalent to imposing the
linearized Einstein
equation.
  One can therefore derive the Einstein equation by assuming that entanglement
entropy is maximized at fixed volume.  This is the origin of the name ``entanglement equilibrium,''
because equilibrium states are ones of maximal entropy.  
Although the above setup applies to linearized perturbations to maximally
symmetric spaces, it has implications for a much wider class of spacetimes.  The reason is 
that any smooth spacetime looks flat on small enough scales, so that the entanglement equilibrium
argument can be applied locally to each point in a spacetime.  The small ball limit has the 
added advantage that the metric perturbation can be chosen to coincide with the first corrections
to the locally flat metric by employing Riemann normal coordinates (RNC).  The RNC expansion
parameter is $r/R_c$, where $r$ is the ball radius, and $R_c$ this local radius of curvature, and
so it can be made arbitrarily accurate as $r$ is taken to zero.  Furthermore, the metric
perturbation depends on the fully nonlinear Riemann tensor evaluated at the center of the ball,
so one finds that the linearized equations applied in the small ball limit actually require that 
the nonlinear Einstein equation holds at the center of the ball.   

Implicit in the derivation of equation (\ref{eqn:dSEEtotintro}) is that the matter fields coupled
to gravity are conformally-invariant.  While this is clearly not true in general, 
it should be approximately true in the small ball limit in which the matter should flow to its 
conformal fixed point.  However, one still must check whether all aspects of the entanglement
equilibrium argument hold to a good enough approximation in this limit to conclude
the Einstein equation holds.  This is the subject of chapter 
\ref{ch:EEinCPT}, where
explicit calculations of entanglement entropies are made for excited states  in
non-conformal field theories.  It is found that for certain classes of states, the 
matter entanglement entropy is {\it not} sufficiently well-approximated by the conformal
boost Hamiltonian to apply the entanglement equilibrium argument in its present form.
One modification, suggested in \cite{Jacobson2015a} and elaborated on 
in section \ref{sec:EE=EE}, is to allow for a local 
cosmological constant to absorb the extra term in the entanglement entropy coming from
the non-conformality of the matter.  Other possible resolutions of this issue are 
discussed in section \ref{sec:EEimps}.

A natural generalization of the entanglement equilibrium argument is to apply it to higher 
curvature theories of gravity, which is the topic of chapter \ref{ch:EEhigher}.  
As mentioned in section \ref{sec:SgenSEE}, these higher curvature
corrections are naturally associated with the subleading divergences in the entanglement entropy.
Hence, whenever such subleading divergences are present (such as in $d=4$, when 
there are logarithmic divergences in addtional to the leading area term), the entanglement
equilibrium argument should be modified to include higher curvature corrections.  
This requires a higher curvature generalization of the 
first law of causal diamonds, given in equation (\ref{titis}).  The area term generalizes 
straightforwardly to a Wald entropy, but there is a question of how to generalize
 the fixed-volume constraint.  As shown in section \ref{sec:localgeo}, 
 the appropriate functional to hold
fixed can be derived by applying the Iyer-Wald formalism \cite{Iyer1994a} 
to the conformal Killing vector of the ball,
and this leads to a generalized notion of volume for the ball.  
One difference in the higher curvature entanglement equilibrium argument is that the 
small ball limit is not as useful as it is for general relativity.  In particular, 
even after taking the small ball limit and employing Riemann normal coordinates,
one can only conclude the linearized higher curvature field equations hold from the 
entanglement equilibrium requirement, see section \ref{sec:equations}.  


\section{Edge modes} \label{sec:edgemodes}
The discussion up to this point has been reticent about how gauge fields factor into
the identification of black hole entropy with entanglement entropy.  This is a subtle point because
the definition of entanglement entropy of a subregion is ambiguous when gauge 
constraints are present.  The definition of entanglement entropy begins with the assumption
that the Hilbert space under consideration splits, $\Hi=\Hi_\Sigma\otimes \Hi_{\bar\Sigma}$,
into tensor factors $\Hi_\Sigma$ and $\Hi_{\bar\Sigma}$ 
associated with a subregion $\Sigma$ and its complement $\bar\Sigma$,
and the observables are assumed to exhibit a similar factorization.
In a theory with gauge symmetry, this factorization no longer occurs because the gauge
constraints relate observables on $\Hi_\Sigma$ to those on $\Hi_{\bar\Sigma}$.  
This nonfactorization then leads to an ambiguity when tracing out the $\bar\Sigma$ degrees
of freedom, which roughly corresponds to how one chooses to deal with nonlocal observables 
such as Wilson loops that are cut by the entangling surface.

On the other hand, the replica trick method for computing the entanglement entropy 
seems to give a definite answer, even when including gauge fields.  
A question arises in how to give a Hilbert space interpretation of entropy calculated by the 
replica trick, and in particular how to understand what 
choice the replica trick makes in factorizing 
the Hilbert space. 
The solution proposed by Donnelly and Wall for abelian gauge fields
\cite{Donnelly2012, Donnelly2015E, Donnelly2016W} is that 
the Hilbert space is extended by degrees of freedom living on the entangling surface, and 
these {\it edge modes} give an additional contribution to the entanglement entropy.  
This contribution is essential in matching the renormalization of Newton's constant to 
entanglement entropy divergences, so the edge modes play a key role 
in the interpretation of $S_\text{gen}$ as entanglement entropy.  As such, they are also relevant
for understanding how gauge fields and gravitons factor into the entanglement equilibrium 
program described in section \ref{sec:enteq}.

One can see more explicitly how the edge modes contribute to the entanglement entropy
by examining the form of the reduced density matrix in the extended phase space 
\cite{Donnelly2012a, Donnelly2014a}.  The 
edge modes are labeled by representations of the {\it surface symmetry algebra}, which
arises as a remnant of the gauge symmetry that was broken by the presence of the 
entangling surface.  Each representation defines a superselection sector for the fields in the bulk, 
and the density matrix is just a sum over these sectors,
\beq \label{eqn:rhoSig}
\rho_\Sigma = \sum_i p_i \rho^i_\Sigma\otimes \rho^i_\text{edge},
\eeq
where $p_i$ labels the probability of being in a given representation.  The fact that
the density matrix arose from a global state satisfying the gauge constraint allows one to conclude
that each edge mode density matrix must be maximally mixed in its representation $R_i$,
\beq
\rho_\text{edge}^i = \frac{\mathbbm{1}}{\dim R_i }.
\eeq
The entropy simply follows from  plugging the density matrix (\ref{eqn:rhoSig}) into 
the formula for the von Neumann entropy (\ref{eqn:trrlogr}), giving
\beq\label{eqn:S3terms}
S = \sum_i\left( p_i S_i - p_i \log p_i + p_i \log \dim R_i\right).
\eeq
The first term gives the expectation value of the bulk entropies associated with the $\rho_\Sigma^i$,
and the second term is the Shannon entropy associated with the uncertainty of being 
in a given superselection sector.  This Shannon term is responsible for the additional entropy 
that appears for the abelian gauge field.  The final term is special to nonabelian theories
(since all representations of an abelian surface symmetry algebra are one dimensional), 
and represents entanglement between the edge modes themselves.  

This ``$\log\dim R$'' term takes the form of an expectation value of some operator at the 
entangling surface, and in the case of gravity, there is a proposal that this operator
simply gives the Bekenstein-Hawking contribution to the generalized entropy
\cite{Donnelly2016F, Harlow2016, Lin2017}.  This is necessarily a regulator-dependent statement,
since the splitting of the generalized entropy into $S_\text{BH}$ and $S_\text{out}$ depends on 
the cutoff for the effective description.  One should therefore expect the separation of the entropy
into three
distinct types of terms in (\ref{eqn:S3terms}) to similarly depend on the regulator.  
In a certain sense, it does not even make sense to consider the last two terms of (\ref{eqn:S3terms})
separately in the gravitational case.  This is because the surface symmetry group 
for gravity is non-compact, which means its representations are infinite-dimensional and labeled
by continuous parameters, as opposed to being finite-dimensional and discrete.  The density
matrix would then take the form of a direct integral over all possible representations, and $p_i$ and 
$\dim R_i$ would generalize to measures on the space of representations.  However, because
they are measures, their logarithm is not invariant under reparameterization of the space of 
representations. On the other hand, the combination $\log\dim R_i - \log p_i = \log(\dim R_i / p_i)$ 
that appears in (\ref{eqn:S3terms}) is reparameterization-invariant, suggesting that these two
terms should be considered together.\footnote{\renewcommand{\baselinestretch}{1}\footnotesize I 
thank Will Donnelly for discussion of this point.}  We should also expect that 
the operator corresponding
to $\dim R$ will depend on the parameters in the gravitational action, and hence 
the regulator-dependence of these parameters will produce regulator-dependent operators,
so that changing the regulator will cause entropy to shift between the first  and final two 
terms in (\ref{eqn:S3terms}).  If these properties could be demonstrated explicitly, it would confirm
the conjecture that all black hole entropy is entanglement entropy, once edge mode degrees of 
freedom are properly accounted for.



Chapter \ref{ch:lps} of this thesis is devoted to studying edge modes for arbitrary
diffeomorphism-invariant theories, using the {\it extended phase space} construction
 of Donnelly and Freidel \cite{Donnelly2016F}.  This phase space provides a classical construction
 of the edge modes as a first step toward obtaining their quantum description and 
 calculating the entropy.  The classical description has the advantage of preserving 
 diffeomorphism symmetry (which would be broken in certain choices of regularizations, 
 such as a lattice), and allows the surface symmetry algebra to be identified.  
 The algebra turns out to be universal for all diffeomorphism-invariant theories (for a given
 choice of Noether charge ambiguities), and can include transformations that move the surface
 if the fields satisfy appropriate boundary conditions.  The identification 
 of this symmetry algebra and the symplectic structure for the edge modes are the main 
 results of this chapter, while the quantization of these degrees of freedom
 is left to future work.

\section{Summary}
A driving motivation behind this thesis is the idea that gravity tends to act as a universal
regulator, following from the underlying diffeomorphism symmetry.  This statement 
suggests that gravity renders finite the divergences appearing in entanglement entropy.  
Applying this observation to black holes leads to the identification of the 
generalized entropy with entanglement entropy, with the leading divergence fulfilling the role of the 
Bekenstein-Hawking entropy, $S_\text{BH} = A/4G$.  Its interpretation as entanglement entropy 
allows generalized 
entropy to be assigned to surfaces other than black hole horizons, and when this is done, certain
entanglement identities reproduce the  gravitational field equations.  Finally, 
attempting to define entanglement entropy when gauge symmetry is present leads to the 
notion of edge modes, and these may provide a statistical interpretation for the
Bekenstein-Hawking entropy within the low energy effective theory.  

The picture that emerges is one in which entanglement supplants 
Riemannian geometry in the quantum regime of gravity.  
This viewpoint has already offered many insights about the nature quantum gravity, and the 
pages below explore just a few of the conclusions that derive from this perspective.  
It is clear that,
 going forward, 
 entanglement has a role to play in resolving the many enigmas  of quantum gravity.


\chapter{Excited state entanglement entropy in conformal perturbation theory}

\label{ch:EEinCPT}

This chapter is based on my paper ``Entanglement entropy of excited states in conformal 
perturbation theory and the Einstein equation,'' published in the Journal of High Energy 
Physics in 2016 \cite{Speranza2016}.  
\section{Introduction}

The entanglement equilibrium argument, outlined in section \ref{sec:enteq}, proceeds by 
replacing geometrical quantities that appear in the first law of causal diamond mechanics
(\ref{eqn:flcd}) with an equivalent expression related to entanglement.  
The discussion of section \ref{sec:SgenSEE} motivates interpreting the area term in
this equation with the leading divergence in the entanglement entropy.  It remains to provide
an entanglement interpretation for $H_\text{matter}$. As described in 
section \ref{sec:EE=EE}, when the matter fields under consideration are conformally invariant, 
the density matrix for the fields restricted to the ball has a simple expression in terms of an
integral of the matter stress energy tensor.  This expression is precisely what is needed
to write $H_\text{matter}$ in terms of a variation of entanglement entropy, leading 
to equation (\ref{eqn:dSEEtotintro}) and completing the argument.  This chapter explores 
how the argument needs to be modified when including fields that are not conformally
invariant.

Extending the argument for the equivalence between Einstein's equations and 
maximal vacuum entanglement  to  non-conformal  fields
requires taking the  ball to be much smaller than any length scale appearing in the 
field theory.  Since the theory will flow to an ultraviolet (UV) fixed point at 
short length scales, one expects to recover CFT behavior in this limit.
Jacobson made a conjecture about the form of the entanglement entropy for excited states
in small spherical regions that allowed the argument to go through.  
The purpose of the present chapter is to check this conjecture 
using conformal perturbation theory (see also \cite{Carroll2016a} for 
alternative ideas for checking the conjecture).  

In this chapter, we will consider a CFT deformed by a relevant operator $\op$ of dimension $\D$, 
and examine the
entanglement entropy for a class of excited states formed by a path integral over
Euclidean space.  The entanglement entropy in this case may be evaluated using recently
developed perturbative techniques \cite{Rosenhaus2014, Rosenhaus2014a, 
Rosenhaus:2014ula,Rosenhaus:2014zza,Faulkner2015,Faulkner2015a} which express
the entropy in terms of correlation functions, and notably do not rely on the replica 
trick \cite{Susskind:1993ws, Callan1994a}.  In particular, one knows from the expansion
in \cite{Rosenhaus2014,Rosenhaus:2014ula} that the first correction to the 
CFT entanglement entropy comes from the $\op\op$ two-point function and the
$K\op\op$ three point function, where $K$ is the CFT vacuum modular Hamiltonian.  
However, those works did not account for the noncommutativity of the density matrix
perturbation $\delta\rho$ with the original density matrix $\rho_0$, so the results
cannot be directly applied to find the finite change in entanglement entropy between the perturbed
theory excited state and the CFT ground state.\footnote{However, references 
\cite{Rosenhaus:2014ula,
Rosenhaus:2014zza} are able to reproduce universal logarithmic divergences when they
are present. \renewcommand{\baselinestretch}{1}
\footnotesize }  Instead, we will apply the technique developed by Faulkner \cite{Faulkner2015} 
to compute
these finite changes to the entanglement entropy, which we review in section
\ref{sec:EEballs}.  The result for the change in entanglement entropy between the excited
state and vacuum is
\beq  \label{eqn:Dneqdt}
\delta S =  \frac{2\pi\Omega_{d-2} }{d^2-1} \left[R^d\left(\delta\vev{T^g_{00}}  
-\frac{1}{2\D-d} \delta
\vev{T^g}  \right) - R^{2\D} \vev{\op}_g\delta\vev{\op} \frac{\D\Gamma(\dt+\frac32) 
\Gamma(\D-\dt+1)}{(2\D-d)^2 \Gamma(\D+\frac32)}  \right],
\eeq
which holds to first order in the variation of the state and for $\D\neq\dt$.  Here, $\Omega_{d-2}
= \frac{2\pi^{\dt-\frac12}}{\Gamma(\dt-\frac12)}$ is the volume of the unit $(d-2)$-sphere, $R$
is the radius of the ball,
$T^g_{\mu\nu}$ is the stress tensor of the deformed theory with trace $T^g$, $\vev{\op}_g$
stands for the vacuum expectation value of $\op$, and the $\delta$ refers to the change in 
each quantity relative to the vacuum value.  

The case $\D=\dt$ requires special attention, since the 
above expression degenerates at that value of $\D$.  The result for $\D=\dt$ is 
\beq\label{eqn:Deqdt} 
\delta S = 2\pi \frac{\Omega_{d-2}}{d^2-1} R^d \left[ \delta\fvev{T^g_{00}} + \delta \vev{T^g} 
\left(\frac2d-\frac12H_{\frac{d+1}{2}} + \log\frac{\mu R}{2} \right)-\dt \vev{\op}_g\delta\vev{\op}
\right],
\eeq
where $H_{\frac{d+1}{2}}$ is a harmonic number, defined for the integers by $H_n = 
\sum_{k=1}^n \frac1k$ and for arbitrary values of $n$ by $H_n = \eul+\psi_0(n+1)$ with
$\eul$ the Euler-Mascheroni constant, and $\psi_0(x) = \frac{d}{dx}\log\Gamma(x)$ the
digamma function.  
This result depends on a renormalization scale $\mu$ which arises due to an ambiguity 
in defining a renormalized value for the vev $\vev{\op}_g$.  The above result only superficially
depends on $\mu$, since this dependence cancels between the $\log\frac{\mu R}{2}$  and 
$\vev{\op}_g$ terms.
These results agree with the holographic calculations \cite{Casini2016a}, and this 
chapter therefore establishes that those results extend beyond holography.

In both equations (\ref{eqn:Dneqdt}) and (\ref{eqn:Deqdt}), the first terms scaling as $R^d$ 
take the  form required for Jacobson's argument.  However, when $\D\leq\dt$, 
the terms scaling as $R^{2\D}$ or $R^d \log R$ dominate over this term in the small $R$ limit.  
This leads to some tension with the argument for the equivalence of the Einstein equation and 
the hypothesis of maximal vacuum entanglement.  We revisit this point in section
\ref{sec:EEimps} and suggest some possible resolutions to this issue.

Before presenting the calculations leading to equations (\ref{eqn:Dneqdt}) and 
(\ref{eqn:Deqdt}), we briefly review Jacobson's argument in section \ref{sec:EE=EE}, where
we describe in more detail the form of the variation of the entanglement entropy that would 
be needed for the derivation of the Einstein equation to go through.
We also provide a review of Faulkner's method for calculating
entanglement entropy in section \ref{sec:EEballs}, since it will be used heavily in the sequel.  
Section \ref{sec:excited} describes the type of excited states considered in this chapter, including
an important discussion of the issue of UV divergences in operator expectation values.  
Following this, we present the derivation of the above result to first order in
$\delta\vev{\op}$ in section \ref{sec:harder}.  Finally, we discuss
the implications of these results for the Einstein equation derivation and avenues for 
further research in section \ref{sec:discussion}.

\section{Background}

\subsection{Einstein equation from entanglement equilibrium} \label{sec:EE=EE}
This section provides a brief overview of Jacobson's argument for
the equivalence of the Einstein equation and 
the maximal vacuum entanglement hypothesis
\cite{Jacobson2015a}.  The hypothesis states that the entropy of a small  geodesic
ball is maximal in 
a vacuum configuration
of quantum fields coupled to gravity, i.e.\ the vacuum is an equilibrium state.  
This implies that as the state is varied at fixed volume away from 
vacuum, the change in the entropy must be zero at first order in the variation.  
In order for this to be possible, the entropy increase of the matter fields must be compensated
by an entropy decrease due to the variation of the geometry.  Demanding that these
two contributions to the entanglement entropy cancel leads directly to the Einstein equation. 

Consider the simultaneous variations of the metric and the state of the quantum fields, $(\delta
g_{ab}, \delta \rho)$.  
The metric variation induces a change $\delta A$ in the surface area of the geodesic ball,
relative to the surface area of a ball with the same volume in the unperturbed metric.
Due to the area law, this leads to a proportional change $\delta S_\text{UV}$
in the entanglement
entropy
\beq \label{eqn:dSUV}
\delta S_\text{UV} = \frac{c_0}{\ep^{d-2}} \delta A.
\eeq
Normally, the coefficient $c_0/\ep^{d-2}$ 
is divergent and regularization-dependent; however, one further
assumes that quantum gravitational effects render it finite and universal.   For small enough
balls, the area variation is expressible in terms of the $00$-component of the Einstein tensor
at the center of the ball.  
Allowing for the background geometry from which the variation is taken to 
be any maximally symmetric space, with Einstein tensor $G_{ab}^\text{MSS} = -\Lambda
g_{ab}$,  (\ref{eqn:dSUV}) becomes \cite{Jacobson2015a}
\beq \label{eqn:dSUVG}
\delta S_\text{UV}=  -\frac{c_0}{\ep^{d-2}} 
\frac{\Omega_{d-2} R^d}{d^2-1} (G_{00} + \Lambda g_{00}) .
\eeq

The variation of the quantum state produces the compensating contribution to the entropy. At
first order in $\delta \rho$, this is given by the change in the modular Hamiltonian $K$,
\beq \label{eqn:dSIR}
\delta S_\text{IR} = 2\pi \delta\vev{K},
\eeq
where $K$ is related to $\rho_0$, the reduced density matrix of the vacuum restricted to the ball, 
via
\beq \label{eqn:rho0}
\rho_0 = e^{-2\pi K}/Z, 
\eeq
with the partition function $Z$ providing the normalization.  Generically, $K$ is a complicated,
nonlocal operator; however, in the case of a ball-shaped region of a CFT, it is given by 
a simple integral of the energy density over the ball \cite{Hislop1982, Casini2011},
\beq \label{eqn:K}
K = \int_\Sigma d\Sigma^a \zeta^b T_{ab} = \int_\Sigma d\Omega_{d-2} dr\, r^{d-2} \left(
\frac{R^2-r^2}{2R}\right) T_{00}.
\eeq
In this equation, $\zeta^a$ is the conformal Killing vector in Minkowski 
space\footnote{The conformal Killing vector is different 
for a general maximally symmetric space \cite{Casini2016a}.  
However, the Minkowski space vector is sufficient as long as $R^2\ll\Lambda^{-1}$.
\renewcommand{\baselinestretch}{1} \footnotesize} that fixes the 
boundary $\partial \Sigma$ of the ball.  With the standard Minkowski time $t=x^0$ and spatial
radial coordinate $r$, it is given by
\beq \label{eqn:CKV}
\zeta = \left(\frac{R^2-r^2-t^2}{2R}\right) \partial_t - \frac{rt}{R}\partial_r.
\eeq
If $R$ is taken small enough such that $\vev{T_{00}}$ is approximately constant throughout
the ball, equation (\ref{eqn:dSIR}) becomes
\beq\label{eqn:dSIRT}
\delta S_\text{IR} = 2\pi \frac{\Omega_{d-2}R^d}{d^2-1} \delta \vev{T_{00}}. 
\eeq

The assumption of vacuum equilibrium states that $\delta S_\text{tot} = \delta S_\text{UV}
+\delta S_\text{IR} = 0$, and this requirement, along with the expressions (\ref{eqn:dSUVG}) 
and (\ref{eqn:dSIRT}), leads to the relation
\beq
G_{00}+\Lambda g_{00} = \frac{2\pi}{c_0/\ep^{d-2} }\delta\vev{ T_{00}},
\eeq
which is recognizable as a component of the Einstein equation with $G_N = \frac{\ep^{d-2}}{4c_0}$.  
Requiring that this hold for 
all Lorentz frames and at each spacetime 
point leads to the full tensorial equation, and conservation of $T_{ab}$
and the Bianchi identity imply that $\Lambda(x)$ is a constant.  

The expression of $\delta S_\text{IR}$ in (\ref{eqn:dSIRT}) is  
special to a CFT, and cannot be expected to hold for
more general field theories.  However, it is enough if, in the small $R$ limit, it takes the 
following form
\beq\label{eqn:dSIRmod}
\delta S_\text{IR} = 2\pi \frac{\Omega_{d-2} R^d}{d^2-1} \left(\delta\vev{T_{00}} + C g_{00} \right).
\eeq
Here, $C$ is some scalar function of spacetime, formed from expectation values of operators
in the quantum theory.  With this form of $\delta S_\text{IR}$, 
the requirement that $\delta S_\text{tot}$
vanish in all Lorentz frames and at all points now leads to the tensor equation
\beq\label{eqn:EEtensor}
G_{ab}+\Lambda g_{ab} = \frac{2\pi}{c_0/\ep^{d-2} }\left(\delta\vev{T_{ab}}+ C g_{ab}\right).
\eeq
Stress tensor conservation and the Bianchi identity now impose that $\frac{2\pi}{c_0/\ep^{d-2} }
C(x) = \Lambda(x)+
\Lambda_0$, and once again the Einstein equation with a cosmological constant is recovered. 

The purpose of the present chapter is to evaluate $\delta S_{\text{IR}}$ appearing in equation
(\ref{eqn:dSIRmod}) in a CFT deformed by a relevant operator of dimension $\D$.  
It is crucial in the above derivation that $C$ transform as a scalar under 
a change of Lorentz frame.  As long as this requirement is met, complicated dependence on 
the state or operators in the theory is allowed.  
In the simplest case, $C$ would be given by the variation of some 
scalar operator expectation value, $C = \delta\vev{X}$, with $X$ independent of the quantum 
state, since such an object has trivial transformation properties under Lorentz boosts.  
We find this to be the case for the first order state variations we considered; however, 
the operator $X$ has the peculiar feature that it depends explicitly on the radius of the ball.  
The constant $C$ is found to have a term scaling with the ball size as $R^{2\D-d}$
(or $\log R$ when $\D=\dt$), and 
when $\D\leq\dt$, this term dominates over the stress tensor term as $R\rightarrow 0$.  
Furthermore, as pointed out in \cite{Casini2016a}, even in the CFT where the first order
variation of the entanglement entropy vanishes, the second order piece contains the 
same type of term scaling as $R^{2\D-d}$, which again dominates for small $R$.  
This leads to the conclusion that the local curvature scale $\Lambda(x)$ must be 
allowed to depend on $R$.  
This proposed resolution will be discussed further in section 
\ref{sec:EEimps}.

\subsection{Entanglement entropy of balls in conformal perturbation theory} \label{sec:EEballs} 
Checking the conjecture (\ref{eqn:dSIRmod}) requires  a method for calculating the 
entanglement entropy of balls in a non-conformal theory.  Faulkner has recently 
shown how to perform this calculation in a CFT deformed by a relevant operator, 
$\int f(x)\op(x)$ \cite{Faulkner2015}.  This deformation may be split into two parts, 
$f(x) = g(x) + \lambda(x)$, where the coupling $g(x)$ represents the deformation of the 
theory away from a CFT, while the function $\lambda(x)$ produces a variation of the 
state away from vacuum.  The change in entanglement relative to the CFT vacuum will then
organize into a double expansion in $g$ and $\lambda$,
\beq
\delta S = S_g + S_\lambda+ S_{g^2} + S_{g\lambda} + S_{\lambda^2} +\ldots.
\eeq
The terms in this expansion that are $O(\lambda^1)$ and any order in $g$ are the ones relevant
for $\delta S_\text{IR}$ in equation (\ref{eqn:dSIRmod}).  Terms that are $O(\lambda^0)$ are part
of the vacuum entanglement entropy of the deformed theory, and hence are not of interest 
for the present analysis.  Higher order in $\lambda$ terms may also be relevant,  especially
in the case that
the $O(\lambda^1)$ piece vanishes, which occurs, for example, in a CFT.

We begin with the Euclidean path integral 
representations of the reduced density matrices in the ball $\Sigma$
for the CFT vacuum $\rho_0$ and for
the deformed theory excited state $\rho = \rho_0+\delta\rho$.  The matrix
elements of the vacuum density
matrix are
\beq \label{eqn:rho0pi}
\langle\phi_-| \rho_0 |\phi_+ \rangle = \frac1Z \int_{\substack{\phi(\Sigma_+) = \phi_+\\
\phi(\Sigma_-) = \phi_-}}\pd \phi\, e^{-I_0 }.
\eeq
Here, the integral is over all fields satisfying the boundary conditions  $\phi = \phi_+$
on one side of the surface $\Sigma$, and $\phi=\phi_-$ on the other side.  
The partition function $Z$ 
is represented by an unconstrained path integral,
\beq
Z= \int \pd\phi\, e^{-I_0}.
\eeq
It is useful to think of the path integral (\ref{eqn:rho0pi}) as evolution along an angular
variable $\theta$ from the $\Sigma_+$ 
surface at $\theta=0$ to the $\Sigma_-$ surface at $\theta=2\pi$
\cite{Kabat:1994vj, Holzhey:1994we, Wong2013}.
When this evolution follows the flow of the conformal Killing
vector (\ref{eqn:CKV}) (analytically continued to Euclidean space), it is generated by 
the conserved Hamiltonian $K$ from equation (\ref{eqn:K}).  This leads  to the operator
expression for $\rho_0$ given in equation (\ref{eqn:rho0}).

The path integral representation for $\rho$ is given in a similar manner, 
\begin{align}
\label{eqn:rhopi}
\langle\phi_-| \rho |\phi_+ \rangle &= \frac1N \int_{\substack{\phi(\Sigma_+) = \phi_+\\
\phi(\Sigma_-) = \phi_-}}\pd \phi\, e^{-I_0 -\int f\op}  \\
\label{eqn:rhoexpand}
&=\frac1{Z+\delta Z} \int_{\substack{\phi(\Sigma_+) = \phi_+\\
\phi(\Sigma_-) = \phi_-}}\pd \phi\, e^{-I_0}\left(1-\int f\op +\frac12\iint f\op f\op  -\ldots\right)
\end{align}
Again viewing this path integral as an evolution from $\Sigma_+$ to $\Sigma_-$, with 
evolution operator
$\rho_0 = e^{-2\pi K}/Z$, we can extract the operator expression of $\delta \rho = \rho-\rho_0$, 
\beq\label{eqn:drho}
\delta\rho = -\rho_0\int  f\op +\frac12\rho_0\iint  T\left\{f\op f\op \right\} - \ldots -\text{traces},
\eeq
where $T\{\}$ denotes angular ordering in $\theta$. 
The ``-traces'' terms in this expression arise from  $\delta Z$ in (\ref{eqn:rhoexpand}).  
These terms ensure that $\rho$ is normalized, or equivalently
\beq\label{eqn:traceless}
\tr(\delta\rho) = 0.
\eeq
We suppress writing these terms explicitly since they will play no role in the remainder of this
work.  

Using these expressions for $\rho_0$ and $\delta\rho$, we can now develop the perturbative
expansion of the entanglement entropy, 
\beq
S = -\tr\rho \log\rho .
\eeq
It is useful when expanding out the logarithm to write this 
in terms of the resolvent integral,\footnote{One can also expand the logarithm
using the Baker-Campbell-Hausdorff formula, see e.g.\ \cite{Kelly:2015mna}.
 \renewcommand{\baselinestretch}{1} \footnotesize}
\begin{align}
S &= \int_0^\infty 
d\beta\left[\tr\left(\frac{\rho}{\rho+\beta}\right) - \frac1{1+\beta}\right]\\
\label{eqn:drhobetaint}
&= S_0 +\tr\int_0^\infty d\beta \frac{\beta}{\rho_0+\beta}\left[\delta\rho\frac{1}{\rho_0+\beta}
-\delta\rho\frac{1}{\rho_0+\beta}\delta\rho\frac{1}{\rho_0+\beta}+\ldots\right].
\end{align}
The first order term in $\delta \rho$ is straightforward to evaluate.  Using the 
cyclicity of the trace and 
equation (\ref{eqn:traceless}), the $\beta$ integral is readily evaluated, and applying 
(\ref{eqn:rho0}) one finds
\beq \label{eqn:dS1}
\delta S^{(1)} = 2\pi \tr(\delta \rho\, K) = 2\pi\delta\vev{K}.
\eeq
Note when $\delta\rho$ is a first order variation, this is simply the first law of entanglement
entropy \cite{Blanco2013a} (see also \cite{Bhattacharya:2012mi}). 

The second order piece of (\ref{eqn:drhobetaint}) is more involved, and much of reference
\cite{Faulkner2015} is devoted to evaluating this term.  
The surprising result is that this term may
be written holographically as the 
flux through an emergent  AdS-Rindler horizon of a conserved energy-momentum
current for a scalar field\footnote{Reference 
\cite{Faulkner2015} further showed that this is equivalent
to the Ryu-Takayanagi prescription for calculating the entanglement entropy
\cite{Ryu:2006bv, Ryu:2006ef}, using
an argument similar to the one employed in \cite{Faulkner:2013ica} 
deriving the bulk linearized Einstein equation from
the Ryu-Takayanagi formula. \renewcommand{\baselinestretch}{1} \footnotesize }  
(see figure \ref{fig:AdSRindler}).
The bulk scalar field $\phi$ satisfies the free Klein-Gordon equation in AdS with mass
$m^2 = \D(\D-d)$, as is familiar from the usual holographic dictionary \cite{Witten:1998qj}. 
The  specific AdS-Rindler horizon that is used is the one with
a bifurcation surface that asymptotes near the boundary 
to the entangling surface $\partial\Sigma$
in the CFT.
This result holds for {\it any} CFT, including those which are not normally considered 
holographic. 

\begin{figure}
\centering
\includegraphics[width=0.65\textwidth]{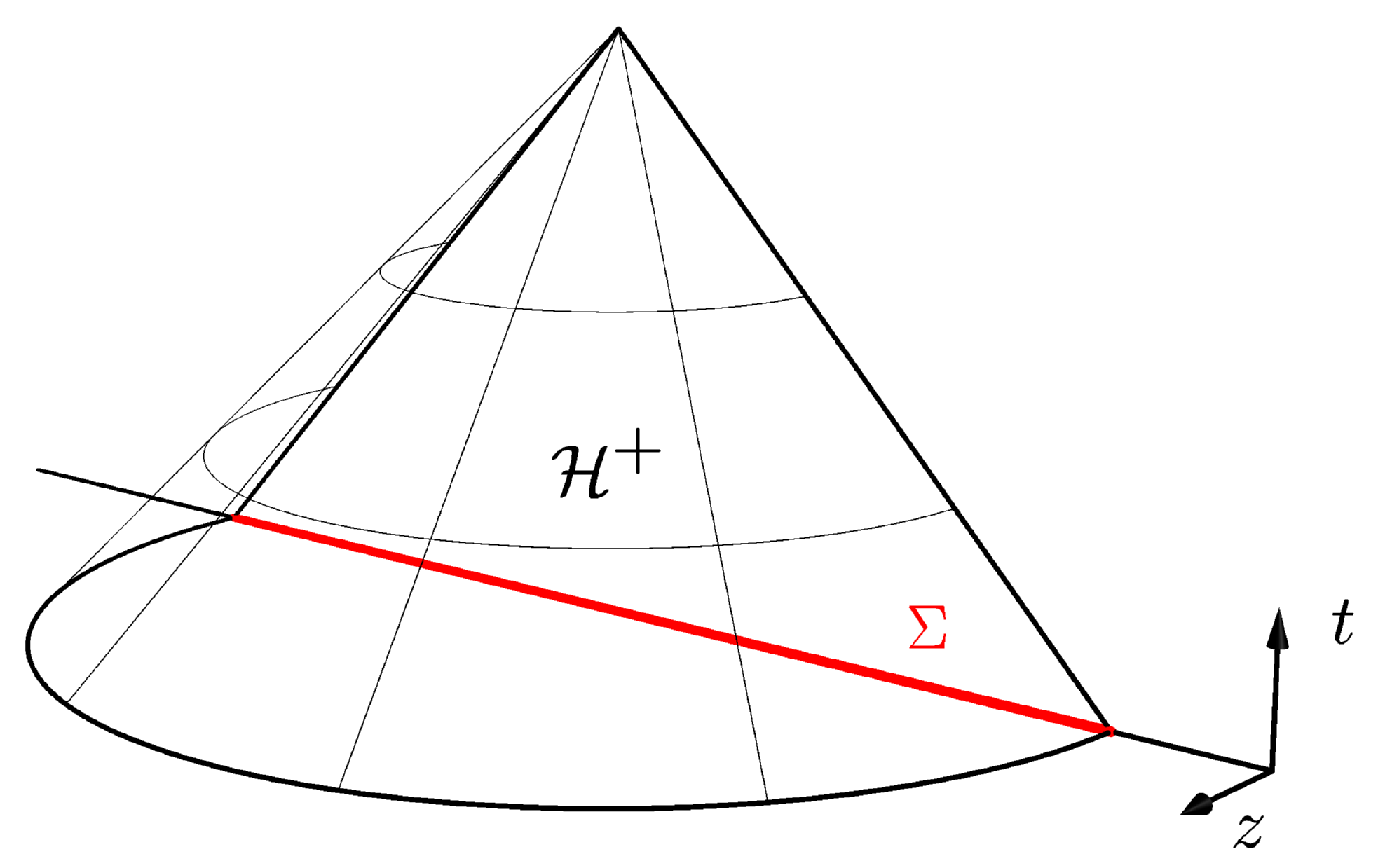}
\renewcommand{\baselinestretch}{1}
\small\normalsize
\begin{quote}
\caption[AdS Rindler horizon]{\small
Bulk AdS-Rindler horizon $\mathcal{H}^+$.  The horizon extends from the 
bifurcation surface in the bulk at $t=0$ along the cone to the tip at $z=0$, $t=R$.  
The ball-shaped surface $\Sigma$ in the boundary CFT shares a boundary
with the bifurcation surface at $t=z=0$. \label{fig:AdSRindler} }
\end{quote}
\renewcommand{\baselinestretch}{2}
\small\normalsize
\end{figure}

We now describe the bulk calculation in more detail.  Poincar\'{e} coordinates are used
in the bulk, where the metric takes the form
\beq
ds^2 = \frac1{z^2} \left( -dt^2+dz^2+dr^2 + r^2 d\Omega_{d-2}^2\right).
\eeq
The coordinates $(t,r,\Omega_i)$ match onto the Minkowski coordinates of the CFT at the 
conformal boundary $z=0$.  The conformal Killing vector $\zeta^a$ of the CFT, defined in
equation (\ref{eqn:CKV}), extends to a Killing vector in the bulk,
\beq
\xi = \left(\frac{R^2-t^2-z^2-r^2}{2R}\right)\partial_t - \frac{t}{R}(z\partial_z + r\partial_r).
\eeq
The Killing horizon $\mathcal{H^+}$ 
of $\xi^a$ defines the inner boundary of the AdS-Rindler patch for $t>0$, and sits at
\beq
r^2+z^2 = (R-t)^2.
\eeq

The contribution of the second order piece of (\ref{eqn:drhobetaint}) to the entanglement entropy
is 
\beq \label{eqn:H+int}
\delta S^{(2)} = -2 \pi\int_\mathcal{H^+} d\Sigma^a \xi^b T^B_{ab},
\eeq
where the integral is over the horizon to the future of the bifurcation surface at $t=0$. 
The surface element on the horizon is $d\Sigma^a = \xi^a d\chi dS$, where $\chi$ is a
parameter for $\xi^a$ satisfying $\xi^a\nabla_a\chi=1$,
and $dS$ is the area element in the transverse space.  $T^B_{ab}$
is the stress tensor of a scalar field $\phi$ satisfying the Klein-Gordon equation,
\beq \label{eqn:bulkeom}
\nabla_c\nabla^c \phi -\D(\D-d) \phi = 0.
\eeq
Explicitly, the stress tensor is
\beq
T^B_{ab} = \nabla_a\phi\nabla_b\phi-\frac12(\D(\D-d)\phi^2+\nabla_c\phi \nabla^c\phi) g_{ab},
\eeq
which may be rewritten when $\phi$ satisfies the field equation (\ref{eqn:bulkeom}) as 
\beq\label{eqn:TBab}
T^B_{ab} = \nabla_a\phi \nabla_b\phi - \frac14 g_{ab}\nabla_c\nabla^c \phi^2.
\eeq

The boundary conditions for $\phi$ come about
from its defining integral,
\beq \label{eqn:phiint}
\phi(x_B) = \frac{\Gamma(\D)}{\pi^{\dt} \Gamma(\D-\dt)} 
\int_{C(\delta)} d \tau \int d^{d-1} \vec{x} \frac{z^\D f(\tau,\vec{x})}{
\left(
z^2+(\tau-i t_B)^2+(\vec{x}-\vec{x}_B)^2\right)^\D},
\eeq
where $x_B = (t_B, z, \vec{x}_B)$ are the real-time bulk coordinates, and $(\tau,\vec{x})$ are
coordinates on the boundary Euclidean section.  The normalization of this field arises from
a particular choice of the normalization for the $\op\op$ two-point function,
\beq \label{eqn:cD}
\vev{\op(x)\op(0)} = \frac{c_\D}{x^{2\D}},\qquad c_\D = \frac{(2\D-d)\Gamma(\D)}{\pi^{\dt}\Gamma
(\D-\dt)},
\eeq
which is chosen so that the relationship (\ref{eqn:phibdyEuc}) holds.  Note that 
sending $c_\D\rightarrow \alpha^2 c_\D$ multiplies $\phi$ by a single factor of $\alpha$.  
The integrand in (\ref{eqn:phiint}) has branch points at $\tau = i\left(t_B\pm\sqrt{z^2+
(\vec{x}-\vec{x}_B)^2}\right)$, and the branch cuts extend along the imaginary axis to 
$\pm i\infty$.  The notation $C(\delta)$ on the $\tau$ integral refers to the $\tau$ contour
prescription, which must lie along the real axis and be cut off near $0$ at $\tau=\pm\delta$.  
This can lead to a divergence in $\delta$ when the contour is close to the branch point (which
can occur when $t_B\sim \sqrt{z^2+(\vec{x}-\vec{x}_b)^2}$), and this ultimately cancels
against a divergence in  $\vev{T_{00} \op \op}$ from $\delta S^{(1)}$.  More details about 
these divergences and the origin of this contour and branch 
prescription can be found in \cite{Faulkner2015}.

From equation (\ref{eqn:phiint}), one can now read off the boundary conditions
as $z\rightarrow 0$.  The solution
should be regular in the bulk, growing at most like $z^{d-\D}$ for large $z$ if $f(\tau,\vec{x})$ is 
bounded.  On the Euclidean section $t_B=0$, it behaves for $z\rightarrow 0$ as
\beq \label{eqn:phibdyEuc}
\phi\rightarrow f(0,\vec{x_B}) z^{d-\D} +\beta(0,\vec{x_B}) z^\D,
\eeq
where the function $\beta$ may be determined by the integeral (\ref{eqn:phiint}), but also 
may be fixed by demanding regularity of the solution in the bulk. This is consistent with 
the usual holographic dictionary \cite{Klebanov1999, Balasubramanian:1998de}, 
where $f$ corresponds to the coupling, and $\beta$ is related
to $\vev{\op}$ by\footnote{The minus sign appearing here is due to the source in the 
generating functional being $-\int f\op$ as opposed to $\int f\op$
 \renewcommand{\baselinestretch}{1}\footnotesize}
\beq\label{eqn:beta}
\beta(x) = \frac{-1}{2\D-d} \vev{\op(x)}.
\eeq
This formula follows from defining the renormalized expectation value $\vev{\op}$ using 
a holographically renormalized two-point function,
\beq
\mvev{\big}{\op(0)\op(x)}^{z,\text{ren.}} 
= \frac{c_\D}{(z^2 + x^2)^\D} - (2\D-d)z^{d-2\D}  \delta^d(x).
\eeq  
The $\delta$ function in this formula subtracts off the divergence near
$x= 0$.\footnote{Additional subleading divergences are present when $\D\geq\dt+1$,
which involve subtractions proportional to derivatives of the $\delta$-function.
 \renewcommand{\baselinestretch}{1} \footnotesize}  
Using the renormalized two-point function, the expectation value of $\op$ at 
first order in $f$ is
\beq
\vev{\op(x)} = -\int d^d y f(y) \mvev{\Big}{\op(x) \op(y)}^{z,\text{ren.}},
\eeq  
and by comparing this formula to (\ref{eqn:phiint}) at small values $z$ and $t_B=0$, 
one arrives at 
equation (\ref{eqn:beta}).

In real times beyond $t_B >z$, $\phi(x_B)$ has only a $z^\D$ component near $z=0$.  The
integral effectively shuts off the coupling $f$ in real times.  This follows from the 
use of a Euclidean path integral to define the state; other real-time behavior may be 
achievable using the Schwinger-Keldysh formalism.
When $t_B\sim z$, there are divergences associated with switching off the coupling in real
times, and these are regulated with the $C(\delta)$ contour prescription.

Returning to the flux equation (\ref{eqn:H+int}), 
since $\xi^a$ is a Killing vector, this integral defines a conserved quantity, and 
may be evaluated on any other surface homologous to 
$\mathcal{H}^+$.  The choice which is
most tractable is to push the surface down to $t_B=0$, where the Euclidean AdS solution can 
be used to evaluate the stress tensor.  The $t_B=0$ surface $\mathcal{E}$ 
covers the region between the 
horizon and $z=z_0$, where it must be cut off to avoid a divergence in the integral.  To remain
homologous to $\mathcal{H^+}$, this must be supplemented by a timelike surface $\mathcal{T}$ 
at the 
cutoff $z=z_0$ which extends upward to connect back with $\mathcal{H}^+$.  In the limit
$z_0\rightarrow 0$, the surface $\mathcal{T}$ approaches the domain of dependence
$D^+(\Sigma)$ of the ball-shaped region in the CFT (see figure \ref{fig:ET}).  Finally, there
will  be a contribution from a 
region along the original surface $\mathcal{H}^+$ between $z_0$ and $0$, 
but in the limit $z_0\rightarrow 0$, the contribution to the integral from this surface will  
vanish.\footnote{
This piece may become important in the limiting case $\D=\dt-1$, which requires special 
attention.  We will not consider this possibility further here.
 \renewcommand{\baselinestretch}{1} \footnotesize }

\begin{figure}
\centering
\includegraphics[width=0.4\textwidth]{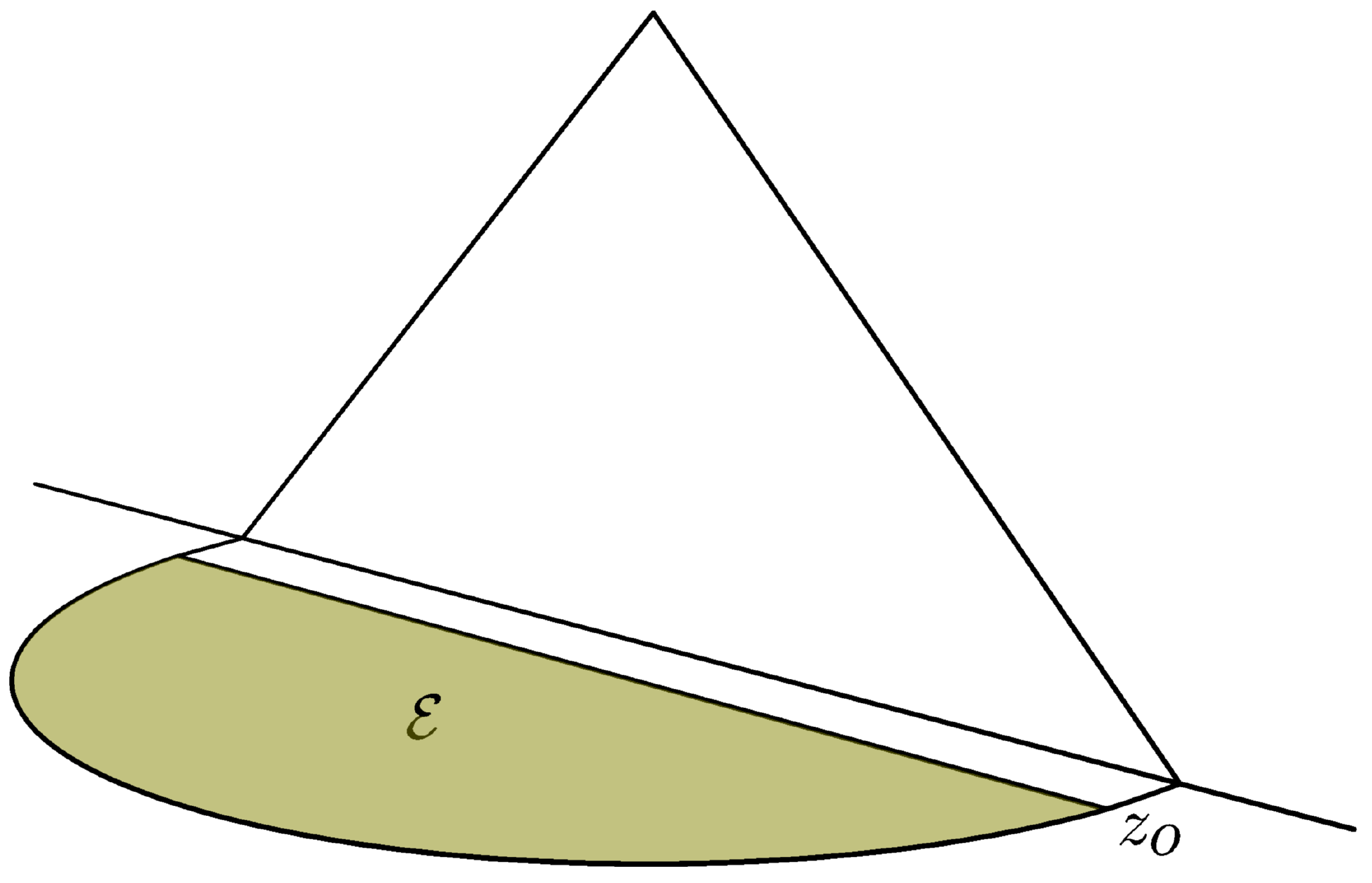}\qquad
\includegraphics[width=0.4\textwidth]{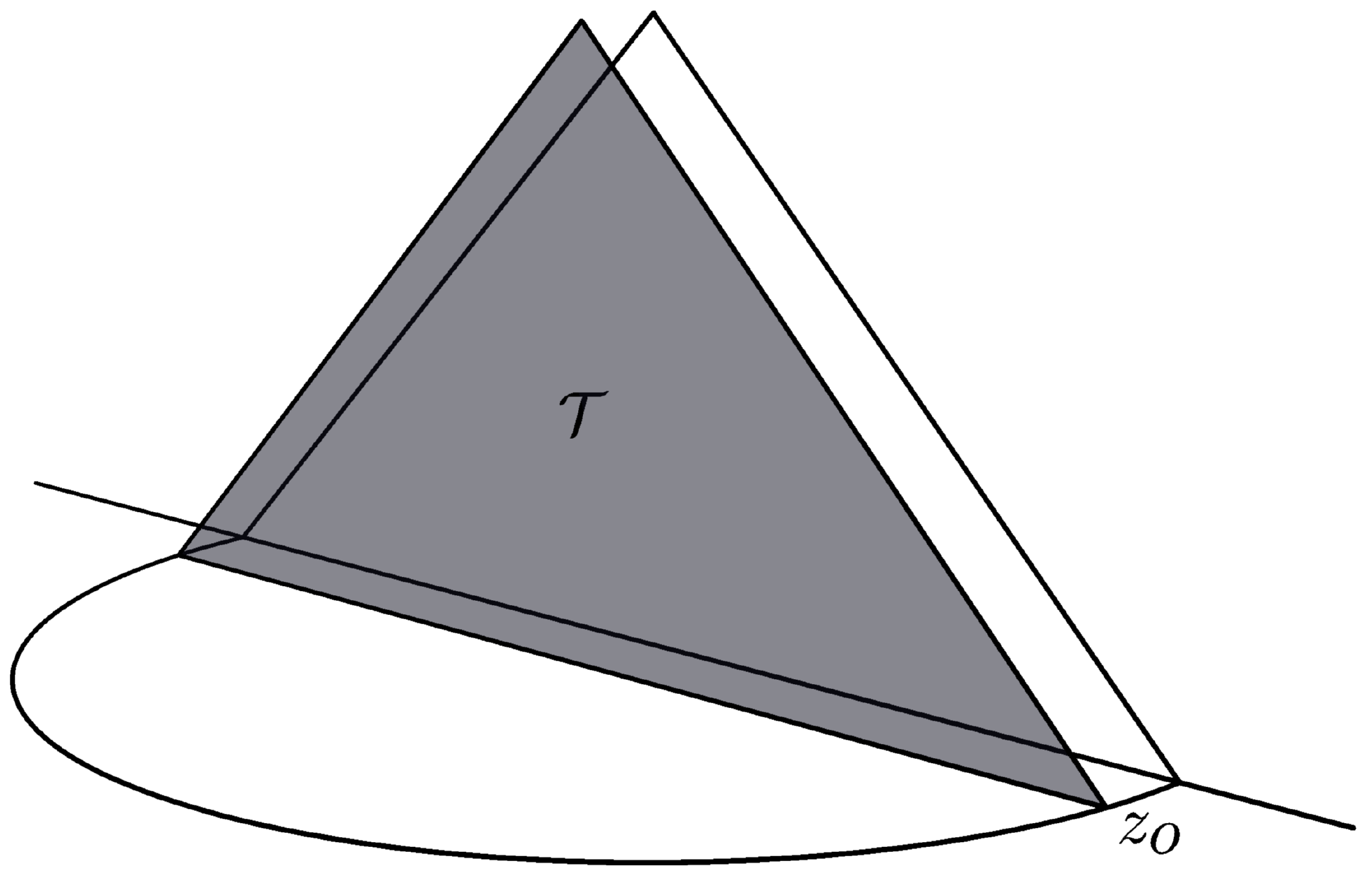}
\renewcommand{\baselinestretch}{1}
\begin{quote}
\caption[Bulk surfaces $\mathcal{E}$ and $\mathcal{T}$]{\small 
$\mathcal{E}$ and $\mathcal{T}$ surfaces over which the flux integrals (\ref{eqn:Eint})
and (\ref{eqn:Tint}) are computed. \label{fig:ET}}
\end{quote}
\end{figure}
\renewcommand{\baselinestretch}{2}

Using equation (\ref{eqn:TBab}), the integral on the surface $\mathcal{E}$ 
can be written out more explicitly: 
\begin{align}
&-2\pi\int_\mathcal{E}d\Sigma^a\xi^b T^B_{ab} \nonumber \\
\label{eqn:Eint}
&= 2\pi\int d\Omega_{d-2}\int_{z_0}^R 
\frac{dz}{z^{d-1}} \int_0^{\sqrt{R^2-z^2}} dr\, r^{d-2} \left[\frac{R^2-r^2-z^2}{2R}\right]\left[
(\partial_\tau \phi)^2-\frac{ \nabla^2_E\phi^2}{4z^2}\right].
\end{align}
This formula uses the solution on the Euclidean section in the bulk, with Euclidean 
time $\tau_B = i t_B$.  This is acceptable on the $t_B=0$ surface since the stress tensor
there satisfies $T_{\tau\tau}^B = -T^B_{tt}$.  The Laplacian $\nabla_E^2$ is hence the 
Euclidean AdS Laplacian.
The  
$\mathcal{T}$ surface integral
is 
\begin{align}
&2\pi \int_{\mathcal{T}} d\Sigma^a \xi^bT^B_{ab} \nonumber \\
\label{eqn:Tint}
&= \frac{2\pi}{z_0^{d-1}} \int d\Omega_{d-2} \int_0^R dt\int_0^{R-t} dr\, r^{d-2}\left\{
\left[\frac{R^2-r^2-t^2}{2R}\right]\partial_z\phi\partial_t\phi - \frac{z_0t}{R} \left[
(\partial_z\phi)^2-\frac{ \nabla^2\phi^2 }{4z_0^2}\right]
\right\}. 
\end{align}
Here, note that the limits of integration have been set to coincide with $D^+(\Sigma)$, which 
is acceptable when taking $z_0\rightarrow0$.

\section{Producing excited states} \label{sec:excited}
This section describes the class of states that are formed from the Euclidean path integral
prescription, and also discusses restrictions on the source function $f(x)$.  One requirement
is that the density matrix be Hermitian.  For a density matrix constructed from a path integral
as in (\ref{eqn:rhopi}), this translates to the condition that the deformed action $I_0 + \int f\op$
be reflection symmetric about the $\tau=0$  surface on which the state is evaluated.  When
this is satisfied, $\rho$ defines a pure state \cite{Cooperman:2013iqr}.  
Since this imposes  $f(\tau, \vec{x}) = f(-\tau,\vec{x})$, it gives
the useful condition
\beq
\partial_\tau f(0,\vec{x})=0,
\eeq
which simplifies the evaluation of the bulk integral (\ref{eqn:Eint}).  

Another condition on the state is that the stress tensor $T^g_{ab}$ of the deformed theory
and the operator $\op$ have non-divergent expectation values, compared to the vacuum.  
These divergences are not independent, but are related to each other through Ward identities.  
The $\vev{\op}$ divergence is straightforward to evaluate, 
\begin{align}
\vev{\op(0)} &= \frac{1}{N}\int \pd\phi e^{-I_0}\left(1-\int f\op+\ldots\right)\op(0) \\
&= -\int_{C(\delta)} d^dx f(x)\mvev{\Big}{\op(0)\op(x)}_0 ,
\end{align}
where the $0$ subscript indicates a CFT vacuum correlation function.  $C(\delta)$
refers to the regularization of this correlation function, which  is a  point-splitting
cutoff for  $|\tau| <\delta$.   Note that $\delta$ is the same regulator appearing in 
the 
definition of the bulk scalar field, equation (\ref{eqn:phiint}).  

Only the change  $\delta\vev{\op}$  in this correlation
function
relative to the deformed theory vacuum must be free of divergences. From the 
decomposition $f(x) = g(x)+\lambda(x)$, with $g(x)$ representing the 
deformation of the theory and $\lambda(x)$ the state deformation, one finds that 
the divergence in $\delta\vev{\op}$ comes from the coincident limit $x\rightarrow 0$.  It can
be extracted by expanding $\lambda(x)$ around $x=0$.  The leading divergence is then
\begin{align}
\delta\vev{\op(0)}_\text{div} &= -\lambda(0)\int_{C(\delta)}d\tau \int d\Omega_{d-2}\int_0^\infty dr 
\frac{r^{d-2}c_\D}{
(\tau^2 +r^2)^\D} \nonumber\\
&= 
\label{eqn:dOdiv}
-\lambda(0)\frac{2\Gamma(\D-\dt+\frac12)}{\sqrt{\pi} \,\Gamma(\D-\dt)} 
\delta^{d-2\D}
\end{align}
When $\D\geq\dt$, a divergence in $\delta\vev{\op}$ exists unless $\lambda(0)=0$.\footnote{When
$\D=\dt$, after appropriately redefining $c_\D$ (see equation (\ref{eqn:cD'})), 
it becomes a $\log\delta$ divergence.  \renewcommand{\baselinestretch}{1}\footnotesize}  
Further, this must hold at every point on the 
$\tau=0$ surface, which leads to the requirement that $\lambda(0,\vec{x})=0$.  Additionally, there
can be subleading divergences proportional to $\delta^{d-2\D+2n} \partial_\tau^{2n}\lambda(0,
\vec{x})$ for all integers $n$ where the $\delta$ 
exponent is negative or zero.\footnote{Divergences
proportional to the spatial derivative of $\lambda$ are not present since the condition from the
 leading divergence
already set these to zero.  \renewcommand{\baselinestretch}{1}\footnotesize }
Thus, the requirement on $\lambda$ is that its first $2q$ $\tau$-derivatives should vanish at
$\tau=0$, where 
\beq\label{eqn:q}
q=\left \lfloor \D-\dt\right\rfloor.
\eeq

We can also check that this condition leads to a finite value expectation value for the stress 
tensor, which for the deformed theory is 
\beq \label{eqn:Tgabdef}
T^g_{ab} = \frac{2}{\sqrt{g}} \frac{\delta I}{\delta g^{ab}} = T^0_{ab}-g\op g_{ab},
\eeq
where $T^0_{ab}$ is the stress tensor for the CFT.  For the $T^0_{\tau\tau}$ component,  the 
expectation value is 
\begin{align}
\vev{T^0_{\tau\tau}(0)} 
&=\frac12\iint_{C(\delta)} d^d x \,d^d y f(x) f(y)\mvev{\Big}{T^0_{\tau\tau}(0)\op(x)\op(y)}_0.
\end{align}
The divergence in this correlation function comes from $x,y\rightarrow 0$ simultaneously.
It can be evaluated by expanding $f$ around $0$, and then employing Ward identities to 
relate it to the $\op\op$ two-point function (see, e.g.\  section \ref{sec:D=d2calc} of this 
chapter or 
Appendix D of \cite{Faulkner2015}).  
The first order in $\lambda$ piece, which gives $\delta\vev{T^0_{\tau\tau}}$, is
\beq
\delta\vev{T^0_{\tau\tau}}_\text{div}
 = -g \lambda(0) 2^{d-2\D}\frac{2\Gamma(\D-\dt+\frac12)}{\sqrt{\pi}\,
\Gamma(\D-\dt)} \delta^{d-2\D}.
\eeq
The divergence in the actual energy density also receives a contribution from the $\op$ 
divergence (\ref{eqn:dOdiv}).  Using (\ref{eqn:Tgabdef}), this is found to be
\beq
\delta\vev{T^g_{\tau\tau}}_\text{div} = -g\lambda(0)\frac{2\Gamma(\D-\dt+\frac12)}{\sqrt{\pi}\,
\Gamma(\D-\dt)} (2^{d-2\D}-1) \delta^{d-2\D}.
\eeq
As with the $\delta\vev{\op}$ divergence,  requiring that $\lambda(0,\vec{x})=0$ ensures that 
the excited state has finite energy density.\footnote{Curiously, the divergences in $T^g_{ab}$
cancel without imposing $\lambda(0)=0$ when $\D=\dt$.  \renewcommand{\baselinestretch}{1}\footnotesize }  
Subleading divergences and other components of 
$T^g_{ab}$ can be evaluated in a 
similar way, and lead to the same requirements on $\lambda$ as were found for the $\op$ 
divergences.

\section{Entanglement entropy calculation} \label{sec:harder} 
Now we compute the change in entanglement entropy for the state formed by the path
integral with the deformed action $I=I_0+\int f\op$, with $f(x) = g(x)+\lambda(x)$ being
a sum of the theory deformation $g$ and the state deformation $\lambda$.
The bulk term $\delta S^{(2)}$ in 
plays an important role in this case.\footnote{A slightly simpler situation would be to
consider the deformed action $I = I_0 + \int g \op + \int \lambda\op_s$, with $\D\neq\D_s$.
Then $\delta S^{(2)}$ gives no contribution at first order in $\lambda$, since this term
arises from the $\op\op_s$ two point function, which vanishes.  However, in this case, the 
term at second order in $\lambda$ would receive a contribution from $\delta S^{(2)}$, and 
it is computed in precisely the same way as described in this section.  Hence we do not 
focus on this case where $\D\neq\D_s$.  \renewcommand{\baselinestretch}{1}\footnotesize }    
To evaluate this term, we need the solution for the scalar field in the bulk subject to the 
boundary conditions described in section \ref{sec:EEballs}.  Since $\phi$ satisfies
a linear field equation, so we may solve separately for the solution corresponding to $g$ and 
the solution corresponding to $\lambda$.  The function $g(x)$ is taken to be spatially constant, 
and either constant in Euclidean time or set to zero at some IR length scale $L$.  Its solution
is most readily found by directly evaluating the integral (\ref{eqn:phiint}), and we will discuss
it separately in each of the cases $\D>\dt$, $\D<\dt$ and $\D=\dt$ considered below.  

The solution for $\lambda(x)$ takes the same form in all three cases, so we begin by describing
it.  On the Euclidean section in Poincar\'{e} coordinates, the field equation (\ref{eqn:bulkeom}) 
is
\beq\label{eqn:eompoin}
\left[ z^{d+1}\partial_z(z^{-d+1}\partial_z)+z^2\left(\partial_\tau^2 +
r^{-d+2}\partial_r(r^{d-2}\partial_r) +r^{-2}\nabla^2_{\Omega_{d-2}} \right)\right]\phi -
\D(\D-d) \phi = 0,
\eeq
where $\nabla^2_{\Omega_{d-2}}$ denotes the Laplacian on the $(d-2)$-sphere.  
Although one may consider arbitrary spatial dependence for the function $\lambda(x)$,
the present calculation is concerned with the small ball limit, where the state may be taken
uniform across the ball.  We therefore restrict to $\lambda = \lambda(\tau)$.  One can 
straightforwardly generalize to include corrections due to spatial dependence in $\lambda$,
and these will produce terms suppressed in powers of $R^2$.  

Equation (\ref{eqn:eompoin}) 
may be solved by separation of variables.  The $\tau$ dependence is given by
$\cos(\omega\tau)$, since it must be $\tau$-reflection symmetric.  
This leads to the equation for the $z$-dependence,  
\beq
\partial_z^2\phi-\frac{d-1}{z}\partial_z\phi -\left(\omega^2+\frac{\D(\D-d)}{z^2}\right)\phi = 0.
\eeq
This has modified Bessel functions as solutions, and regularity as $z\rightarrow\infty$ selects 
the solution proportional to $z^{\dt} K_\alpha(\omega z)$, with
\beq
\alpha = \dt-\D.
\eeq
Hence, the final  bulk solution is
\beq \label{eqn:phio}
\phi_\omega = \lambda_{\omega}\left(\frac{\omega}{2}\right)^{\D-\dt}
\frac{2z^{\dt}K_\alpha(\omega z)}{\Gamma(\D-\dt)} \, \cos\omega \tau.
\eeq
where the normalization has been chosen so that the coefficient of $z^{d-\D}$
in the near-boundary expansion is
\beq \label{eqn:lambda}
\lambda = \lambda_\omega \cos(\omega\tau).
\eeq
A single frequency solution will not satisfy the requirement derived in section \ref{sec:excited}
that $\lambda(0,\vec{x})$ and its first $2q$ $\tau$-derivatives vanish (where $q$ was
given in 
(\ref{eqn:q})).  
Instead, $\lambda$ must be constructed from a wavepacket of several frequencies,
\beq
\lambda(\tau) = \int_0^\infty d\omega \lambda_\omega \cos(\omega\tau),
\eeq
with Fourier components $\lambda_\omega$ satisfying
\beq \label{eqn:lambdaoint}
\int_0^\infty d\omega\, \omega^{2n} \lambda_\omega = 0
\eeq
for all nonnegative integers $n\leq q$.  Finally, the coefficients $\lambda_\omega$ should 
fall off rapidly before $\omega$ becomes larger than $R^{-1}$, since such a state would be 
considered highly excited relative to the scale set by the ball size.

Using these solutions, we may proceed with the entanglement entropy calculation.  The answer
for $\D>\dt$ in section \ref{sec:D>dt}
comes from a simple application of the formula derived in \cite{Faulkner2015}.  
In section \ref{sec:D<dt} when considering 
$\D<\dt$, we must introduce a new element into the calculation to deal with IR divergences
that arise.  This is just a simple IR cutoff in the theory deformation $g(x)$, 
which allows a finite answer to emerge, although a new set of divergences along the timelike
surface $\mathcal{T}$ must be shown to cancel. A similar story emerges in section \ref{sec:D=dt}
for $\D=\dt$,
although extra care must be taken due to the presence of logarithms in the solutions.   

\subsection{$\D >\dt$} \label{sec:D>dt}
The full bulk scalar field separates into two parts, 
\beq
\phi = \phi_0 + \phi_\omega,
\eeq
with $\phi_\omega$ from (\ref{eqn:phio}) describing the state deformation, while $\phi_0$ 
corresponds to the theory deformation $g(x)$. Since no IR divergences arise at this order in
perturbation theory when $\D>\dt$, we can take $g$ to be constant everywhere.  The solution
in the bulk on the Euclidean section then takes the simple form
\beq
\phi_0 = g z^{d-\D}.
\eeq

Given these two solutions, the bulk contribution to $\delta S^{(2)}$ may be computed using 
equation (\ref{eqn:Eint}).  Note that $\partial_\tau\phi=0$ on the $\tau=0$ surface, so we 
only need the $\nabla^2 \phi^2$ term in the integrand.  Before evaluating this term, it is useful
to expand $\phi_\omega$ near $z=0$.  This expansion takes the form
\beq\label{eqn:phioseries}
\phi_\omega = \left[\lambda_\omega z^{d-\D}\sum_{n=0}^{\infty} a_n (\omega z)^{2n}
+ \beta_\omega z^{\D}\sum_{n=0}^\infty b_n (\omega z)^{2n}\right]\cos(\omega\tau),
\eeq
where 
\beq\label{eqn:bo}
\beta_\omega = \lambda_\omega \left(\frac\omega2\right)^{2\D-d} \frac{\Gamma(\dt-\D)}{\Gamma(
\D-\dt)},
\eeq
and the coefficients $a_n$ and $b_n$ are given in appendix \ref{sec:Bessel}.  The $O(\lambda^1)$
term in $\phi^2$ is $2\phi_0\phi_\omega$, and this modifies the power series
(\ref{eqn:phioseries}) by changing the leading powers to $z^{2(d-\D)}$ and $z^d$.  The 
Laplacian in the bulk is 
\beq
\nabla^2 = z^2\partial_\tau^2+z^{d+1}\partial_z(z^{-d+1}\partial_z).
\eeq
Acting on the $\phi_0\phi_\omega$ series, the effect of the $\tau$ derivative is to multiply
by $-\omega^2z^2$, which shifts each term to one higher term in the series.  The $z$ derivatives
do no change the power of $z$, but rather multiply each term by a constant, 
$2(d-\D+n)(d-2\D+2n)$
for the $a_n$ series and $2  n(d+2n)$ for the $b_n$ series (note in particular it annihilates the 
first term in the $b_n$ series).  After this is done, the series may be reorganized for $\tau=0$ as
\beq \label{eqn:cndn}
2\nabla^2
\phi_0\phi_\omega = 2g\lambda_\omega z^{2(d-\D)}\sum_{n=0}^\infty c_n(\omega z)^{2n} +
2g\beta_\omega z^d\sum_{n=1}^\infty d_n(\omega z)^{2n},
\eeq
with the coefficients $c_n$ and $d_n$ computed in appendix \ref{sec:Bessel}.

From this, we simply need to evaluate the integral (\ref{eqn:Eint}) for each term in the series.
For a given term of the form $A z^\eta$, the contribution to $\delta S^{(2)}$ is 
\begin{align}
\delta S^{(2)}_\eta &
= -\frac\pi2 \Omega_{d-2} \int_{z_0}^R \frac{dz}{z^{d+1}} \int_0^{\sqrt{R^2-z^2}}
dr\, r^{d-2} \left[\frac{R^2-r^2-z^2}{2R}\right] A z^\eta \\
\label{eqn:Eresult}
&=-A\frac{\pi\Omega_{d-2}}{4(d^2-1)} \left[R^\eta\,\frac{\Gamma(\dt+\frac32)\Gamma(\frac\eta2
-\dt)}{\Gamma(\frac\eta2+\frac32)}  + \frac{R^d z_0^{\eta-d}
\tensor[_2]{F}{_1}\left(-\frac{d+1}{2}, \frac{\eta-d}{2}; \frac{\eta-d}{2}+1;\frac{z_0^2}{R^2} \right)}
{\frac\eta2-\dt}  
 \right].
\end{align}
The second term in this expression contains a set of divergences at $z_0\rightarrow0$ for all
values of $\eta<d$.  These arise exclusively from the $c_n$ series in (\ref{eqn:cndn}). 
In general, the expansion of the hypergeometric function near $z_0=0$ can produce 
subleading divergences, which mix between different terms from the series (\ref{eqn:cndn}).  
These divergences
eventually must cancel against compensating divergences that arise from the $\mathcal{T}$
surface integral in (\ref{eqn:Tint}).  Although we do not undertake a systematic study of 
these divergences, we may assume that they cancel out because the cutoff surface at $z_0$ was 
chosen arbitrarily, and the original integral (\ref{eqn:H+int}) made no reference to it.  Thus,
we may simply discard these $z_0$ dependent divergences, and are left with only the 
first term in   (\ref{eqn:Eresult}).\footnote{When $\eta=d+2j$ for an integer $j$, there are 
subtleties
related to the appearance of $\log z_0$ divergences.  These cases arise when $\D=\dt+m$
with $m$ an integer.  We leave analyzing this case for future work.  \renewcommand{\baselinestretch}{1}\footnotesize}  

There is another reason for discarding the $z_0$ divergences immediately: they only arise
in states with divergent energy density.  The coefficient of a term with a $z_0$ divergence
is $2gc_n \omega^{2n} \lambda_\omega$.  The final answer for the entanglement entropy 
will involve integrating over all values of $\omega$. But the requirement of finite energy 
density (\ref{eqn:lambdaoint}) shows that all terms with $n\leq q$, corresponding to
$\eta\leq2d-2\D+2q$, will vanish from the final result.  Given the definition of $q$ in 
(\ref{eqn:q}), these are precisely the terms in (\ref{eqn:Eresult}) that have divergences
in $z_0$.  Note that since $\beta_\omega\propto \omega^{2\D-d}$, which is generically
a non-integer power, the integral over $\omega$ will not vanish, so all the $\beta_\omega$
terms survive.

The resulting bulk contribution to the entanglement entropy at order $\lambda g$ is
\begin{align}
\delta S^{(2)}_{\mathcal{E},\lambda g} = -\frac{g \pi^{\dt+\frac12} }{4} \int_0^\infty d\omega
\left[
\lambda_\omega R^{2(d-\D)} \sum_{n=q+1}^\infty \right. 
& c_n \frac{\Gamma(\dt -\D +n)}{\Gamma(d-\D +\frac32+n)}(\omega R)^{2n}  \nonumber  \\
\label{eqn:dS2Elg}
+\; \beta_\omega R^d \sum_{\hphantom{+} n=1 \hphantom{+} }^\infty 
& \left.   \vphantom{\sum_{n=1}^\infty }
d_n \frac{\Gamma(\dt + n) }{\Gamma(\dt + \frac32+n)} (\omega R)^{2n}
\right].
\end{align}
This expression shows that the lowest order pieces scale as $R^{2(d-\D+q+1)}$ and 
$R^{d+2}$, which both become subleading with respect to the $R^d$ scaling 
of the $\delta S^{(1)}$ piece for small ball size.  Note that a similar technique could 
extend this result to spatially dependent $\lambda(x)$, and simply would amount to an
additional series expansion.

One could perform a similar analysis for the $O(\lambda^2)$ contribution from $\delta S^{(2)}$.  
The series of $\nabla^2\phi_\omega\phi_{\omega'}$ would organize into three series, with 
leading coefficients $\lambda_\omega\lambda_{\omega'} z^{2(d-\D)}$, $(\beta_\omega 
\lambda_{\omega'} +\lambda_\omega\beta_{\omega'})z^d$, and $\beta_\omega \beta_{\omega'}
z^{2\D}$. After integrating over $\omega$ and $\omega'$, and noting which terms vanish 
due to the requirement (\ref{eqn:lambdaoint}), one would find the leading contribution
going as $\beta^2R^{2\D}$.  The precise value of this term is 
\beq\label{eqn:R2Dlambda2}
\delta S^{(2)}_{\lambda^2} = -\frac{\pi\Omega_{d-2}}{d^2-1} R^{2\D}\big(\delta\vev{\op}\big)^2 
\,\frac{\D \Gamma(\dt 
+\frac32) \Gamma(\D-\dt+1)}{(2\D -d)^2\Gamma(\D+\frac32)},
\eeq
which is quite similar to the $R^{2\D}$ term in equation (\ref{eqn:Dneqdt}).
This is again subleading when $\D>\dt$, but the same terms 
show up for $\D\leq\dt$
in sections \ref{sec:D<dt} and \ref{sec:D=dt}, where they become 
the dominant contribution when $R$ is taken small enough.  The importance of these
second order terms in the small $R$ limit was first noted in \cite{Casini2016a}.

The remaining pieces to calculate come from the integral over $\mathcal{T}$ given by 
(\ref{eqn:Tint}), and $\delta S^{(1)}$ in (\ref{eqn:dS1}), 
which just depends on $\delta\vev{T^0_{00}}$.  When
$\D>\dt$, the only contribution from the $\mathcal{T}$ surface integral is near $t_B\sim z
\rightarrow 0$.  These terms were analyzed in appendix E of \cite{Faulkner2015}, and were
found to give two types of contributions.  The first were counter terms that cancel against
the divergences in the bulk as well as the divergence in $\delta S^{(1)}$.  Although subleading
divergences were not analyzed, these can be expected to cancel in a predictable way.  We
also already argued that such terms are not relevant for the present analysis, due to the 
requirement of finite energy density.  The second type of term is finite, and takes the form
\beq\label{eqn:intDb}
\delta S^{(2)}_{\mathcal{T},\text{finite}} = -2\pi\D\int_{\Sigma} \zeta^t g \beta.
\eeq
The relation between $\beta$ and $\delta\vev{\op}$ identified in (\ref{eqn:beta}) implies
from equation (\ref{eqn:bo}),
\beq \label{eqn:dvevoD>dt}
\delta\vev{\op} = \lambda_\omega\frac{2\Gamma(\dt-\D+1)}{\Gamma(\D-\dt)}
\left(\frac{\omega}{2}\right)^{2\D-d},
\eeq
and assuming the ball is small enough so that this expectation value may be considered 
constant, (\ref{eqn:intDb}) evaluates to
\beq \label{eqn:dS2Tfin}
\delta S^{(2)}_{\mathcal{T},\text{finite}} = 2\pi \frac{\Omega_{d-2} R^d}{d^2-1} \left[\frac{\D}{2\D-d}
g\delta\vev{\op}\right].
\eeq
Similarly, taking $\delta \vev{T^0_{00}}$  to be constant over the ball, the final contribution is
the variation of the modular Hamiltonian piece, given by
\beq\label{eqn:dS1gl}
\delta S^{(1)} = 2\pi \int_\Sigma \zeta^t\delta\vev{T^0_{00}} = 2\pi \frac{\Omega_{d-2} R^d}{d^2-1}
\delta\vev{T^0_{00}}.
\eeq

Before writing the final answer, it is useful to write $\delta\vev{\op}$ in terms of the trace of 
the stress tensor of the deformed theory, 
$T^g$.  The two are related by the dilatation Ward identity, which gives \cite{Osborn1994}
\beq\label{eqn:dTg}
\delta\vev{T^g} = (\D-d)g\delta\vev{\op}.
\eeq
Then, using the definition of the deformed theory's stress tensor (\ref{eqn:Tgabdef}) and 
summing up the contributions (\ref{eqn:dS2Elg}), (\ref{eqn:dS2Tfin}), and (\ref{eqn:dS1gl}),
the total variation of the entanglement entropy at $O(\lambda^1 g^1)$ is 
\beq \label{eqn:dSlg}
\delta S_{\lambda g} = 2\pi \frac{\Omega_{d-2} R^d}{d^2-1}\left[\delta\vev{T^g_{00}}
-\frac1{2\D-d} \delta\vev{T^g}\right] +\delta S^{(2)}_{\mathcal{E},\lambda g} .
\eeq
Since $\delta S^{(2)}_{\mathcal{E},\lambda g}$ is subleading, 
this matches the result (\ref{eqn:Dneqdt}) quoted in the introduction, apart from the $R^{2\D}$
term, which is not present because we have arranged for the renormalized vev 
$\vev{\op}_g$ to vanish.  However, as noted in equation (\ref{eqn:R2Dlambda2}), we
do find such a term at second order in $\lambda$.

\subsection{$\D < \dt$} \label{sec:D<dt}
Extending the above calculation to $\D<\dt$ requires the introduction of one novel
element: a modification of the coupling $g(x)$ to include an IR cutoff.  It is straightforward to 
see why this regulator is needed.  The perturbative calculation of the entanglement entropy
involves integrals of the two point correlator over all of space, schematically of the form
\beq
\int d^dx g(x)\mvev{\Big}{\op(0)\op(x)}_0 = \int d^d x \frac{c_\D g(x)}{x^{2\D}}.
\eeq
If this is cut off at a large distance $L$, the integral scales as $L^{d-2\D}$ (or $\log L$ for 
$\D=\dt$) when the coupling $g(x)$ is constant.  This clearly diverges for $\D\leq\dt$.  

The usual story with IR divergences is that resumming the higher order terms remedies
the divergence, effectively imposing an IR cut off.  Presumably this cut off is set by the scale
of the coupling $L_\text{eff}\sim g^{\frac1{\D-d}}$, but since it arises from higher order 
correlation functions, it may also depend on the details of the underlying CFT.  Although
it may still be possible to compute these IR effects in perturbation theory 
\cite{Zamolodchikov1991, Guida1996, Guida1997}, this goes beyond the techniques employed in the present work.  
However, if we work on length scales small compared to the IR scale, it is possible to
capture the qualitative behavior by simply putting in an IR cut off by hand
 (see \cite{Berenstein2014} for a related approach).  We implement this 
IR cutoff by setting the coupling $g(x)$ to zero when $|\tau|\geq L$.\footnote{This will
work only for $\D>\dt-\frac12$.  For lower operator dimensions, a stronger regulator is needed,
such as a cutoff in the radial direction,
but the only effect this should have is to change the value of $\vev{\op}_g$.
 \renewcommand{\baselinestretch}{1}\footnotesize}  We may then
express the final answer in terms of the vev $\vev{\op}_g$, which implicitly depends on
the IR regulator $L$.  

The bulk term $\delta S^{(2)}$ involves a new set of divergences from the $\mathcal{T}$ surface
integral that were not present in the original calculation for $\D>\dt$ \cite{Faulkner2015}.  
To compute these divergences and show that they cancel, we will need the real time 
behavior of the bulk scalar fields, in addition to its behavior at $t=0$.  
These are described in appendix \ref{sec:realt}.
The important features are that $\phi_0$ on the $t=0$ surface takes the form
\beq\label{eqn:phi0Etext}
\phi_0 = -\frac{\vev{\op}_g}{2\D-d} z^\D + g z^{d-\D},
\eeq
and the vev $\vev{\op}_g$ is determined in terms of the IR cutoff $L$ by
\beq
\vev{\op}_g = 2gL^{d-2\D}\frac{\Gamma(\D-\dt+\frac12)}{\sqrt{\pi}\,\Gamma(\D-\dt)}.
\eeq
For $t>0$, the time-dependent solution is given by
\beq\label{eqn:phi0Ttext}
\phi_0 = -\frac{\vev{\op}_g}{2\D-d} z^\D + g z^{d-\D} F(t/z),
\eeq
where the function $F$ is defined in equation (\ref{eqn:F}).  To compute the divergences
along $\mathcal{T}$, the form of this function is needed in the region $t\gg z$, where it simply becomes 
\beq
F(t/z)\xrightarrow{t\gg z} B \left(\frac{t}{z}\right)^{d-2\D},
\eeq
with the proportionality constant $B$ given in equation (\ref{eqn:Fasym}).
The field $\phi_\omega$ behaves similarly as long as $\omega^{-1}\gg z,t$.  In particular,
it has the same form as $\phi_0$ in equations (\ref{eqn:phi0Etext}) and (\ref{eqn:phi0Ttext}),
but with $g$ replaced by $\lambda_\omega$, and $\vev{\op}_g$ replaced with $\delta\vev{\op}$,
given by
\beq
\delta\vev{\op} = \lambda_\omega\frac{2\Gamma(\dt-\D+1)}{\Gamma(\D-\dt)}
\left(\frac{\omega}{2}\right)^{2\D-d},
\eeq
which is the same relation as for $\D>\dt$, equation (\ref{eqn:dvevoD>dt}).

Armed with these solutions, we can proceed to calculate
$\delta S^{(2)}$.  In this calculation, the contribution from the timelike surface $\mathcal{T}$
now has a novel role.  Before, when $\D>\dt$, the integral from this surface died
off as $z\rightarrow 0$ in the region $t_B>z$, and hence the integral there did not need
to be evaluated.  For $\D<\dt$, rather than dying off, this integral is now leads to divergences
as $z\rightarrow0$.  These divergences either cancel among themselves, or cancel against
divergences coming from bulk Euclidean surface $\mathcal{E}$, so that a finite answer is 
obtained in the end.  These new counterterm divergences seem to be related to the 
alternate quantization in holography \cite{Klebanov1999, Casini2016a}, which
invokes a different set of boundary counterterms when defining the bulk AdS action.  It 
would be interesting to explore this relation further.  

At first order in $g$ and $\lambda$, three types of terms will appear,  proportional to each of
$\vev{\op}_g\,\delta\vev{\op}$,  $(g\delta\vev{\op} +\lambda(0)\vev{\op}_g)$, or 
$g \lambda(0)$.  Here, we allow $\lambda(0)\neq0$ because there are no UV divergences 
arising in the energy density or $\op$ expectation values when $\D<\dt$. The descriptions
of the contribution from each of these terms are given below, and the details
of the surface integrals over $\mathcal{E}$ and $\mathcal{T}$  are
contained in appendix \ref{sec:D<dtcalc}.  

The $\vev{\op}_g\delta\vev{\op}$ term has both a finite and a divergent piece coming 
from the integral over $\mathcal{E}$ (see equation (\ref{eqn:dSEz2D})).  This divergence
is canceled by the $\mathcal{T}$ integral in the region $t_B\gg z_0$.  This is interesting since
it differs from the $\D>\dt$ case, where the bulk divergence was canceled by the $\mathcal{T}$
integral in the region $t_B \lesssim z_0$.   The final finite contribution from this term is 
\beq
\delta S^{(2)}_{\mathcal{E},1} = -2\pi \vev{\op}_g\,\delta\vev{\op}
\frac{\Omega_{d-2}}{d^2-1} R^{2\D}
\frac{\D \Gamma(\dt+\frac32)\Gamma(\D-\dt+1)}{ (2\D-d)^2 \Gamma(\D+\frac32)} .
\eeq
It is worth noting that we can perform the exact same calculation with $\vev{\op}_g\delta
\vev{\op}$ replaced by $\frac12\delta\vev{\op}^2$ to compute the second order in $\lambda$
change in entanglement entropy.  The value found in this case agrees with  holographic
results \cite{Casini2016a}.  

The $g\delta\vev{\op}+\lambda(0)\vev{\op}_g$ term receives no contribution from the 
$\mathcal{E}$ surface at leading order since this term in $\phi^2$ scales as $z^d$ 
in the bulk, and the $z$-derivatives in the Laplacian $\nabla^2$ annihilate such a term.  
The surface $\mathcal{T}$ produces a finite term, plus a collection of divergent terms 
from both regions $t\sim z$ and $t\gg z$, which cancel among themselves.  The finite 
term is given by
\beq
\delta S^{(2)}_{\mathcal{T},2} = 2\pi \frac{\Omega_{d-2} R^d\D}{(d^2-1)(2\D-d)}(g\delta\vev{\op}
+\lambda(0)\vev{\op}_g), 
\eeq
which is exactly analogous to the term (\ref{eqn:dS2Tfin}) found for the case $\D>\dt$.  

Finally, the term with coefficient $\lambda(0) g$ produces subleading terms, scaling as 
$R^{2(d-\D+n)}$ for positive integers $n$.  Since these terms are subleading, we do not
focus on them further.  In this case, it must also be shown that the divergences appearing
in the $\mathcal{T}$ cancel amongst themselves, since no divergences arise from the 
$\mathcal{E}$ integral.  The calculations in appendix \ref{sec:D<dtcalc} verify that 
this indeed occurs.

We are now able to write down the final answer for the change in entanglement entropy for
$\D<\dt$.  The contribution from $\delta S^{(1)}$ is exactly the same as the $\D>\dt$ case,
and is given by (\ref{eqn:dS1gl}).  Following the same steps that led to equation 
(\ref{eqn:dSlg}), the contributions from the finite piece of $\delta S^{(2)}_{\mathcal{E},1}$ 
in (\ref{eqn:dSEz2D}) and $\delta S^{(2)}_{\mathcal{T},2}$ in (\ref{eqn:dS2T2}) combine
with $\delta S^{(1)}$ to give
\beq
\delta S_{\lambda g} = \frac{2\pi \Omega_{d-2}}{d^2-1} \left[R^d\left(
\vev{T^g_{00}} -\frac{1}{2\D-d}
\vev{T^g} \right)-R^{2\D} \vev{\op}_g\delta\vev{\op} \frac{\D\Gamma(\dt+\frac32) 
\Gamma(\D-\dt+1)}{(2\D-d)^2 \Gamma(\D+\frac32)}   \right],
\eeq
where we have set $\lambda(0)=0$ for simplicity and to match the expression for 
$\D>\dt$, which required $\lambda(0)=0$.  

\subsection{$\D = \dt$} \label{sec:D=dt}
Similar to the $\D<\dt$ case, there are IR divergences that arise when $\D=\dt$.  These 
are handled as before with an IR cutoff $L$, on which the final answer explicitly 
depends.  A new feature arises, however, when expressing the answer in terms of  $\vev{\op}_g$
rather than $L$: the appearance of a renormalization scale $\mu$.  The need 
for this renormalization scale can be seen by examining the expression for $\vev{\op}_g$, 
which depends on the $\op\op$ two-point function with $\D=\dt$:
\beq
\vev{\op}_g = -\int d^d x \frac{g c'_\D}{x^d}  =
-gc'_\D\frac{\pi^{\dt}}{\Gamma(\dt)} \int \frac{d\tau}{\tau}. 
\eeq
This has a logarithmic divergence near $x=0$ which must be regulated.  The UV-divergent
piece can be extracted using the point-splitting cutoff for $|\tau|<\delta$; however, there is
an ambiguity in identifying this divergence since the upper bound of this integral
cannot be sent to $\infty$.  The appearance of the renormalization scale is related to 
matter conformal anomalies that exist for special values of $\D$ \cite{Osborn:1991gm,
Osborn1994, Petkou1999}.   Thus we must impose an upper cutoff on the integral, which
introduces the renormalization scale $\mu^{-1}$.  The divergent piece of $\vev{\op}_g$
is then
\beq
\vev{\op}_g^\text{div.} = g c'_\D \frac{\pi^{\dt}}{\Gamma(\dt)} 2\log\mu\delta.
\eeq
Now we can determine the renomalized vev of $\op$, using the IR-regulated $\tau$
integral,
\begin{align}
\vev{\op}_g^\text{ren.} = \vev{\op}_g-\vev{\op}_g^\text{div.} &= - \int^L d\tau\int d^{d-1}x
\frac{g c'_\D}{x^d} - gc'_\D \frac{\pi^{\dt}}{\Gamma(\dt)} 2\log\mu\delta \\
&=
\label{eqn:vevOren}
-gc'_\D\frac{\pi^{\dt}}{\Gamma(\dt)} 2\log \mu L.
\end{align}
The final answer we derive for the entanglement entropy when $\D=\dt$ will depend on 
$\log L$ but not on explicitly $\mu$ or $\vev{\op}_g$.  Only after rewriting it in terms of 
$\vev{\op}^\text{ren.}_g$ does the $\mu$ dependence appear.  

One other small modification is necessary when $\D=\dt$.  The normalization  $c_\D$
for the $\op\op$ two point function defined in (\ref{eqn:cD}) has a double zero at $\D=\dt$ 
which must be removed.  
This is easily remedied by dividing by $(2\D-d)^2$ \cite{Klebanov1999, Freedman:1998tz}, 
so that 
the new constant appearing in the two point function is 
\beq \label{eqn:cD'}
c'_\D = \frac{\Gamma(\D)}{2\pi^{\dt} \Gamma(\D-\dt+1)} \;\xrightarrow{\hphantom{n}
\D\rightarrow\dt \hphantom{n}} \;
\frac{\Gamma(\dt)}{2\pi^{\dt}} .
\eeq
This change affects the normalization of the bulk field $\phi$ by dividing by a single
factor of $1/(2\D-d)$, so that
\beq\label{eqn:phicD'}
\phi(x_B) = \frac{\Gamma(\dt)}{2\pi^{\dt}}\int_{C(\delta)} d\tau
\int d^{d-1}\vec{x}
\frac{z^\D f(\tau,\vec{x})}{(z^2+(\tau - i t_B)^2+ (\vec{x}-\vec{x}_B)^2)^\D}.
\eeq

These are all the components needed to proceed with the calculation of the entanglement 
entropy.  As before, we solve for the bulk field $\phi_0$ associated with a constant 
coupling $g$, set to zero for $|\tau|>L$.  The $\phi_\omega$ field associated with the 
state deformation $\lambda = \lambda_\omega \cos\omega\tau$ is again given by 
a modified Bessel function on the Euclidean section.  Its form along the timelike surface
$\mathcal{T}$ is derived from the integral representation (\ref{eqn:phicD'}), and particular
care must be taken in the region $t_B\sim z$, where a divergence in $\delta$ appears.
Although this divergence is not present if we require $\lambda(0)=0$, we analyze the 
terms that it produces for generality.  This $\delta$ divergence is shown to cancel
against a similar divergence in $\delta S^{(1)}$ related to the divergence in the $\vev{T_{00}
\op\op}$ three-point function.   

The full real-time solutions for $\phi_0$ and $\phi_\omega$ are given 
in appendix \ref{sec:D=dtrt}.
The $\phi_0$ solution from equation (\ref{eqn:gzd2G}) takes the form
\beq \label{eqn:phi0gG}
\phi_0 = gz^{\dt} G(t_B/z, \delta / z, L/z),
\eeq 
with the function $G$ defined in equation (\ref{eqn:G}).  The dependence of this 
function on $\delta$ is needed only in the region $t_B\sim z$; everywhere else it can
safely be taken to zero.  On the $\mathcal{E}$ surface where 
$t_B=0$, the solution  in the limit $L\gg z$ is 
\beq \label{eqn:phi0vevOren}
\phi_0 = g z^{\dt} \log\frac{2L}{z}  = -\vev{\op}_g^\text{ren.} - g z^{\dt} \log \frac{\mu z}{2},
\eeq 
where the second equality uses the value of $\vev{\op}_g^{\text{ren.}}$ derived in 
(\ref{eqn:vevOren}).  
We also need $\phi_0$ in the region $t_B\gg z$, given by
\beq
\phi_0 = g z^{\dt}\log \frac{L}{t_B} .
\eeq
For $\phi_\omega$, the solution on the $\mathcal{E}$ surface is still given by a modified Bessel
function as in equation (\ref{eqn:phio}), but must be divided by $(2\D-d)$ according 
to our new normalization (\ref{eqn:phicD'}), 
\beq \label{eqn:phiD=dtz0}
\phi_\omega = \lambda_{\omega} z^\dt K_0(\omega z)\xrightarrow{z\rightarrow 0}-
\lambda_\omega z^{\dt} \left(
\gamma_E + \log\frac{\omega z}{2}\right). 
\eeq  
By writing the argument of the $\log$ term as in equation (\ref{eqn:phi0vevOren}), one can
read off the renormalized operator expectation value,
\beq \label{eqn:dOrentext}
\delta\vev{\op}^\text{ren.} = \lambda_\omega \left(\gamma_E + \log\frac{\omega}{\mu}\right).
\eeq
Beyond $t_B=0$, as long as $\omega^{-1}\gg t_B$, the solution can be written in a similar
form as (\ref{eqn:phi0gG}).  When $t_B\gg z$, this is given by 
\beq
\phi_\omega = -\lambda_\omega z^{\dt}(\gamma_E + \log \omega t_B). 
\eeq

Now that we have the form of the solutions on the surfaces $\mathcal{E}$ and $\mathcal{T}$, 
the entanglement calculation contains four parts.  The first is the integral over $\mathcal{E}$,
where a $\log z_0$ divergence appears.  This cancels against a collection of divergences
from the $\mathcal{T}$ surface.  The second part is the $\mathcal{T}$ surface near 
$t_B\sim z$.  This region produces  more divergences
in $z_0$ and $\delta$, some of which cancel the bulk divergence.  
The third part is the integral over $\mathcal{T}$ for $t_B\gg z$, 
which eliminates the remaining $z_0$ divergences.  Finally, an additional divergence from
the stress tensor in $\delta S^{(1)}$ cancels the  $\delta$ divergence, producing a finite
answer.

Appendix \ref{sec:D=d2calc} describes the details of these calculations.  In the end, 
the contributions from 
equations (\ref{eqn:dS2Efinlog}), (\ref{eqn:dS2Ediv}),
(\ref{eqn:dS2Tdivlog}), (\ref{eqn:dS2T1+2}) and (\ref{eqn:deltact}) combine
together to give the following total change in entanglement entropy, at $O(\lambda^1 g^1)$,
\begin{align}
\delta S_{\lambda_\omega g}&=2\pi  \frac{\Omega_{d-2}R^d}{d^2-1}\left\{
\delta\vev{T^0_{00}}^\text{ren.} +g\lambda_\omega \left[ 
\dt \logp{\frac{2L}{R}}
\left(\gamma_E + \log\frac{\omega R}{2}\right)    \right. \right. \nonumber \\
&\left.\left. + \frac{d}{4}H_{\frac{d+1}{2}}
\left(\gamma_E+\log\frac{R^2\omega}{4L}\right)
 -  \log\mu R  -\frac18
\left(H^{(2)}_{\frac{d+1}{2}} +H_{\frac{d+1}{2}} (H_{\frac{d+1}{2}}-2)\right) \right]
\right\}.    \label{eqn:D=d2final}
\end{align}
This is the answer for a single frequency $\omega$ in the state deformation function 
$\lambda(x)$.  Since $\lambda(0)\neq 0$, this result cannot be immediately interpreted as 
the entanglement entropy of an excited state, since the state has a divergent expectation 
value for $\op$.\footnote{However, viewing $\omega$ as an IR regulator, this equation can be 
adapted to express
the change in \emph{vacuum} 
entanglement entropy between a CFT and the deformed theory.  \renewcommand{\baselinestretch}{1}\footnotesize}
To get the entanglement entropy for an excited state, we should integrate over all 
frequencies, and use the fact that $\int d\omega \lambda_\omega=0$.  When this is 
done, all terms with no $\log\omega$ dependence drop out.  Also, we no longer need to specify
that operator expectation values are renormalized, since the change in expectation values
between two states is finite and scheme-independent.  

We would like to express the answer in terms of $\delta\vev{\op}$.  By integrating equation 
(\ref{eqn:dOrentext}) over all frequencies and using that $\lambda(0)=0$, we find
\beq
\delta\vev{\op} = \int_0^\infty d\omega\,\lambda_\omega\log\omega.
\eeq
With this, the total change in entanglement entropy for nonsingular states coming from 
integrating \ref{eqn:D=d2final} over all frequencies is 
\beq
\delta S_{\lambda g} = 2\pi\frac{\Omega_{d-2} R^d}{d^2-1} \left[
\delta\vev{T^0_{00}} + g\dt\delta\vev{\op} \left(\frac12 H_{\frac{d+1}{2}}+\log\frac{2L}{R}\right)
\right].
\eeq
This can be expressed in terms of the deformed theory's stress tensor $T^g_{00}$ and 
trace $T^g$ using equations (\ref{eqn:Tgabdef}) and (\ref{eqn:dTg}),
\beq \label{eqn:D=d2logL}
\delta S_{\lambda g} = 2\pi\frac{\Omega_{d-2} R^d}{d^2-1} \left[
\delta\vev{T^g_{00}} + \delta\vev{T^g}\left(\frac2d-\frac12 H_{\frac{d+1}{2}} + \log\frac{R}{2L}
\right)
\right].
\eeq
Although the answer is 
scheme-independent in the sense that $\mu$ does not explicitly appear, there is a dependence
on the IR cutoff $L$.  This cutoff is related to the renormalized vev $\vev{\op}^\text{ren.}_g$
via (\ref{eqn:vevOren}), which does depend on the renormalization scheme.  
Thus the dependence on $L$ in
the above answer can be traded for $\vev{\op}_g^\text{ren.}$, at the cost of introducing 
(spurious) $\mu$-dependence, 
\beq \label{eqn:dSlgvevO}
\delta S_{\lambda g} = 2\pi\frac{\Omega_{d-2} R^d}{d^2-1} \left[
\delta\vev{T^g_{00}} + \delta\vev{T^g}\left(\frac2d-\frac12 H_{\frac{d+1}{2}} + \log\frac{\mu R}{2}
\right) -\dt \vev{\op}_g\delta\vev{\op} 
\right],
\eeq
which is the result quoted in the introduction, equation (\ref{eqn:Deqdt}).

\section{Discussion} \label{sec:discussion}

The equivalence between the Einstein equation and maximum vacuum entanglement 
of small balls relies on 
a conjecture about the behavior of the entanglement entropy of excited states, equation
(\ref{eqn:dSIRmod}).  This chapter has sought to check the conjecture in CFTs deformed
by a relevant operator.  In doing so, we have derived new results on the behavior of
excited state entanglement entropy in such theories, encapsulated by equations
(\ref{eqn:Dneqdt}) and (\ref{eqn:Deqdt}).  These results agree with holographic calculations
\cite{Casini2016a}
that employ the Ryu-Takayanagi formula.  Thus, this chapter extends those results to 
\emph{any} CFT, including those which are not thought to have holographic duals.  

For deforming operators of dimension $\D>\dt$ considered in section \ref{sec:D>dt}, 
the calculation is a straightforward 
application of Faulkner's method for computing entanglement entropies \cite{Faulkner2015}.  
One subtlety in this case is the presence of UV divergences in 
$\delta\vev{\op}$ and $\delta\vev{T^0_{00}}$ unless
the state deformation function $\lambda(x)$ is chosen appropriately.  As discussed in 
section \ref{sec:excited}, this translates to the condition that $\lambda$ and sufficiently
many of its $\tau$-derivatives vanish on the $\tau=0$ surface.  When the entanglement
entropy of the state is calculated, this condition implies that terms scaling with the ball
radius as $R^{2(d-\D+n)}$, which are present for generic $\lambda(x)$, vanish, 
where $n$ is a positive integer less than or equal to
$\left\lfloor\D-\dt\right\rfloor$.  As $R$ approaches
zero, these terms  
dominate over the energy density term, which scales as $R^d$.  This shows that regularity 
of
the state translates to the dominance of the modular Hamiltonian term 
in the small ball limit when $\D>\dt$.  The subleading terms arising from 
this calculation are  given in equation (\ref{eqn:dS2Elg}).

Section \ref{sec:D<dt} then extends this result to operators of dimension $\D<\dt$.  In this 
case, IR divergences present a novel facet to the calculation.  To deal with these divergences,
we impose an IR cutoff on the coupling $g(x)$ at  scale $L$.  A more complete 
treatment of the IR divergences would presumably involve resumming higher order 
contributions, which then would effectively impose an IR cutoff in the lower order terms.  
This cutoff should be of the order $L_\text{eff.}\sim g^{\frac{1}{\D-d}}$, but can depend
on other details of the CFT, including any large parameters that might be present.  Note
this nonanalytic dependence of the IR cutoff on the coupling signals nonperturbative
effects are at play \cite{Nishioka:2014kpa,Herzog2013}.  
After the IR cutoff is imposed, the calculation
of the entanglement entropy proceeds as before.  In the final answer, equation 
(\ref{eqn:Dneqdt}), the explicit dependence on the IR cutoff is traded for the renormalized
vacuum expectation value $\vev{\op}_g$.  This expression agrees with the holographic
calculation to first order in $\delta\vev{\op}$ in the case that $\vev{\op}_g$ is nonzero
\cite{Casini2016a}.  

Finally, the special case of $\D=\dt$ is addressed in section \ref{sec:D=dt}.  Here, both
 UV and IR divergences arise, and these are dealt with in the same manner as the 
$\D>\dt$ and $\D<\dt$ cases.  The answer before imposing that the state is nonsingular
is given in equation (\ref{eqn:D=d2final}), and it depends logarithmically on an arbitrary
renormalization scale $\mu$.   This scale $\mu$ arises when renormalizing the 
stress tensor expectation value $\delta\vev{T^0_{00}}$,  
as is typical of logarithmic UV divergences.  
Note that the dependence on $\mu$ in the final answer is only superficial, since the 
combination $\delta\vev{T^0_{00}}^\text{ren.} - \log\mu{R}$ appearing there is
independent of the choice of $\mu$.  Furthermore, for regular states, $\delta\vev{T^0_{00}}$
is UV finite, and hence the answer may be written without reference to the renormalization scale as in (\ref{eqn:D=d2logL}),
although it explicitly depends on the IR cutoff.  In some cases, such as free field theories, 
the appropriate IR cutoff may be calculated exactly \cite{Casini2009,Blanco2011, 
Casini2016a}.  
Re-expressing the answer in terms of  $\vev{\op}_g$ instead of the IR cutoff, 
as in equation (\ref{eqn:Deqdt}),
re-introduces the renormalization scale $\mu$, since the 
vev requires renormalization and hence is $\mu$-dependent.  Again, this dependence
on $\mu$ is superficial; it cancels between $\vev{\op}_g$ and the $\log\frac{\mu R}{2}$ 
terms.

\subsection{Implications for the Einstein equation} \label{sec:EEimps}
We now ask whether the results (\ref{eqn:Dneqdt}) and (\ref{eqn:Deqdt}) are consistent
with the conjectured form of the entanglement entropy variation (\ref{eqn:dSIRmod}).  The
answer appears to be yes, with the following caveat: the scalar function $C$ explicitly
depends on the ball size $R$.  This comes about from the $R^{2\D}$ in equation 
(\ref{eqn:Dneqdt}), in which case $C$ contains
a piece scaling as $R^{2\D-d}$,  and from the $R^d \log R$ term in (\ref{eqn:Deqdt}),
which gives $C$ a $\log R$ term.  When $\D\leq\dt$, these terms are the dominant
component of the entanglement entropy variation when the ball size is taken to be small.  

The question now shifts to whether $R$-dependence in the function $C$ still allows the 
derivation of the Einstein equation to go through.  As long as $C(R)$ transforms as a 
scalar under Lorentz boosts for fixed ball size $R$, the tensor equation (\ref{eqn:EEtensor})
still follows from the conjectured form of the entanglement entropy variation (\ref{eqn:dSIRmod})
\cite{Jacobson2015a}.  
One then concludes from stress tensor conservation and the Bianchi identity that the
curvature scale of the maximally symmetric space characterizing the local vacuum
is dependent on the size of the ball, $\Lambda=\Lambda(x,R)$.\footnote{This idea 
was proposed by Ted Jacobson, and I thank him for 
for discussions regarding this point.  \renewcommand{\baselinestretch}{1}\footnotesize}  
There does not seem to be an 
immediate reason  disallowing an $R$-dependent $\Lambda$.  

There are two requirements on $\Lambda(R)$ for this to be a valid interpretation.  First, 
$\Lambda^{-1}$ should remain much larger than $R^2$ in order to justify using the 
flat space conformal Killing vector (\ref{eqn:CKV}) for the CFT modular Hamiltonian, and 
also to justify keeping only the first order correction to the area due to 
curvature in equation (\ref{eqn:dSUVG}).  
Since $C(R)$ is dominated by the $R^{2\D}$ for $\D\leq\dt$ as $R
\rightarrow 0$, it determines $\Lambda(R)$ by
\beq
\Lambda(R) = \frac{2\pi}{\eta} C\sim \ell_P^{d-2}\vev{\op}_g\delta\vev{\op} R^{2\D-d}.
\eeq
The the requirement that $\Lambda(R) R^2\ll1$ becomes
\beq\label{eqn:Rupper}
 \frac{R}{\ell_P} \ll \left(\frac{1}{\ell_P^{2\D}\vev{\op}_g\delta\vev{\op}}\right)^{\frac{1}{2\D-d+2}}.
\eeq
Since $2\D-d+2\geq0$ by the CFT unitarity bound for scalar operators, this 
inequality can always be satisfied
by choosing $R$ small enough.  Furthermore, since $\vev{\op}_g\delta\vev{\op}$ should 
be  small in Planck units, the right hand side of this inequality is large, and hence can be 
satisfied for $R\gg\ell_P$.   A second requirement is that $\Lambda$ remain sub-Planckian
to justify using a semi-classical vacuum state when discussing the variations.  
This means $\Lambda(R)
\ell_P^2 \ll1$, which then implies
\beq\label{eqn:Rlower}
\frac{R}{\ell_P} \gg \Big(\ell_P^{2\D}\vev{\op}_g\delta\vev{\op}\Big)^{\frac{1}{d-2\D}}
\eeq
This now places a lower bound on the size of the ball for which the derivation is valid.  However,
the $R$-dependence in $\Lambda(R)$ is only significant when $d-2\D$ is positive, and hence
the right hand side of this inequality is small.  Thus, there should be a wide range of $R$ values
where both (\ref{eqn:Rupper}) and (\ref{eqn:Rlower}) are satisfied.  
The implications of such an $R$-dependent 
local curvature scale merit further investigation.  Perhaps it is related to 
a renormalization group flow of the cosmological constant \cite{Shapiro2009}.

A second, more speculative possibility is 
that the $R^{2\D}$ and $\log R$ terms are resummed due to higher
order corrections into something that is subdominant in the $R\rightarrow 0$ limit.  One reason
for suspecting that this may occur is that the $R^{2\D}$ at second order in the state variation
can dominate over the lower order $R^d$ terms at small $R$, possibly hinting at a break
down of perturbation theory.\footnote{However, reference \cite{Casini2016a} found that terms
at third order in the state variation are subdominant to this term for small values of $R$.
 \renewcommand{\baselinestretch}{1}\footnotesize } 
As a trivial example, 
suppose the $R^{2\D}$ term arose from a function of the form
\beq
\frac{R^d}{1+(R/R_0)^{2\D-d}}.
\eeq
Since $\D<\dt$, this behaves like $R^d-R^{2\D}R_0^{d-2\D}$ 
when $R\gg R_0$.  
However, about $R=0$, 
it becomes
\beq
\frac{R^d}{1+(R/R_0)^{2\D-d}}\xrightarrow{R\rightarrow 0} R_0^d\left(\frac{R}{R_0}\right)^{2
(d-\D)},
\eeq
which is subleading with respect to a term scaling as $R^d$.  Note however that something must 
determine the scale $R_0$ in this argument, and it is difficult to find a scale that is free of
nonanalyticities in the coupling or operator expectation values.  It would be interesting to
analyze whether these sorts of nonperturbative effects  play a role in the entanglement
entropy calculation.  

One may also view the $R$ dependence in $\Lambda$ as evidence that the relation
between maximal vacuum entanglement and the Einstein equation does not hold for some 
states.  In fact, there is some evidence that the relationship must not hold for some states for
which the entanglement entropy is not related to the energy density of the state.  A 
particular example is a coherent state, which has no additional entanglement entropy 
relative to the vacuum despite possessing energy \cite{Varadarajan2016}.  

A final possibility is that these terms scaling as $R^{2\D-d}$ are coming from a nonlocal term
in the entanglement entropy, and that the entanglement equilibrium argument
should be applied only to local terms in the modular Hamiltonian \cite{Arias2017}.

\subsection{Future work}
This work leads to several possibilities for future investigations.  
First is the question
of how the entanglement entropy changes under a change of Lorentz frame.  The 
equivalence between vacuum equilibrium and the Einstein equation rests crucially on 
the transformation properties of the quantity $C$ appearing in equation (\ref{eqn:dSIRmod}).  
Only if it transforms as a scalar can it be absorbed in to the local curvature scale $\Lambda(x)$.  
The calculation in this chapter was done for a large class of states 
defined by a Euclidean path integral.
For a boosted state, one could 
simply repeat the calculation using the Euclidean space relative to the boosted frame, and 
the same form of the answer would result.  For states considered here that were stationary
on time scales on the order $R$ (since $\omega R\ll1$), it seems plausible that the states 
constructed in the boosted Euclidean space contain the boosts of the original states. However,
this point should be investigated more thoroughly.  
Another possibility for checking how the entanglement entropy changes under boosts is to use
the techniques developed in 
\cite{Faulkner2015a}, which provides perturbative methods for evaluating the 
change in entanglement entropy under a deformation of the region $\Sigma$.  In particular, 
a formula is derived  that applies to timelike deformations of the surface, and hence 
could be used to investigate the behavior under boosts.  

Performing the calculation to the next order in perturbation theory would also provide 
new nontrivial checks on the conjecture, in addition to providing new insights for the 
general theory of perturbative entanglement entropy calculations.  This has been 
done in holography \cite{Casini2016a}, so it would be interesting to see if the holographic
results continue to match for a general CFT.  The entanglement entropy at the next order in 
perturbation theory depends on the $\op\op\op$ three point function \cite{Rosenhaus:2014ula}. 
One reason for suspecting that the holographic results will continue to work 
stems from the universal form of this three point function in CFTs. 
For
scalar operators, it is completely fixed by conformal invariance up to an overall constant.  
Thus, up to this multiplicative constant, 
there is nothing in the calculation distinguishing 
between holographic and non-holographic theories.  
At higher order, one would eventually expect the 
holographic calculation to differ from the general case.  For example, the four point function
has much more freedom, depending on an arbitrary function of two conformally invariant
cross-ratios.  It is likely that universal statements about the entanglement entropy would 
be hard to make at that order.

The IR divergences when $\D\leq\dt$ were dealt with using an IR cutoff, which  
captures the qualitative behavior of the answer, but misses out on the precise details of 
how the coupling suppresses the IR region.  It may be possible to improve on this calculation
at scales above the IR scale using established techniques for handling IR divergences 
perturbatively \cite{Zamolodchikov1991,Guida1996,Guida1997}, or by examining
specific cases that are exactly solvable \cite{Casini2009,Blanco2011,Zamolodchikov1991}.  
IR divergences continue to plague the calculations at 
 higher order in perturbation theory.  This   
can be seen by examining the $\op\op\op$ three point function,
\begin{align}
\iint d^d x_1 d^d x_2 \mvev{\big}{\op(0)\op(x_1)\op(x_2)} = \iint d^d x_1 d^d x_2
 \frac{c}{|x_1|^\D |x_2|^\D|x_1-x_2|^\D}. 
\end{align}
By writing this in spherical coordinates, performing the angular integrals, and 
defining $u = \frac{r_2}{r_1}$, this may be written 
\beq
c\Omega_{d-1} \Omega_{d-2} 2 \pi \int_0^1 du \int_0^\infty dr_1\, r_1^{2d-3\D-1} u^{d-\D-1}
(1+u)^{-\D} \tensor[_2]{F}{_1}\left(\frac12,\D;1;\frac{2u}{1+u}\right),
\eeq
This is clearly seen to diverge in the IR region $r_1\rightarrow\infty$ when $\D\leq\frac{2d}{3}$,
so that some operators that produced IR finite results in the two-point function now 
produce IR divergences.

Finally, one may be interested in extending Jacobson's derivation to include higher
order corrections to the Einstein equation.  There are two possibilities for pursuing this. 
First, one may consider higher order in $R^2$ corrections to the entanglement entropy.  
On the geometrical side, this involves considering
additional terms in the Riemann normal coordinate expansion of the metric about a point.  
This could also lead to deformations of the entangling surface $\partial \Sigma$, and these effects
could be computed perturbatively using the techniques of \cite{Rosenhaus2014,
Rosenhaus:2014ula, Rosenhaus:2014zza, Faulkner2015a}.  Additional corrections would 
come about in the computation of $\delta S_{\text{IR}}$ from spatial variation of the state
across the ball, as well as subleading contributions in the energy of the state.  
It may be interesting to 
see whether these expansions can be carried out further to compute the higher curvature
corrections to Einstein's equation.
Another approach would be to compute the Wald entropy associated with the ball 
\cite{Wald1993a, Jacobson1994b, Iyer1995b}, with 
additional corrections added to account for the nonzero extrinsic curvature of the surface 
\cite{Dong2014}.
This is the appropriate generalization of the area terms to the entanglement entropy when
the gravitational theory contains higher curvature corrections.  
In this case, care has to be taken in order to determine what is held fixed during the 
variation.\footnote{I thank Rob Myers for suggesting this approach for incorporating higher
curvature terms in the Einstein equation.  \renewcommand{\baselinestretch}{1}\footnotesize}  
This higher curvature generalization is the topic of chapter \ref{ch:EEhigher}.

\section*{Appendices}
\renewcommand\thesection{\arabic{chapter}.\Alph{section}}
\setcounter{section}{0}

\section{Coefficients for the bulk expansion} \label{sec:Bessel}
This appendix lists the coefficients appearing in section \ref{sec:D>dt} for the expansion of
$\phi_\omega$ and $\nabla^2 \phi_0\phi_\omega$.  Given its definition (\ref{eqn:phio}), the
coefficients appearing in the expansion (\ref{eqn:phioseries}) follow straightforwardly from
known expansions of the modified Bessel functions \cite{NIST:DLMF}:
\begin{align}
a_n &= \frac{\Gamma(\dt-\D+1)}{4^n\, n!\Gamma(\dt-\D+n+1)} \\
b_n &=\frac{\Gamma(\D-\dt+1)}{4^n\, n!\Gamma(\D-\dt+n+1)}.
\end{align}
When acting with $\nabla^2$ on the series $\phi_0\phi_\omega$, the $\tau$ and $z$ derivatives
mix adjacent terms in the series.  The relation this gives is 
\beq
c_n = 2(d-\D+n)(d-2\D+2n)a_n-a_{n-1},
\eeq
which, given the properties of the $a_n$, simplifies to
\beq
c_n = 2(d-\D)(d-2\D+2n)a_n.
\eeq
Similarly, for the $d_n$ series,
\beq
d_n = 2n(d+2n) b_n - b_{n-1},
\eeq
which implies
\beq
d_n = 4n(d-\D) b_n.
\eeq

\section{Real-time solutions for $\phi(x)$}

\subsection{$\D<\dt$} \label{sec:realt}

This appendix derives the real time behavior of the fields $\phi_0$ and $\phi_\omega$.  Starting
with $\phi_0$, the coupling $g(x)$ is a constant $g$ for $|\tau|$ less than
the IR cutoff $L$, 
and zero otherwise. The bulk solution  found by evaluating (\ref{eqn:phiint}) is
\begin{align}
\label{eqn:phi0rtint}
\phi_0 &= g z^{d-\D}\frac{\Gamma(\D-\dt+\frac12)}{\sqrt{\pi}\,\Gamma(\D-\dt)}\left[\int_0^{L/z} dy\,
\left(1+(y-it_B/z)^2\right)^{\dt-\D-\frac12} + \text{c.c.}\right]  \\
&=  g z^{d-\D}\frac{\Gamma(\D-\dt+\frac12)}{\sqrt{\pi}\,\Gamma(\D-\dt)}\left[ 
\frac{L-i t_B}{z} \; \tensor[_2]{F}{_1}\left(\frac12, \D-\dt+\frac12; \frac32;
\frac{-(L-i t_B)^2}{z^2} 
\right)   \right. \nonumber \\
&\qquad\qquad \left.  +\frac{ it_B}{z}\; \tensor[_2]{F}{_1}\left(\frac12, \D-\dt +\frac12; \frac32;
\frac{t_B^2}{z^2} \right)  +\text{c.c.}\right].  \label{eqn:Lisigma}
\end{align}
Here, notice that no cut off near  $y=0$
was needed, since the $\op\op$ two point function
has no UV divergences.  However, one still has to be mindful of the branch prescription,
which is appropriately handled by adding the complex conjugate as directed in the expressions
above (denoted by ``c.c.'').  When $t_B>z$, the branch in the hypergeometric function 
along the real axis is dealt with by replacing $t_B\rightarrow t_B+i \delta$, and taking the 
$\delta\rightarrow 0$ limit.  

This solution can be simplified in the two regimes of interest, namely on 
$\mathcal{E}$ with $t_B=0$ and on $\mathcal{T}$ in the $z\rightarrow 0$ limit.  In the first 
case, $\phi_0$ reduces to
\beq\label{eqn:phi0euclidean}
\phi_0\big|_{t_B=0} = g z^{d-\D} - z^\D\, \frac{g L^{d-2\D} \Gamma(\D-\dt+\frac12)}{
\sqrt{\pi}\, \Gamma(\D-\dt+1)} 
\; \tensor[_2]{F}{_1}\left(\D-\dt, \D-\dt+\frac12; \D-\dt +1;
\frac{-z^2}{L^2}\right),
\eeq
and since we are assuming $R\ll L$, we only need this in the small $z$ limit, 
\beq\label{eqn:phi0slz}
\phi_0\rightarrow g z^{d-\D} - z^\D\, \frac{g L^{d-2\D} \Gamma(\D-\dt+\frac12)}{
\sqrt{\pi}\, \Gamma(\D-\dt+1)} .
\eeq
From this, one immediately reads off the vev of $\op$,
\beq
\vev{\op}_g = 2 gL^{d-2\D}\frac{\Gamma(\D-\dt+\frac12)}{\sqrt{\pi}\,\Gamma(\D-\dt)}.
\eeq
The real time behavior near $z\rightarrow0$ and with $t_B\ll L$ takes the form 
\beq \label{eqn:phi0realt}
\phi_0 = -\frac{\vev{\op}_g}{2\D-d}z^\D + g z^{d-\D} F(t_B/z),
\eeq
with 
\beq \label{eqn:F}
F(s) = \begin{cases}
\hfil 1&s<1 \\
 \frac{\sqrt{\pi}\, (s^2-1)^{\dt-\D +\frac12}}{s\, \Gamma(\D-\dt+1)\, \Gamma(\dt-\D+\frac12)}\;\;
\tensor[_2]{F}{_1}\left(1,\frac12;\D-\dt+1;\frac{1}{s^2}\right) & s>1
\end{cases} .
\eeq
In particular, for large argument, this function behaves as
\beq \label{eqn:Fasym}
F(s\rightarrow\infty) =Bs^{d-2\D};\qquad 
B= \frac{\sqrt{\pi}}{\Gamma(\D-\dt+1)\Gamma(\dt-\D+\frac12)}.
\eeq

We also need the solution for the field corresponding to the state deformation $\lambda(x)$.  
The oscillatory behavior for the choice (\ref{eqn:lambda}) for this function serves to
regulate the IR divergences, and hence no additional IR cutoff is needed.  Thus the 
bulk solution on the Euclidean section (\ref{eqn:phio}) is still valid.  The real time
behavior of the solution is given by the following integral,
\beq
\phi_\omega = \lambda_\omega z^{d-\D}\frac{\Gamma(\D-\dt+\frac12)}{\sqrt{\pi}\,
\Gamma(\D-\dt)}\left[\int_0^{\infty} dy\,\cos(\omega z y)
\left(1+(y-it_B/z)^2\right)^{\dt-\D-\frac12} + \text{c.c.}\right] .
\eeq
To make further progress on this integral, we note that we only need the solution
up to times $t_B\sim R\ll\omega^{-1}$.  In this limit, the solution 
should not be sensitive to the details
of the IR regulator. Therefore, 
the  answer should be the same as for $\phi_0$ in (\ref{eqn:phi0realt}),
the 
only difference being the numerical value for the operator expectation value.  
This behavior can be seen by breaking the integral into two regions, 
$(0,\frac{a}{z})$ and $(\frac{a}{z},\infty)$, with
$t_B\ll a \ll \omega^{-1}$.  In the first region, the cosine can be set to $1$ since
its argument is small.  The resulting integral is identical to (\ref{eqn:phi0rtint}), with $L$
replaced by $a$.  
In the second region, the integration variable $y$ is large compared to $1$ and $t_B/z$,
so the integral reduces to
\begin{align}
\label{eqn:asplit}
&\lambda_\omega z^{d-\D}\frac{2\Gamma(\D-\dt+\frac12)}{\sqrt{\pi}\,
\Gamma(\D-\dt)} \int_{a/z}^{\infty} dy\,\cos(\omega z y) y^{d-2\D-1}  \\
\label{eqn:oa}
=&\; \lambda_\omega z^\D \left(\frac{\omega}{2}\right)^{2\D-d}\, 
\frac{\Gamma(\dt-\D)}{\Gamma(\D-\dt)}  + \lambda_\omega z^{d-\D} \left(\frac{a}{z}\right)^{d-2\D}
\frac{\Gamma(\D-\dt+\frac12)}{\sqrt{\pi}\,
\Gamma(\D-\dt+1)},
\end{align}
valid for $a\ll\omega^{-1}$.  The second term in this expression cancels against the same
term appearing in the first integration region, effectively replacing it with the first term in
(\ref{eqn:oa}).  The final answer for the real time behavior of $\phi_\omega$ near $z=0$
is
\beq \label{eqn:phiort}
\phi_\omega =-\frac{\delta\vev{\op}}{2\D-d}z^\D
+\lambda_\omega z^{d-\D} F(t_B/z).
\eeq
where we have identified $\delta\vev{\op}$ as 
\beq
\delta\vev{\op} = 
\lambda_\omega
\frac{2\,\Gamma(\dt-\D+1)}{\Gamma(\D-\dt)} \left(\frac{\omega}{2}\right)^{2\D-d}.
\eeq

\subsection{$\D=\dt$}  \label{sec:D=dtrt}
Here we derive the real-time behavior of $\phi_0$ and $\phi_\omega$ when $\D=\dt$. 
We begin with $\phi_0$.  The integral (\ref{eqn:phicD'}) can be evaluated, with $\tau$-cutoffs at
$\delta$ and $L$ to give
\begin{align}
\label{eqn:phi0d2int}
\phi_0 &=  \frac{g z^{\dt}}2 \left[\int_{\delta/z}^{L/z} dy \left(1+(y-it_B/z)^2\right)^{-\frac12} 
+\text{c.c.} \right] \\
\label{eqn:gzd2G}
&= g z^{\dt} G(t_B/z, \delta/z, L/z),
\end{align}
where 
\beq \label{eqn:G}
G(s,\vep, l) = \frac12\left(\sinh^{-1} (l-is)-\sinh^{-1} (\vep-i s) + \text{c.c.} \right). 
\eeq
The $\delta$-dependence in (\ref{eqn:gzd2G}) is needed in the region $t_B\sim z$ where 
it is necessary for regularizing a divergence.  Everywhere else the limit $\delta\rightarrow 0$ may
be taken.
Also, since we will need this solution in the 
regions where $z$ and $t_B$ are at most on the order of $R\ll L$, we often use the limiting
form of this function taking $L\gg z,t_B$.  In particular, on the surface $\mathcal{E}$ with
$t_B=0$, it evaluates to
\beq\label{eqn:phi0Elog}
\phi_0\rightarrow g z^{\dt} \log{\frac{2L}{z}},
\eeq
plus terms suppressed by $\frac{z^2}{L^2}$. It is useful to express this in terms of the 
renormalized vev for $\op$ calculated in (\ref{eqn:vevOren}):
\beq
\phi_0\rightarrow -\vev{\op}^\text{ren.}_g z^{\dt} - g z^{\dt} \log\frac{\mu z}{2}.
\eeq
The $\log$ term in this expression is what would have resulted if we had cut the integral
(\ref{eqn:phi0d2int}) off at $\mu^{-1}$ rather than $L$.  
Finally, it is also useful to have the form of the function (\ref{eqn:gzd2G})
along $\mathcal{T}$, where $t_B\gg z$,
\beq\label{eqn:phi0Tlog}
\phi_0\rightarrow gz^{\dt} \log\frac{L}{t_B}.  
\eeq

At $t_B=0$, 
the modified Bessel function solution for $\phi_\omega$ is still valid, and the appropriate
normalization is given in equation (\ref{eqn:phiD=dtz0}).
We also need expressions for the behavior of $\phi_\omega$ along the surface
$\mathcal{T}$.  When 
$t_B\ll \omega^{-1}$, the same arguments that led to equation (\ref{eqn:phiort}) for 
$\D<\dt$ can be applied to the defining integral for $\phi_\omega$ to show it takes
the form
\beq\label{eqn:phiobzlzG}
\phi_\omega = 
 \beta_\omega z^{\dt}+\lambda_\omega z^{\dt} G(t_B/z,\delta/z, a/z );
 \qquad \beta_\omega = -\gamma_E-\log\omega a,
\eeq
where $a$ is the intermediate scale introduced in the integral, as in equation (\ref{eqn:asplit}),
and satisfies $t_B
\ll a\ll \omega^{-1}$.  Note that this answer does not actually depend on $a$ since 
it will cancel between the $\log$ and $G$ terms, but it is convenient to make this 
separation when evaluating the $\mathcal{T}$ surface integrals in 
section \ref{sec:D=d2calc}.  From this, the form of $\phi_\omega$ can be read off for $t_B\gg z$:
\beq \label{eqn:phiologot}
\phi_\omega\rightarrow -\lambda_\omega z^{\dt} \left(\gamma_E + \log \omega t_B\right).
\eeq

\section{Surface integrals} 
This appendix gives the details of the $\mathcal{E}$ and $\mathcal{T}$ surface integrals
for $\D<\dt$ (section \ref{sec:D<dtcalc}) and for $\D=\dt$ (section \ref{sec:D=d2calc}).

\subsection{$\D<\dt$} \label{sec:D<dtcalc}
Each integral in this case will be proportional to one of $\vev{\op}_g\delta{\vev{\op}}$, 
$(g\delta\vev{\op}+\lambda(0)\vev{\op}_g)$, or $\lambda(0) g$.    In
each case, we show explicitly that the possibly divergent terms coming from the $z_0
\rightarrow 0$ limit cancel, as they must to give an unambiguous answer.

\paragraph{1. $\vev{\op}_g\, \delta\vev{\op} $ term.}
This term arises from the piece of $\phi_0$ and $\phi_\omega$ that goes like $\frac{-z^\D}{2\D
-d}$.  In particular, it has no dependence on $t_B$ anywhere.  On the surface $\mathcal{E}$, 
since $\partial_\tau\phi=0$, the integrand in (\ref{eqn:Eint}) 
only depends on $\nabla^2\phi^2$.  Working to leading
order in $R$ means only keeping the $z$ derivatives in the Laplacian.  The term in this 
expression with coefficient $\vev{\op}_g\,\delta\vev{\op}$ is $\frac{2 z^{2\D}}{(2\D-d)^2}$,
and acting with the Laplacian on this gives $\frac{4\D z^{2\D}}{2\D-d}$.  Then the $\mathcal{E}$
integral is
\begin{align}
\delta S^{(2)}_{\mathcal{E},1} &= -2 \pi \vev{\op}_g\,\delta\vev{\op}
\frac{\D \Omega_{d-2} }{2\D-d} \int_{z_0}^R dz\, z^{2\D-d-1}
\int_0^{\sqrt{R^2-z^2} }dr\, r^{d-2}\left[\frac{R^2-r^2-z^2}{2R}\right] \\
&=
\label{eqn:dSEz2D}
-2\pi \vev{\op}_g\,\delta\vev{\op}
\frac{\D\Omega_{d-2}}{d^2-1}\left[ R^{2\D}
\frac{\Gamma(\dt+\frac32)\Gamma(\D-\dt+1)}{ (2\D-d)^2 \Gamma(\D+\frac32)} 
- \frac{R^dz_0^{2\D-d} }{(2\D-d)^2}\right]. 
\end{align}
Note this consists of a finite term scaling as $R^{2\D}$ and a divergence in $z_0$.  

The divergence must cancel against the integral over $\mathcal{T}$, given by 
(\ref{eqn:Tint}).  Unlike the case $\D>\dt$, this integral has a vanishing contribution from
the region $t_B\sim z$, but instead a divergent contribution from $t_B\gg z$.  Again picking
out the $\vev{\op}_g\,\delta\vev{\op}$ term in the integrand (\ref{eqn:Tint}), we find
\begin{align}
\delta S^{(2)}_{\mathcal{T},1} &=  -2\pi 
\vev{\op}_g\,\delta\vev{\op} \frac{\Omega_{d-2} z_0^{-d+1}}{(2\D-d)^2}
\int_0^Rdt\int_0^{R-t}dr\,r^{d-2} \frac{t}{R} \left[2
(\D z_0^{\D-1})^2-\frac{\D}{z^2}(2\D-d)z^{2\D}\right]  \\
&=
-2\pi \vev{\op}_g\,\delta\vev{\op} \frac{\D \Omega_{d-2} R^d z_0^{2\D-d}  }{(d^2-1)(2\D-d)^2} .  
\end{align}
Here, we see this cancels the divergence in (\ref{eqn:dSEz2D}), and thus we are left with only 
the finite term in that expression.  

\paragraph{2. $g\delta\vev{\op}+ \lambda(0)\vev{\op}_g$ term. }
On the surface $\mathcal{E}$, this term comes from the part of one field going like $z^\D$, and 
the other going like $z^{d-\D}$.  Hence, when we evaluate this term in $\nabla^2\phi^2$ for the 
bulk integral, we will be acting on a term proportional to $z^d$, which is annihilated by the 
Laplacian.  So the bulk will only contribute terms that are subleading to $R^d$ terms from
$\delta S^{(1)}$. The calculation of these subleading terms would be similar to the 
calculation for in section \ref{sec:D>dt}, but we do not pursue this further here.

Instead, we examine the integral over $\mathcal{T}$, which can produce finite contributions.  
Along this surface, the fields are now time dependent, and hence all terms in equation
(\ref{eqn:Tint}) are important.  We start by focusing on the terms involving time derivatives
of $\phi$.  The $z$-derivative acts on the term going as $\frac{-z^\D}{2\D-d}$, 
and the $t$ derivative on
$z^{d-\D} F(t/z)$.  To properly account for the behavior of $F$ when $t\sim z$, it is useful
to split the $t$ integral into two regions, $(0,c)$ and $(c,R)$ with $z\ll c\ll R$.  In the 
first region this gives
\beq \label{eqn:0c}
-2\pi \frac{\D\Omega_{d-2}}{2\D-d} 
\int_0^c dt \int_0^R dr\, r^{d-2} \left(\frac{R^2-r^2}{2R}\right) \partial_t
F(t/z_0)  = \frac{-2\pi \D \Omega_{d-2} R^d}{(2\D-d)(d^2-1)} F(t/z_0)\Big|^{c}_0.
\eeq
From $(\ref{eqn:F})$, we see that $F(0)=1$, and the value at $t=c$ can be read off using the 
asymptotic form for $F$ in equation $(\ref{eqn:Fasym})$.  This form is also useful for 
evaluating the integral in the second region, where the integral is
\begin{align}
&\frac{-2\pi\D\Omega_{d-2}  (d-2\D)}{(2\D-d)} B
z_0^{2\D-d} \int_c^R dt\int_0^{R-t} dr\, r^{d-2}\left(\frac{R^2-r^2-t^2}{2R}\right)t^{d-2\D-1}
\nonumber \\
\label{eqn:z2D-d}
&= \frac{-2\pi\D\Omega_{d-2}  }{(2\D-d)}
Bz_0^{2\D-d} \left[ R^{2(d-\D)} \frac{d\,\Gamma(d-1)\Gamma(d-2\D+2)}{\Gamma(2d-2\D
+2)}   - \frac{c^{d-2\D} R^d }{d^2-1} \right],
\end{align}
where this equality holds for $c\ll R$. 
The second term cancels the $c$-dependent term of (\ref{eqn:0c}), while the first term is 
a remaining divergence which must cancel against the other piece of the $\mathcal{T}$ 
integral.  This is the piece coming from the second bracketed expression in equation
(\ref{eqn:Tint}).  This term receives no contribution from the region $t\sim z$, so we can
evaluate it  in the region $t\gg z$, using the asymptotic form for $F(t/z)$.  Evaluating 
the derivatives in this expression (and recalling that only the $z$-derivatives in the Laplacian
will produce a nonzero contribution at $z\rightarrow 0$), this leads to
\begin{align}
& \frac{2\pi\Omega_{d-2}}{(2\D-d)}B z_0^{2\D-d}\int_0^Rdt\int_0^{R-t} dr\, r^{d-2} 
\frac{d\D}{R} t^{d-2\D+1} \nonumber\\
&=
\frac{2\pi\D\Omega_{d-2}}{2\D-d} Bz_0^{2\D-d}\frac{d\,\Gamma(d-1) \Gamma(d-2\D+2)}{\Gamma(
2d-2\D+2)},
\end{align}
which cancels the remaining term in (\ref{eqn:z2D-d}).  

Hence the only contribution remaining comes from (\ref{eqn:0c}) at $t=0$, and gives
\beq\label{eqn:dS2T2}
\delta S^{(2)}_{\mathcal{T},2} = \frac{2\pi  \Omega_{d-2} R^d\D}{(d^2-1)(2\D-d)}(g\delta\vev{\op}
+\lambda(0)\vev{\op}_g).
\eeq

\paragraph{3. $g\lambda(0) $ term.} 
The final type of term arises when both fields behave as $z^{d-\D} F(t/z)$.  
The $\mathcal{E}$ surface term will go like $R^{2(d-\D)}$, and hence will be subleading 
compared to the $R^d$ terms.  In fact, this calculation is essentially the same as the 
change in vacuum entanglement when deforming by a constant source, and the form
of this term is given in equation (4.34) of \cite{Faulkner2015} (although 
that calculation was originally performed only for $\D>\dt$).  
Also there is no divergence in $z_0$ in these terms.

On the other hand, the integral over $\mathcal{T}$ does lead to potential divergences, but 
we will show that these all cancel out as expected.  We may focus on the region $t\gg z$ 
since there is no contribution from $t\sim z$.  Using the asymptotic form (\ref{eqn:Fasym})
for $F$, the part of the integral (\ref{eqn:Tint}) involving $t$ derivatives becomes
\begin{align}
&2\pi\Omega_{d-2} 2\D(d-2\D) B^2 z_0^{2\D-d} 
\int_0^Rdt\int_0^{R-t} dr\, r^{d-2} \left(\frac{R^2-r^2-t^2}{2R}
\right) t^{2d-4\D-1} \nonumber \\
&=
\label{eqn:R3d4D}
2\pi\Omega_{d-2} B^2 z_0^{2\D-d} R^{3d-4\D} 
\frac{\D d\,\Gamma(d-1) \Gamma(2d-4\D+2)}{\Gamma(3d-4\D+2)} .
\end{align}
Similarly, the second bracketed term in (\ref{eqn:Tint}) evaluates to 
\begin{align}
&-2\pi \Omega_{d-2} \D d B^2 z_0^{2\D-d} \int_0^Rdt\int_0^{R-t} dr\, r^{d-2} 
\frac{t^{2d-4\D+1}}{R} \nonumber \\
&= 
-2\pi \Omega_{d-2} B^2 z_0^{2\D-d} R^{3d-4\D} 
\frac{\D d\,\Gamma(d-1) \Gamma(2d-4\D+2)}{\Gamma(3d-4\D+2)} ,
\end{align}
perfectly canceling against (\ref{eqn:R3d4D}).  Hence, the $\mathcal{T}$ surface integral 
gives no contribution, and the full $g\lambda(0)$ contribution, coming entirely from the
$\mathcal{E}$ surface, is subleading.

\subsection{$\D=\dt$} \label{sec:D=d2calc}
Here we compute the surface integrals  and divergence in $\delta S^{(1)}$ 
when $\D=\dt$.  The calculation is divided into
four parts: the $\mathcal{E}$ surface integral, the $\mathcal{T}$ surface integral for 
$t_B\sim z_0$, the $\mathcal{T}$ surface integral for $t_B\gg z_0$, and the 
$\delta S^{(1)}$ divergence.

\paragraph{1. $\mathcal{E}$ surface integral.}
Equation (\ref{eqn:Eint}) shows that we need to compute the Laplacian acting on $(\phi_0
+\phi_\omega)^2$ .  At leading order, only the $z$-derivatives from the Laplacian contribute
since the other derivatives are suppressed by a factor of $z^2$.  Using the bulk solutions 
found for $\phi_0$ (\ref{eqn:phi0Elog}) and $\phi_\omega$ (\ref{eqn:phiD=dtz0}), the 
$\mathcal{E}$ surface integral at $O(\lambda^1 g^1)$ is
\begin{align}
\delta S^{(2)}_{\mathcal{E}} &= -4\pi \Omega_{d-2}  g\lambda_\omega
\int_{z_0}^R \frac{dz}{z} \int_0^{\sqrt{R^2-z^2}} dr\, r^{d-2} \left[\frac{R^2-r^2-z^2}{8R}\right]
\left[ 2 + d\gamma_E + d\log \frac{\omega z^2}{4L} \right] \nonumber \\
&=
-2\pi g\lambda_\omega\frac{\Omega_{d-2} R^d}{d^2-1}\int_{z_0/R}^1 \frac{dw}{w} 
(1-w^2)^{\frac{d+1}{2}}\left(1+\frac{d}{2}\gamma_E +\frac{d}{2} \log \frac{w^2 R^2\omega}{4L}
\right).
\end{align}
The divergence in $z_0$ comes from $w$ near zero, and so can be extracted by 
setting the $(1-w^2)$ term in the integrand to $1$, its value at $w=0$. The divergent integral
evaluates to
\beq \label{eqn:dS2Ediv}
\delta S^{(2)}_{\mathcal{E},\text{div.}} 
= -2\pi g\lambda_\omega\frac{\Omega_{d-2} R^d}{d^2-1}
\logp{\frac{R}{z_0}}\left(1+\frac{d}{2}\gamma_E+\frac{d}{2}\log\frac{\omega R z_0}{4L}\right),
\eeq
and the remaining finite piece with $z_0\rightarrow 0$ is 
\beq
\delta S^{(2)}_{\mathcal{E},\text{fin.}} 
= -2\pi g\lambda_\omega\frac{\Omega_{d-2} R^d}{d^2-1}
\int_0^1\frac{dw}{w} \left[(1-w^2)^{\frac{d+1}{2}} -1\right]\left(1+\frac{d}{2}\gamma_E+
\frac{d}{2}\log w^2\frac{R^2\omega}{4L}\right). \label{eqn:finitebulk}
\eeq
The following two identities are needed to evaluate this,
\begin{align}
&\int_0^1\frac{dw}{w}\left[(1-w^2)^{\frac{d+1}{2}}-1\right] = -\frac12 H_{\frac{d+1}{2}} 
\label{eqn:nologii}\\
&\int_0^1\frac{dw}{w}\left[(1-w^2)^{\frac{d+1}{2}}-1\right]\log w = 
\frac18\left(H_{\frac{d+1}{2}}^{(2)} + H_{\frac{d+1}{2}}^2\right), \label{eqn:logii}
\end{align}
where the harmonic number $H_n$ was defined below equation (\ref{eqn:Deqdt}), and 
$H^{(2)}_n$ is a second order harmonic number, defined for the integers by $H^{(2)}_n = 
\sum_{k=1}^n \frac{1}{k^2}$, and  for arbitrary complex $n$ by $H^{(2)}_n = \frac{\pi^2}{6}
-\psi_1(n+1)$, where $\psi_1=\frac{d^2}{dx^2}\log \Gamma(x)$.  
With these, the finite piece (\ref{eqn:finitebulk}) becomes
\beq\label{eqn:dS2Efinlog}
\delta S^{(2)}_{\mathcal{E},\text{fin.}}  = 2\pi g\lambda_\omega\frac{\Omega_{d-2} R^d}{d^2-1}
\left[\frac{d}{4}H_{\frac{d+1}{2}}\left(\gamma_E + \log\frac{\omega R^2}{4L}\right)
-\frac18\left(H_{\frac{d+1}{2}}^{(2)} + H_{\frac{d+1}{2}}(H_{\frac{d+1}{2}}-2)\right) \right].
\eeq

\paragraph{2. $\mathcal{T}$ surface near $t_B\sim z$.}
This region contains several divergences in $z_0$ and $\delta$.  
The specific range of $t_B$ will be $t_B\in(0,c)$, with $z\ll c\ll R$.  Only the first 
bracketed term in (\ref{eqn:Tint}) contributes in this region, and using the general 
solutions for $\phi_0$ and $\phi_\omega$ from equations (\ref{eqn:gzd2G}) and 
(\ref{eqn:phiobzlzG}), it gives at $O(\lambda^1 g^1)$
\beq \label{eqn:dS2TG}
\delta S^{(2)}_{\mathcal{T},\text{div.}} = 2\pi g \frac{\Omega_{d-2} R^d}{d^2-1} \int_0^c dt \left[
\dt\partial_t\left(\lambda_\omega G_L G_a + \beta_\omega G_L\right)
+\lambda_\omega z_0\left(\partial_z G_L\partial_t G_a + \partial_z G_a\partial_t G_L\right)
\right],
\eeq
having introduced the shorthand $G_L\equiv G(t/z_0,\delta/z_0,L/z_0)$ 
and similarly for $G_a$.  The
first term in this expression is a total derivative so can be integrated directly.  The boundary
term at $t=0$ is
\beq\label{eqn:Tz0div}
 2\pi g \lambda_\omega 
\frac{\Omega_{d-2} R^d}{d^2-1} \dt \logp{\frac{2L}{z_0}} \left(\gamma_E+
\log{\frac{\omega z_0}{2}}
\right).
\eeq
At the other boundary $t=c\gg z_0$, the asymptotic formulas (\ref{eqn:phiologot}) and 
(\ref{eqn:phi0Tlog})  produce the term 
\beq\label{eqn:Tcdiv}
-2\pi g \lambda_\omega 
\frac{\Omega_{d-2} R^d}{d^2-1} \dt\logp{\frac{L}{c}} \left(\gamma_E+\log\omega c\right).
\eeq

The remaining terms in (\ref{eqn:dS2TG}) contain a divergence in $\delta$, coming from $t
\sim z$.  To extract it, we focus specifically on the regions $(z_0-u, z_0+v)$ and $(z_0+v, c)$,
 where 
$u,v\ll z$ and positive.  It is straightforward to show that the integral over the region $(0,z_0-u)$ 
is $O(\delta)$, and so does not contribute when $\delta$ is sent to zero.  The divergence 
in the $(z_0-u,z_0+v)$ region can be evaluated by taking a scaling limit with a 
change of  variables, $t_B
= z_0 + s \delta$, and expanding the integrand about $\delta =0$.  After also taking the limit 
$L/z_0, a/z_0\rightarrow \infty$ in the integrand, the integral in this region becomes
\beq \label{eqn:Tddiv}
-\lambda_\omega \int_{-u/\delta}^{v/\delta} ds\, \frac{s+\sqrt{1+s^2}}{1+s^2} 
\rightarrow-\lambda_\omega \log\frac{2v}{\delta},
\eeq
which holds for $u,v\gg\delta$.  For the region $(z+v,c)$, we can take $\delta/z\rightarrow 0$
and $L/z, a/z\rightarrow\infty$, which produces the integral
\beq\label{eqn:Tddivct}
2 \lambda_\omega \int_{z_0+v}^c dt \left(\frac{1}{\sqrt{t^2-z_0^2}} - \frac{t}{t^2-z_0^2}\right)
\rightarrow \lambda_\omega \log\frac{8v}{z_0},
\eeq
where we have taken the limits $c/z_0\gg1$, $v/z_0\ll 1$.  

The final collection of the four contributions (\ref{eqn:Tz0div}), (\ref{eqn:Tcdiv}), 
(\ref{eqn:Tddiv}) and (\ref{eqn:Tddivct})  is 
\begin{align}
\delta S^{(2)}_{\mathcal{T},\text{div.}} = 2\pi g \lambda_\omega \frac{\Omega_{d-2}R^d}{d^2-1}
\left[\dt \logp{\frac{2L}{z_0}}\left(\gamma_E+\log\frac{\omega z_0}{2}\right)
-\dt\logp{\frac{L}{c}}\left(\gamma_E+\log\omega c\right) 
+ \log \frac{4\delta}{z_0}
\right].
\label{eqn:dS2Tdivlog}
\end{align}

\paragraph{3. $\mathcal{T} $ surface for $t_B\gg z$. }
In this region, $t_B\gg z$, and we can use the asymptotic forms (\ref{eqn:phi0Tlog}) 
and (\ref{eqn:phiologot})
for the fields $\phi_0$ and $\phi_\omega$. We start with the first bracketed term in
equation (\ref{eqn:Tint}),
\begin{align}
\delta S^{(2)}_{\mathcal{T},1} &= 
2\pi g\lambda_\omega \Omega_{d-2} \int_c^R dt\int_0^{R-t} dr\, r^{d-2} 
\left[\frac{R^2-r^2-t^2}{2R}\right] \frac{d}{2t} \left(\gamma_E+ \log\frac{t^2\omega}{L}\right) \\
&= 2\pi g \lambda_\omega\frac{\Omega_{d-2} R^d}{d^2-1} \frac{d}{2} 
\int_{c/R}^1 \frac{ds}{s}(1-s)^d(1+ds) \left(\gamma_E+\log\frac{s^2 R^2\omega}{L}\right).
\end{align}
The divergence in this integral comes from $s=0$, so it can be separated out by setting 
$(1-s)^d(1+ds)$ to $1$ (its value at $s=0$), leading to 
\beq\label{eqn:logRcdiv}
\int_{c/R}^1\frac{ds}{s}\left( \gamma_E + \log\frac{s^2 R^2\omega}{L}\right) = 
\log\left(\frac{R}{c}\right)\left(\gamma_E+ \log\frac{cR\omega}{L}\right).
\eeq
The remaining finite piece of the integral is
\beq\label{eqn:remfin} 
\int_0^1\frac{ds}{s} \left[(1-s)^d(1+ds)-1\right]\left(\gamma_E+\log\frac{s^2 R^2 \omega}{L}\right).
\eeq
Evaluation of this integral involves the following identites,
\begin{align}
&\int_0^1 \frac{ds}{s} \left[(1-s)^d(1+ds)-1\right] = 1-H_{d+1},\\
&\int_0^1\frac{ds}{s}\left[(1-s)^d(1+ds)-1\right]\log s=\frac12\left(H^{(2)}_{d+1} +H_{d+1}
(H_{d+1}-2)\right),
\end{align}
where the harmonic numbers $H_n$ and $H_n^{(2)}$ were defined below equations 
(\ref{eqn:Deqdt}) and (\ref{eqn:logii}). Using these to compute (\ref{eqn:remfin}), and 
combining the answer with equation (\ref{eqn:logRcdiv}) gives
\begin{align}
\delta S^{(2)}_{\mathcal{T},1} =
2\pi g \lambda_\omega& \frac{\Omega_{d-2} R^d}{d^2-1}\dt\left[ 
\logp{\frac{R}{c}}\left(\gamma_E+\log\frac{cR\omega}{L}\right)   \right. \nonumber \\
& \label{eqn:dS2T2log}
 \left. -(H_{d+1}-1)\left(\gamma_E+\log\frac{R^2\omega}{L}\right)
+H^{(2)}_{d+1} +H_{d+1}(H_{d+1}-2)\right].
\end{align}

Finally, we compute the second bracketed term of (\ref{eqn:Tint}).  Only the $z$-derivatives 
in the Laplacian term $\nabla^2 \phi^2$  
contribute in the limit $z\rightarrow 0$. Since $\phi^2$ scales as $z^d$, 
the $z$-derivatives in the Laplacian annihilate it, and hence this piece is zero.  
The integral then becomes
\begin{align}
\delta S^{(2)}_{\mathcal{T},2}&=2\pi g\lambda_\omega \Omega_{d-2}\left(\dt\right)^2 
2 \int_0^R dt \int_0^{R-t} dr r^{d-2}
\frac{t}{R}\logp{\frac{L}{t} } (\gamma_E+\log\omega{t}) \\
&= 2\pi g \lambda_\omega  \frac{\Omega_{d-2} R^d}{d^2-1}  \frac{d}{2} 
\left[ 
-H_{d+1}^{(2)} -H_{d+1}(H_{d+1}-2) +(H_{d+1}-1)\left(\gamma_E +\log\frac{R^2\omega}{L}
\right) 
\right. \nonumber \\
&\hphantom{=2\pi g \lambda_\omega  \frac{\Omega_{d-2} R^d}{d^2-1}  \frac{d}{2} } \left.
-\logp{\frac{R}{L}} \left(\gamma_E+\log R\omega\right)
\right].
\end{align}
The finite terms cancel against those appearing in (\ref{eqn:dS2T2log}), and the final
combined result is 
\beq \label{eqn:dS2T1+2}
\delta S^{(2)}_{\mathcal{T},1+2} = 
2\pi g \lambda_\omega \frac{\Omega_{d-2} R^d}{d^2-1}\dt
\logp{\frac{L}{c}} \left(\gamma_E+\log\omega c\right),
\eeq
which perfectly cancels the $c$-dependent terms in (\ref{eqn:dS2Tdivlog}).  Hence, 
no finite terms result from the integral along $\mathcal{T}$ in the $t_B\gg z$ region.

\paragraph{4. $\delta S^{(1)}$ term. }
The final divergence in $\delta$ comes from the expectation value of the CFT stress tensor,
in $\delta S^{(1)}$.  At order $g \lambda_\omega$, this is given by
\beq
\delta\fvev{T^0_{00}(0)}=  -\int d^d x_a d^d x_b g\lambda_\omega(x_b)
\fvev{T^0_{\tau\tau}(0) \op(x_a)\op(x_b)}.
\eeq
The only divergence in this correlation function comes from when $x_a
\rightarrow x_b\rightarrow 0$, and is logarithmic in the cutoff  $\delta$.  As was the
case for the logarithmic divergence in $\vev{\op}$, regulating this divergence involves 
introducing a renormalization scale $\mu$ that separates the divergence from the 
finite part of the correlation function.  This is done by cutting off the $\tau$ integrals when 
$|\tau_a|\geq\mu^{-1}$ and 
$|\tau_b|\geq\mu^{-1}$.  

The divergence comes from the leading piece in the expansion of $\lambda_\omega(x)$
about $x=0$, 
\beq
\delta\fvev{T^0_{\tau\tau}(0) }_{\text{div.}} =  g \lambda_\omega \int d^d x_a d^d x_b 
\fvev{T^0_{\tau\tau}(0) \op(x_a)\op(x_b)}.
\eeq
This divergence can be evaluated using the same method described in Appendix D of 
\cite{Faulkner2015}.  The translation invariance of the correlation function allows one to 
write it as an integral of the stress tensor averaged over the spatial volume,
\beq
g\lambda_\omega \frac1V \int d^{d-1}\vec{x}
\int_{C(\delta,\mu)} d\tau_a \int_{C(\delta,\mu)} d\tau_b \int d\vec{x}_a d\vec{x}_b
\fvev{T^0_{\tau\tau}(0,\vec{x}) \op(x_a)\op(x_b)}.
\eeq

The stress tensor integrated over $\vec{x}$ is now a conserved quantity, and so the surface 
of integration may deformed away from $\tau=0$.  As long as it does note encounter 
the points $\tau_a$
or $\tau_b$, the surface can be pushed to infinity, so that the correlation function vanishes.  
This is possible if $\tau_a$ and $\tau_b$ have the same sign.  However, when $\tau_a$ and 
$\tau_b$ have opposite signs, one of them will be passed as  the surface is pushed to
infinity.  This leads to a contribution from the operator insertion at that point, as dictated by 
the translation Ward identity.  Let us choose to push past $\tau_a$.  For $\tau_a<0$,
the contribution from the operator insertion is 
\begin{align} 
&\hphantom{=}
-g\lambda_\omega\frac1V \int d\vec{x} d\vec{x_a} d\vec{x_b}\int_\delta^\mu d\tau_b
\int_{-\mu}^{-\delta} d\tau_a \partial_{\tau_a}\fvev{\op(x_a)\op(x_b)} \delta(\vec{x}-\vec{x_a}) \\
&=  -g\lambda_\omega c_\Delta' S_{d-2} \frac{\sqrt{\pi}\,\Gamma(\dt-\frac12)}{2\Gamma(\dt)}
\int_\delta^\mu d\tau_b \left[\frac1{\tau_b+\delta} - \frac1{\tau_b +\mu}\right] \\
&=-\frac12g\lambda_\omega \log{\frac{\mu}{4\delta}},
\end{align}
where in this last equality we have taken $\mu\gg\delta$. 
It is straightforward to check that for $x_a^0>0$, you get the same contribution, 
so that the full divergent
piece of the stress tensor is 
\beq
\delta\fvev{T_{00}(\vec{x})}_{\text{div.}} = 
g\lambda_\omega \log{\frac\mu{4\delta}}.
\eeq
This then defines a renormalized stress tensor expectation value,
\beq
\delta\vev{T_{00}(0)}^{\text{ren.}} = \delta\vev{T_{00}(0)} - g\lambda_\omega\log\frac{\mu}{4\delta}
\eeq

Finally, the contribution to $\delta S^{(1)}$ comes from integrating $\delta\vev{T_{00}(\vec{x})}$
over the ball $\Sigma$ according to equation (\ref{eqn:dS1}).  Since the stress tensor 
expectation value may be assumed constant over a small enough ball, the expression
for $\delta S^{(1)}$ in terms of the renormalized stress tensor expectation value is 
\beq \label{eqn:deltact}
\delta S^{(1)}_{\lambda g} =
2\pi \frac{\Omega_{d-2} R^d}{d^2-1}
\left(\delta\vev{T^0_{00}}^{\text{ren.}} + g\lambda_\omega \logp{\frac{\mu}{4\delta}}  \right).
\eeq

\renewcommand\thesection{\arabic{chapter}.\arabic{section}}


\chapter{Entanglement equilibrium for higher order gravity }
This chapter is based on my paper ``Entanglement equilibrium for higher order gravity,"
published in Physical Review D in 2017, in collaboration with Pablo Bueno, Vincent Min, and 
Manus Visser \cite{Bueno2017}.  
\label{ch:EEhigher}

\section{Summary of results and outline} \label{sec:intro}
\begin{figure}[t]
\begin{center}
\includegraphics[width=0.65\columnwidth]{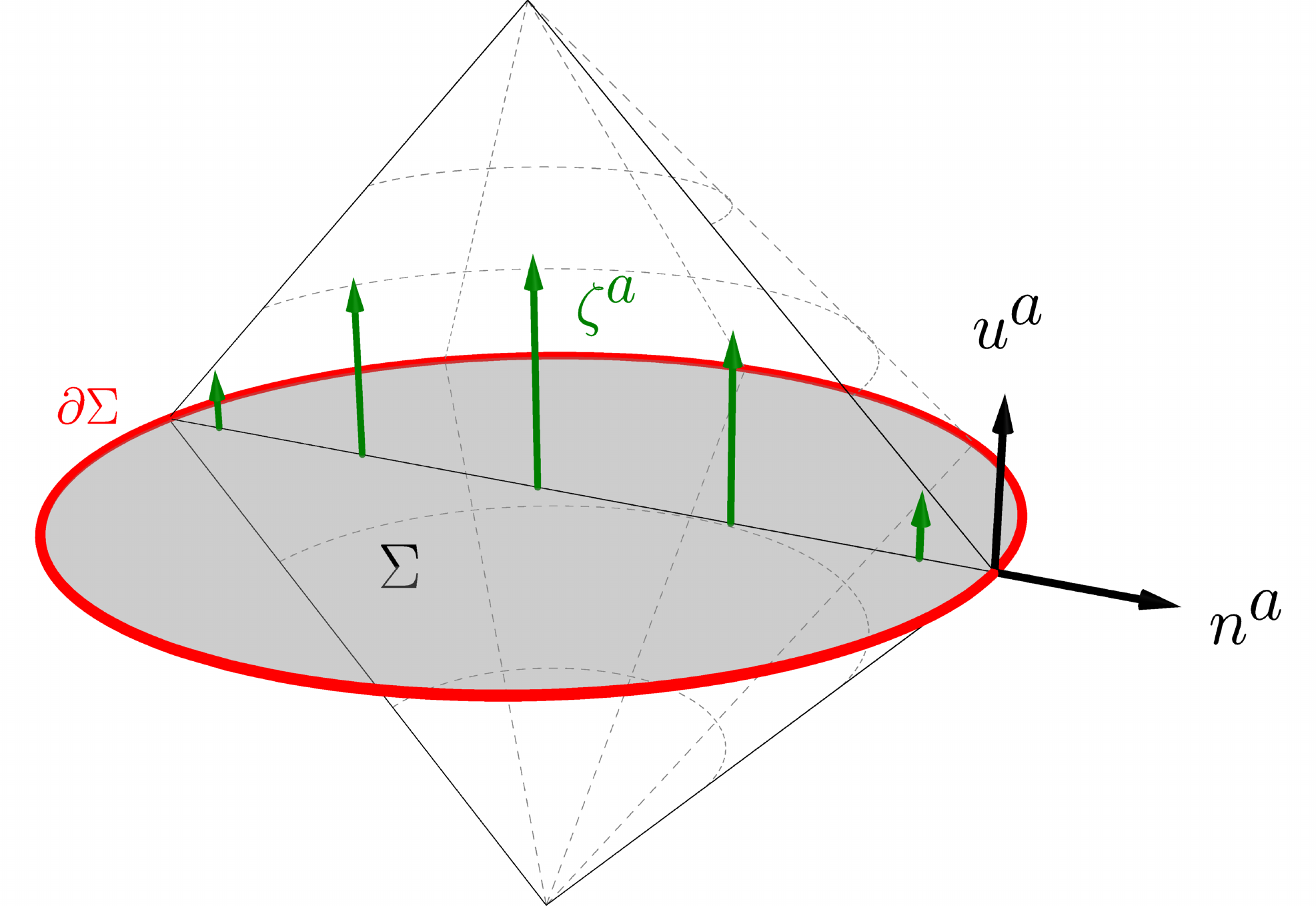}
\end{center}
\renewcommand\baselinestretch{1}
\begin{quote}
\caption[Causal diamond]{The causal diamond consists of the future and past domains
of dependence of a spatial sphere $\Sigma$ in a MSS. 
$\Sigma$ has a unit normal $u^a$, induced metric $h_{ab}$, and volume form $\eta$.  
The boundary  $\partial\Sigma$ has a spacelike unit normal $n^a$, defined
to be orthogonal to $u^a$,  and volume form $\mu$.  
The conformal Killing vector $\zeta^a$ generates a flow within the causal diamond, and 
 vanishes on  the bifurcation surface $\partial\Sigma$. \label{fig:diamond}}
 \end{quote}
\end{figure}

This chapter explores an extension of the entanglement equilibrium argument described in
section \ref{sec:enteq}  to higher curvature theories.  For general relativity, the equilibrium
condition applied to the entanglement was
\beq \label{eqn:dSEEV}
\delta S_\text{EE}\big|_V = \frac{\delta A\big|_V}{4G} + \delta S_\text{mat} = 0 \, .
\eeq
 It is not {\it a priori} clear what the precise statement of the entanglement equilibrium condition
should be for a higher curvature theory, and in particular what replaces the fixed-volume
constraint. 
The  formulation we propose here  is advised by
the {\it first law of causal diamond mechanics}, a purely geometrical identity that holds independently of 
any entanglement considerations.  
It was derived  for Einstein gravity
in the supplemental materials of \cite{Jacobson2015a}, and one of the 
main results of this chapter is to extend it to arbitrary, higher derivative theories.  
As we show in section \ref{sec:firstlaw}, the first law is related to the off-shell identity 
\beq\label{titis}
\frac{\kappa}{2\pi} \delta S_{\text{Wald}}\big|_W + \delta H^m_\zeta = \int_\Sigma \delta C_\zeta \, ,
\eeq   
where $\kappa$ is the surface gravity of $\zeta^a$ \cite{Jacobson1993}, $S_\text{Wald}$ is the 
Wald entropy of $\partial \Sigma$ given in equation (\ref{eqn:SWald}) \cite{Wald1993a, Iyer1994a},
$H_\zeta^m$ is the matter Hamiltonian for flows along $\zeta^a$, defined in equation 
(\ref{potati}), and $\delta C_\zeta=0$ are the linearized constraint equations of the higher
 derivative  theory.  The Wald entropy is varied holding fixed a local geometric 
functional
\beq \label{eqn:Wolume}
W = \frac{1}{(d-2)E_0} \int_\Sigma {\eta}\left(E^{abcd} u_a h_{bc} u_d - E_0\right) \, ,
\eeq
with $\eta$, $u^a$ and $h_{ab}$ defined in Figure \ref{fig:diamond}.  
$E^{abcd}$ is the variation of the gravitational Lagrangian scalar with respect to 
$R_{abcd}$, and $E_0$ is a constant determined by the value of $E^{abcd}$ in a MSS via
$E^{abcd}\overset{\text{MSS}}{=}E_0(g^{ac}g^{bd}-g^{ad}g^{bc})$.
We refer to $W$ as the ``generalized volume'' since it reduces to the volume for
Einstein gravity.

\renewcommand\baselinestretch{2}

The Wald formalism contains ambiguities identified by Jacobson, Kang and Myers (JKM)
\cite{Jacobson1994b}  that modify the  Wald entropy and the 
generalized volume by the terms $S_\text{JKM}$ and $W_\text{JKM}$ 
given in (\ref{eqn:SJKM}) and (\ref{eqn:WJKM}).  Using a modified generalized
volume defined by
\beq \label{eqn:Wp}
W' = W + W_\text{JKM} \, ,
\eeq
the identity (\ref{titis}) continues to hold with $\delta (S_\text{Wald}+S_\text{JKM})\big|_{W'}$
replacing $\delta S_\text{Wald}\big|_W$\,.  As discussed in section \ref{sec:subleading}, the subleading divergences for the 
entanglement entropy involve a particular resolution of the JKM ambiguity, while 
section \ref{sec:fixedflux} argues that
the first law of causal diamond mechanics applies for \emph{any} resolution, as long as the 
appropriate generalized volume is held fixed.

Using the resolution of the JKM ambiguity required for the entanglement entropy calculation,
the first law leads to the following statement of entanglement
equilibrium, applicable to higher curvature theories:
\begin{hypothesis*}[Entanglement Equilibrium]
In a quantum gravitational theory, the entanglement entropy of a spherical 
region with fixed generalized volume $W'$ is maximal in vacuum.
\end{hypothesis*}
This modifies the original equilibrium condition (\ref{eqn:dSEEV}) by replacing the area
variation with 
\beq
\delta(S_\text{Wald} + S_\text{JKM})\big|_{W'} \, .
\eeq
In Section \ref{sec:equilibrium}, this equilibrium condition is shown
to be equivalent to   the linearized higher derivative field equations 
in the case that the matter fields are conformally invariant.
Facts about  
entanglement entropy divergences and 
the reduced density matrix for a sphere in a CFT are used to relate the total variation of 
the entanglement entropy to the left hand 
side of (\ref{titis}).  Once this is done, it becomes clear that imposing the linearized 
constraint equations is 
equivalent to the entanglement equilibrium condition.

In \cite{Jacobson2015a}, this condition was applied in the small
ball limit, in which {\it any} geometry looks like a perturbation of a MSS.  Using Riemann
normal coordinates (RNC), the linearized equations were shown to impose the fully nonlinear
equations for the case of Einstein gravity.  
We will discuss this argument in Section \ref{sec:equations} for higher
curvature theories, and show that the nonlinear equations can \emph{not} be obtained from the 
small ball limit, making general relativity unique in that regard.

In section \ref{sec:conclusion}, we discuss several implications of this work.  First, we describe
how it compares to other approaches  connecting geometry and entanglement.  
Following that, we provide a possible thermodynamic interpretation of the first law of causal 
diamond mechanics derived in section \ref{sec:firstlaw}.  We then comment on a conjectural
relation between our generalized volume $W$ and higher curvature holographic complexity.  
Finally, we lay out several future directions for the entanglement equilibrium program.


\section{First law of  causal diamond mechanics} \label{sec:firstlaw}

Jacobson's 
entanglement equilibrium argument \cite{Jacobson2015a} 
compares the surface area of a 
small spatial ball $\Sigma$ in a curved spacetime to the one that would be obtained in a MSS.  
The comparison is made using balls of equal volume $V$, a choice justified 
by an Iyer-Wald variational identity \cite{Iyer1994a} 
for the conformal Killing vector $\zeta^a$ of the causal diamond in the maximally symmetric background. When the Einstein equation holds, this identity implies the 
{\it first law of causal diamond mechanics} \cite{Jacobson2015a, Manus}
\beq \label{barbecue}
- \delta H_\zeta^m = \frac{\kappa}{8\pi G} \delta A   -\frac{\kappa k}{8\pi G}  \delta V \, ,
\eeq
where $k$ is 
  the trace of the extrinsic curvature of $\partial \Sigma$ embedded in $\Sigma$,
and the matter conformal Killing energy $H_\zeta^m$ is constructed from the  stress 
tensor $T_{ab}$ by
\begin{equation}\label{potati}
 H_{\zeta}^{m}=\int_\Sigma\eta\, u^a \zeta^bT_{ab}\, .
 \end{equation}
 The purpose of this section is to generalize the variational identity to higher derivative
 theories, and to clarify its relation to the equations of motion.  
This is done by focusing on an off-shell version of the identity, which reduces to the first law
when the  linearized  constraint equations for the theory are satisfied.  We begin by reviewing the 
Iyer-Wald formalism in subsection \ref{sec:IyerWald}, which also serves to establish notation.
After describing the geometric setup in subsection \ref{subsec:setup}, we show in subsection
\ref{sec:localgeo} how the quantities appearing in the identity can be written as variations of 
local geometric functionals of the surface $\Sigma$ and its boundary $\partial \Sigma$.  As
one might expect, the area is upgraded to the Wald entropy $S_\text{Wald}$, and we derive 
the generalization of the volume  given in equation (\ref{eqn:Wolume}). 
Subsection \ref{sec:fixedflux} describes how the variational identity can instead be viewed 
as a variation at fixed generalized volume $W$, as quoted in equation (\ref{titis}), 
and describes the effect that JKM
ambiguities have on the setup.

\subsection{Iyer-Wald formalism} \label{sec:IyerWald}
We begin by recalling the Iyer-Wald formalism \cite{Wald1993a, Iyer1994a}.  A general diffeomorphism invariant theory may 
be defined by its Lagrangian $L[\phi]$, a spacetime $d$-form locally constructed 
from the dynamical fields $\phi$, which include the metric and matter fields.  A variation of this
Lagrangian takes the form  
\beq \label{eqn:dL}
\delta L = E\cdot  \delta\phi + \dd\theta[\delta\phi] \, ,
\eeq
where  $E$ collectively denotes the equations of motion for the dynamical fields, and $\theta$
is the symplectic potential $(d-1)$-form.  Taking an antisymmetric variation of $\theta$ yields
the symplectic current $(d-1)$-form
\beq \label{eqn:symplectomega}
\omega[\delta_1\phi, \delta_2\phi] = \delta_1 \theta[\delta_2\phi] - \delta_2\theta[\delta_1\phi] \,,
\eeq
whose integral over a Cauchy surface $\Sigma$ gives the symplectic form for the phase space description
of the theory.  Given an \emph{arbitrary} vector field $\zeta^a$, 
   evaluating the symplectic
form on the Lie derivative  $\lie_\zeta \phi$  
gives the variation of the Hamiltonian $H_\zeta$ that generates the flow of  $\zeta^a$ 
\beq \label{eqn:hamilton}
\delta H_\zeta = \int_\Sigma \omega[\delta\phi, \lie_\zeta\phi] \,.
\eeq
Now consider   a ball-shaped region $\Sigma$, and take $\zeta^a$ to be any future-pointed,
timelike vector that vanishes on the boundary $\partial\Sigma$. Wald's variational identity then
 reads 
\beq \label{eqn:dHz}
\int_\Sigma \omega[\delta\phi, \lie_\zeta\phi] =  \int_\Sigma \delta J_\zeta \,,
\eeq
where the Noether current $J_\zeta$ is defined by
\beq
J_\zeta = \theta[\lie_\zeta \phi]-i_\zeta L \,.
\eeq
Here $i_\zeta$ denotes contraction of the vector $\zeta^a$ on the first index of the differential
form $L$. The identity (\ref{eqn:dHz}) holds when the background geometry satisfies
the field equations $E=0$, and it assumes that $\zeta^a$ vanishes on $\partial \Sigma$.  
Next we note that the 
Noether current can always be expressed as \cite{Iyer1995b}
\beq  \label{eqn:noethercurrent}
J_\zeta = \dd Q_\zeta + C_\zeta,
\eeq
where $Q_\zeta$ is the Noether charge $(d-2)$-form and $C_\zeta$ are the constraint
field equations, which arise as a consequence of the diffeomorphism gauge symmetry. 
For non-scalar matter, these constraints are a combination of the metric and matter field
equations \cite{Seifert2007b, Jacobson2011a}, but, 
assuming the matter equations  are imposed, 
we can take 
\beq \label{eqn:constreq}
C_\zeta = -2 \zeta^a E\indices{_a^b}\epsilon_b,
\eeq 
where $E^{ab}$ is the variation
of the Lagrangian density with respect to the  metric.  
By combining equations (\ref{eqn:hamilton}), (\ref{eqn:dHz}) and (\ref{eqn:noethercurrent}), one finds that 
\beq \label{eqn:FLDM}
-\int_{\partial\Sigma} \delta Q_\zeta+\delta H_\zeta =   \int_\Sigma \delta C_\zeta \, .
\eeq
When the linearized constraints hold, $\delta C_\zeta = 0$, the variation of the Hamiltonian
is a boundary integral of $\delta Q_\zeta$.  This on-shell identity forms the basis for 
deriving the first law of causal diamond mechanics.  Unlike the situation encountered
in black hole thermodynamics, $\delta H_\zeta$ is not zero because below we take $\zeta^a$ to be a conformal Killing vector as opposed to a true Killing vector.

\subsection{Geometric setup}
\label{subsec:setup}

Thus far, the only restriction that has been placed on the vector field $\zeta^a$ is that it 
vanishes on $\partial\Sigma$.  
As such, the quantities $\delta H_\zeta$ and $\delta Q_\zeta$ appearing 
in the identities depend rather explicitly on the fixed vector $\zeta^a$, and therefore these
quantities are not written in terms of only the geometric properties of the surfaces
$\Sigma$ and $\partial\Sigma$.  A purely geometric description is desirable if
the Hamiltonian and Noether charge are 
to be interpreted as  
thermodynamic state functions, which ultimately may be used to define the 
ensemble of geometries in any proposed quantum description of the microstates.  
This situation may be remedied by choosing the vector $\zeta^a$ and the surface
$\Sigma$ to have special properties in the background geometry.  In particular, by choosing
$\zeta^a$ to be a conformal Killing vector for a causal diamond in the MSS, and picking
$\Sigma$ to lie on the surface where the conformal factor vanishes, one finds that 
the perturbations $\delta H_\zeta$ and $\delta Q_\zeta$    have expressions in terms of local
geometric functionals on the surfaces  $\Sigma$ and $\partial\Sigma$, respectively.

Given a causal diamond in a MSS,
there exists a conformal Killing vector $\zeta^a$ which generates a flow within the diamond and 
vanishes at the bifurcation surface $\partial\Sigma$ (see figure \ref{fig:diamond}).  
The metric satisfies the conformal Killing equation
\beq  \label{eqn:lie}
\lie_\zeta g_{ab} =    2\alpha g_{ab} \quad \text{with} \quad \alpha = \frac{1}{d}  \nabla_c \zeta^c  \,.
\eeq
and the conformal factor $\alpha$ vanishes on the spatial ball $\Sigma$.  
The gradient of $\alpha$ is hence proportional to the unit normal to $\Sigma$, 
\beq
u_a = N \nabla_a \alpha  \quad \text{with} \quad N =  \lVert \nabla_a \alpha \rVert^{-1} .
\eeq
Note the vector $u^a$ is future pointing since the conformal factor $\alpha$ decreases to the 
future of $\Sigma$. 
In a MSS, the normalization 
function $N$ has the curious property that it is constant over   $\Sigma$, and is
given by \cite{Manus}
\beq \label{eqn:N}
N =   \frac{d-2}{\kappa  k},
\eeq
where $k$ is the trace of the extrinsic curvature of $\partial \Sigma$ embedded in $\Sigma$, and $\kappa$ is the surface gravity of the conformal Killing horizon, defined momentarily.  
This constancy ends
up being crucial to finding a local geometric functional for $\delta H_\zeta$. 
Throughout this chapter, $N$ and $k$ will respectively
denote  constants equal to the normalization
function and extrinsic curvature trace, both evaluated in the background spacetime.  

Since $\alpha$ vanishes on $\Sigma$, $\zeta^a$ is instantaneously a Killing vector.  On the other hand, the covariant derivative of $\alpha$  is nonzero, so
\beq \label{eqn:covlie}
\nabla_d (\lie_\zeta g_{ab}) \big |_\Sigma = \frac{2}{N} u_d g_{ab} \, .
\eeq
The fact that the covariant derivative is nonzero on  $\Sigma$
is responsible for making   $\delta H_\zeta$ nonvanishing.

A conformal Killing vector with a horizon has a well-defined surface gravity 
$\kappa$ \cite{Jacobson1993},
and since $\alpha$ vanishes on $\partial\Sigma$, we can conclude that 
\beq\label{eqn:dz}
\nabla_a \zeta_b  \big  |_{\partial \Sigma}= \kappa n_{ab}\, , 
\eeq
where $n_{ab} = 2u_{[a} n_{b]}$ is the binormal for the surface $\partial\Sigma$,  
and $n^b$ is the outward
pointing spacelike unit normal to $\partial\Sigma$.  
Since $\partial\Sigma$ is a bifurcation surface of a conformal Killing horizon, 
$\kappa$  is  constant everywhere on 
it.
We provide an example of these constructions in appendix \ref{appkill} where
we discuss the conformal Killing vector for a causal diamond in flat space.

\subsection{Local geometric expressions} \label{sec:localgeo}

In this subsection we    evaluate the Iyer-Wald identity (\ref{eqn:FLDM})   for an arbitrary higher derivative theory of gravity and for the geometric setup described above. The final on-shell 
result is given in (\ref{firstlawhigher}), which is the first law of causal diamond mechanics for higher derivative gravity.

Throughout the computation we   assume that the  matter fields are minimally coupled, so that the Lagrangian splits into a metric and matter piece $L = L^g+L^m$, 
 and we take $L^g$ to be an \emph{arbitrary}, 
diffeomorphism-invariant function of the metric, Riemann tensor, and 
its covariant derivatives.  The symplectic potential and variation of the Hamiltonian  then   exhibit a similar  separation, $\theta  = \theta^g +\theta^m$ and $\delta H_\zeta = \delta H_\zeta^g+ \delta H_\zeta^m$, and so we can write equation (\ref{eqn:FLDM}) as 
\beq \label{newvarid}
-\int_{\partial \Sigma} \delta Q_\zeta+ \delta H_\zeta^g + \delta H_\zeta^m =  \int_\Sigma \delta C_\zeta \,.
\eeq
Below, we explicitly    compute the two terms $\delta H_\zeta^g$ and $\int_{\partial \Sigma} \delta Q_\zeta$ 
for the present geometric context. \\

 \paragraph{Wald entropy.}

By virtue of equation (\ref{eqn:dz})  and the fact that $\zeta^a$ vanishes on $\partial\Sigma$, one can show that the integrated Noether charge
is simply related to the Wald entropy \cite{Wald1993a, Iyer1994a}
\begin{align}
 -\int_{\partial\Sigma} 
Q_\zeta &=   \int_{\partial\Sigma}   \, E^{abcd}  \, \epsilon_{ab} \nabla_c \zeta_d \nonumber\\
&=\frac{\kappa}{2\pi}S_\text{Wald} \,, \label{eqn:Sbar}
\end{align}
where the Wald entropy is defined as
\beq \label{eqn:SWald}
S_\text{Wald} =  -  2\pi \int_{\partial\Sigma} \mu  \, E^{abcd} n_{ab} n_{cd} \,.
\eeq
 $E^{abcd}$ is the variation of the Lagrangian scalar   with respect to the 
Riemann tensor $R_{abcd}$  taken as an independent field, given in (\ref{defEtensor}), and $\mu$ is the 
volume form on $\partial \Sigma$, so that $\epsilon_{ab}
=-n_{ab}\wedge \mu$ there.  
The  equality (\ref{eqn:Sbar}) continues to hold at first order in perturbations, which 
can be shown following the same arguments as   given in \cite{Iyer1994a}, hence,
\begin{equation} \label{QWald}
\int_{\partial \Sigma} \delta Q_\zeta = -  \frac{\kappa}{2\pi } \delta S_\text{Wald} \,.
\end{equation}
 The minus sign is opposite the convention in \cite{Iyer1994a} since the unit normal $n^a$
 is outward pointing for the causal diamond.    \\

\paragraph{Generalized volume.} The gravitational part of  $\delta H_\zeta$ is related to the symplectic current $\omega[\delta g, \lie_\zeta g]$   via (\ref{eqn:hamilton}).  The symplectic
form has been computed 
 on an arbitrary background 
for any higher curvature gravitational theory whose Lagrangian is a function 
of the Riemann tensor, but not its covariant derivatives \cite{Bueno:2016ypa}.  Here,
we  take advantage of the maximal symmetry of the background to compute the symplectic
form and Hamiltonian for the causal diamond in any higher order theory, including those 
with  
derivatives of the Riemann tensor.

Recall that the symplectic current $\omega$ is defined in terms of  the symplectic potential $\theta$ through (\ref{eqn:symplectomega}).
For a     Lagrangian that depends on the Riemann tensor and its covariant derivatives, the symplectic potential $\theta^g$ is given in Lemma 3.1 of \cite{Iyer1994a} 
\begin{align}
\theta^g &= 2 E^{bcd}\nabla_d\delta g_{bc}  +S^{ab}\delta g_{ab} \nonumber\\
&+\sum_{i=1}^{m-1} 
T_i^{abcd a_1\ldots a_i} \delta \nabla_{(a_1}\cdots\nabla_{a_i)} R_{abcd} \, ,
\end{align}
where 
$
E^{bcd} = \epsilon_a E^{abcd}
$
and
the tensors $S^{ab}$ and $T_i^{abcda_1\ldots a_i}$ are locally constructed from the metric, 
its curvature, and covariant derivatives of the curvature.  Due to the antisymmetry of 
$E^{bcd}$ in $c$ and $d$, the symplectic current takes the form 
\begin{align}
&\omega^g = 2\delta_1 E^{bcd}\nabla_d\delta_2 g_{bc}-2E^{bcd}\delta_1\Gamma^{e}_{db}
\delta_2 g_{ec} +\delta_1 S^{ab}\delta_2 g_{ab}   \nonumber\\ 
 &+ \sum_{i=1}^{m-1} 
\delta_1T_i^{abcd a_1\ldots a_i} \delta_2 \nabla_{(a_1}\cdots\nabla_{a_i)} R_{abcd} - 
(1\leftrightarrow2) \label{eqn:og}  .
\end{align}
Next we specialize to the geometric setup described in section \ref{subsec:setup}. We may thus employ the fact that we are perturbing around a maximally symmetric background.
This means the background curvature tensor takes the form
\beq\label{msb}
R_{abcd} = \frac{R}{d(d-1)}(g_{ac}g_{bd}-g_{ad}g_{bc})
\eeq
with a constant Ricci scalar $R$, so that $\nabla_e R_{abcd}=0$, and also $\lie_\zeta R_{abcd} \big|_\Sigma = 0$.
Since the tensors $E^{abcd}$, $S^{ab}$, and $T_i^{abcda_1\ldots a_i}$ are 
all constructed from the  metric and  
curvature, they will also have vanishing Lie derivative along
$\zeta^a$ when evaluated on $\Sigma$. 

Replacing $\delta_2 g_{ab}$ in equation (\ref{eqn:og}) with $\lie_\zeta g_{ab}$ and using (\ref{eqn:covlie}),
we obtain
\begin{align} \label{eqn:olzg}
&\omega^g[\delta g, \lie_\zeta g]  \big |_\Sigma =  \nonumber \\
&\qquad\frac2N\left[2 g_{bc}u_d \delta E^{bcd} + E^{bcd}(u_d \delta g_{bc}
- g_{bd}u^e\delta g_{ec}) \right]   .
\end{align}
We would like to write this as a variation of some scalar quantity.  To do so, 
we split off the background value of $E^{abcd}$ by writing
\beq\label{eqn:Fabcd}
F^{abcd} = E^{abcd} - E_0(g^{ac} g^{bd}-g^{ad}g^{bc}) \,.
\eeq
The second term in this expression is the background value, and, due to maximal symmetry, 
the scalar $E_0$ must be a constant determined by the parameters appearing in the Lagrangian.
By definition, $F^{abcd}$ is zero in the background, so any term in (\ref{eqn:olzg}) that depends
on its variation 
may be immediately written as a total variation, since variations of other tensors appearing
in the formula would multiply the background value of $F^{abcd}$, which vanishes.  Hence,
the piece involving $\delta F^{abcd}$ becomes
\beq\label{eqn:dF}
\frac{4}{N} g_{bc}u_d \delta(F^{abcd}\epsilon_a) = \frac{4}{N} 
\delta(F^{abcd} g_{bc}u_d\epsilon_a) \, . 
\eeq
The remaining terms simply involve replacing $E^{abcd}$ in (\ref{eqn:olzg}) with $E_0
(g^{ac}g^{bd}-g^{ad}g^{bc})$.  These terms then take exactly the same form as the 
terms that appear for general relativity, which we know from 
the appendix of \cite{Jacobson2015a}
combine to give an overall variation of the volume.  The precise form of this variation when
restricted to $\Sigma$ is 
\beq
-\frac{4(d-2)}{N}\delta \eta \, ,
\eeq
where $\eta$ is the induced volume form on $\Sigma$.  Adding this to (\ref{eqn:dF}) 
produces
\beq \label{eqn:symplform}
\omega[\delta g, \lie_\zeta g]  \big |_\Sigma   = -\frac{4}{N} \delta\left[ \eta(E^{abcd} u_a u_d h_{bc} - E_0) \right] \,,
\eeq
where we used that $\epsilon_a = - u_a \wedge \eta$ on $\Sigma$. This leads us to define a generalized volume functional 
\beq \label{eqn:W}
W = \frac{1}{(d-2)E_0} \int_\Sigma{\eta}(E^{abcd}u_a u_d h_{bc} - E_0) \,,
\eeq
and the variation of this quantity is related to the variation of the gravitational Hamiltonian by 
\beq \label{eqn:dJz}
\delta H_\zeta^g = -4E_0 \kappa k\, \delta W \,,
\eeq
where we have expressed $N$ in terms of $\kappa$ and $k$ using (\ref{eqn:N}).
We have thus succeeded in writing $ \delta H^g_\zeta$ in terms of a 
local geometric functional defined on the surface $\Sigma$. 

It is worth emphasizing that $N$ being constant over the ball was crucial to this derivation, 
since otherwise it could not be pulled out of the 
integral over $\Sigma$ and would define a non-diffeomorphism invariant structure on the surface.
Note that the overall normalization of $W$ is arbitrary, since a different
normalization would simply change the coefficient in front of $\delta W$ in (\ref{eqn:dJz}).  As
one can readily check, the normalization in (\ref{eqn:W}) 
was chosen so that $W$  reduces to the volume in the case of Einstein gravity. In appendix \ref{app:W} we provide   explicit expressions for the generalized volume in $f(R)$ gravity and quadratic gravity.

Finally, combining (\ref{QWald}), (\ref{eqn:dJz}) and (\ref{newvarid}), 
we arrive at the off-shell variational identity in terms of local geometric quantities
\beq \label{eqn:offshelllocalgeo}
\frac{\kappa}{2\pi} \delta S_\text{Wald} -4E_0 \kappa k \delta W + \delta H_\zeta^m = 
\int_\Sigma \delta C_\zeta \, .
\eeq
By imposing the linearized constraints $\delta C_\zeta = 0$, this becomes 
the first law of causal diamond mechanics for higher derivative gravity 
\beq \label{firstlawhigher}
- \delta H_\zeta^m = \frac{\kappa}{2 \pi} \delta S_{\text{Wald}} - 4 E_0 \kappa k  \delta W  \, .
\eeq
This reproduces (\ref{barbecue}) for Einstein gravity with Lagrangian $L = \epsilon R/16\pi G$,
for which $E_0 = 1/32\pi G$.

\subsection{Variation at fixed $W$} \label{sec:fixedflux}
We now show that the first two terms in (\ref{eqn:offshelllocalgeo}) can be written in 
terms of the variation of the Wald entropy at fixed $W$, defined as
\beq \label{eqn:dXbarY}
\delta S_\text{Wald}\big|_{W} = \delta S_\text{Wald} - \frac{\partial S_\text{Wald}}{\partial W}  \delta W \, .
\eeq
Here we must specify what is meant by $\frac{\partial S_\text{Wald}}{\partial W}$.  
We will take this partial
derivative to refer to the changes that occur in both quantities when the size of the ball is 
deformed, but the metric and dynamical fields are held fixed.  Take a vector $v^a$ that is 
everywhere tangent to $\Sigma$ that defines an infinitesimal change in the shape of $\Sigma$.
The first order change this produces in $S_\text{Wald}$ and $W$ can be computed by 
holding $\Sigma$ fixed, but varying the Noether current and Noether charge as $\delta J_\zeta
= \lie_v J_\zeta$ and $\delta Q_\zeta = \lie_v Q_\zeta$.  
Since the background field equations are satisfied  and $\zeta^a$ vanishes on
$\partial \Sigma$, we have there that $\int_{\partial \Sigma}   Q_\zeta = \int_\Sigma J_\zeta^g$,
without reference to the matter part of the Noether current.  
Recall that $\delta W$ is related to the variation of the gravitational Hamiltonian, which can be expressed in terms of $\delta J^g_\zeta$ through (\ref{eqn:hamilton}) and (\ref{eqn:dHz}).
Then using the relations (\ref{eqn:Sbar}) and 
(\ref{eqn:dJz}) and the fact that the Lie
derivative commutes with the exterior derivative, we may compute
\beq
\frac{\partial S_\text{Wald}}{\partial W} = \frac{-\frac{2\pi}{\kappa}\int_{\partial\Sigma} 
\lie_v Q_\zeta}{-\frac{1}{4E_0 \kappa k}\int_\Sigma \lie_v J_\zeta^g}
=  8 \pi E_0 k \,.
\eeq
Combining this result with equations  (\ref{firstlawhigher}) and (\ref{eqn:dXbarY}) we arrive at the off-shell variational identity for higher derivative gravity quoted in the introduction
\beq \label{monkey}
\frac{\kappa}{2\pi} \delta S_\text{Wald}\big|_W +   \delta H_\zeta^m
 = \int_\Sigma \delta C_\zeta \,.
\eeq
Finally, we comment on how JKM ambiguities \cite{Jacobson1994b} 
 affect this identity.  The particular ambiguity we are concerned with
comes from the fact that the symplectic potential $\theta$ in equation (\ref{eqn:dL}) 
is defined only up to 
addition of an exact form $\dd Y[\delta \phi]$ that is linear in the field variations and their 
derivatives. This has the effect of changing the Noether current and Noether charge by
\begin{align}
J_\zeta &\rightarrow J_\zeta + \dd Y[\lie_\zeta \phi] \, ,\\
Q_\zeta &\rightarrow Q_\zeta + Y[\lie_\zeta \phi] \,. \label{eqn:QJKM}
\end{align}
This modifies both the entropy and the generalized volume by  surface terms on $\partial\Sigma$
given by
\begin{align}
   S_{\text{JKM}} &= - \frac{2 \pi}{\kappa} \int_{\partial \Sigma} Y[\lie_\zeta \phi]  \, , \label{eqn:SJKM}\\
  W_\text{JKM} &= - \frac{1}{4 E_0 \kappa k} \int_{\partial\Sigma}  Y[\lie_\zeta \phi] \label{eqn:WJKM} \,.
\end{align}
However, it is clear that this combined change in $J_\zeta$ and $Q_\zeta$ leaves the 
left hand side of (\ref{monkey}) unchanged, since the $Y$-dependent terms cancel out.  
In particular,
\beq
\delta S_\text{Wald}\big|_{W} = \delta(S_\text{Wald}+S_\text{JKM})\big|_{W+W_\text{JKM}} \,,
\eeq
showing that any resolution of the JKM ambiguity gives the same first law, provided that 
   the Wald entropy   and generalized volume   are modified by the terms (\ref{eqn:SJKM}) and 
(\ref{eqn:WJKM}).    This should be expected, because the right hand side of
(\ref{monkey}) depends only on the field equations, which are unaffected by JKM 
ambiguities.


\section{Entanglement Equilibrium}\label{sec:equilibrium}

The original entanglement equilibrium argument for Einstein gravity stated that 
the total variation away from the vacuum of the entanglement of a region at fixed volume is zero.  
This statement is encapsulated in equation (\ref{eqn:dSEEV}), which 
shows both an   area variation due to the change in geometry, and a matter piece
from varying the quantum state.  The area variation at fixed volume can 
equivalently be written  
\beq
\delta A\big|_V = \delta A - \frac{\partial A}{\partial V} \delta V 
\eeq
and the arguments of section \ref{sec:fixedflux} relate this combination to the 
terms appearing in the first law of causal diamond mechanics (\ref{barbecue}).  Since $\delta H_\zeta^m$
in (\ref{barbecue})
is related to $\delta S_\text{mat}$ in (\ref{eqn:dSEEV}) for conformally invariant 
matter, the first law
may be interpreted entirely in terms of entanglement entropy variations.  

This section discusses the extension of the argument to higher derivative theories
of gravity.  Subsection
\ref{sec:subleading} explains how  subleading divergences in the entanglement entropy 
are related to a Wald
entropy, modified by a particular resolution of the JKM ambiguity.  Paralleling the 
Einstein gravity derivation, we seek to relate variations of the subleading 
divergences to the higher derivative
first law of causal diamond mechanics (\ref{firstlawhigher}).  Subsection 
\ref{cons} shows that this can be done as long as the generalized volume $W'$ 
[related to 
$W$ by a boundary JKM term as in (\ref{eqn:Wp})] is held fixed.  Then, using the relation of the 
first law to the off-shell identity (\ref{monkey}), we discuss how the entanglement 
equilibrium condition is equivalent to imposing the linearized constraint equations.

\subsection{Subleading entanglement entropy divergences} \label{sec:subleading}

As discussed in section \ref{sec:SgenSEE}, the subleading divergences in the entanglement
entropy are given by a local integral over the entangling surface.  
When the entangling surface is the bifurcation surface
of a stationary horizon, this local integral is simply the Wald entropy 
\cite{Nelson1994, Iyer1995b}. On nonstationary entangling surfaces, 
the computation can be done using the squashed cone techniques of \cite{Fursaev2013a},
which yield terms involving extrinsic curvatures that modify the Wald entropy.  
In holography, the
squashed cone method plays a key role in the proof of the Ryu-Takayanagi
formula \cite{Ryu:2006bv, Lewkowycz2013a}, and its higher curvature generalization
\cite{Dong2014, Camps2013}.  The entropy functionals obtained
in these works seem to also apply outside of holography, giving the 
extrinsic curvature terms in the entanglement entropy for general theories \cite{Fursaev2013a,
Bousso2016}.\footnote{For 
terms involving four or more powers of extrinsic curvature, there are additional
subtleties associated with the so called ``splitting problem'' \cite{Miao2014, Miao2015b,Camps2016}. \renewcommand{\baselinestretch}{1} \footnotesize }

The extrinsic curvature modifications to the Wald entropy in fact take the form of a 
JKM Noether charge ambiguity \cite{Jacobson1994b, Sarkar2013, Wall2015}.  
To see this,  note the vector $\zeta^a$  used to define the Noether charge 
vanishes at the entangling surface and its covariant derivative is antisymmetric and 
proportional to the binormal as in equation (\ref{eqn:dz}).  This means it  
acts like a boost on the normal bundle at the entangling surface.  General covariance
requires that any extrinsic curvature contributions can be written as a sum of boost-invariant
products,
\beq
S_\text{JKM} = \int_{\partial\Sigma} \mu \sum_{n\geq 1} B^{(-n)} \cdot C^{(n)}
\eeq
where the superscript $(n)$ denotes the boost weight of that tensor, so that at the surface:
$\lie_\zeta C^{(n)} = n C^{(n)}$.  
It is necessary that the terms consist of two pieces, each of which has nonzero boost weight,
so that they can be written as
\beq
S_\text{JKM} = \int_{\partial\Sigma} \mu \sum_{n\geq1}\frac1n B^{(-n)}\cdot \lie_\zeta C^{(n)} \, .
\eeq
This is of the form of a Noether charge ambiguity from equation (\ref{eqn:QJKM}), 
with\footnote{This formula defines $Y$ at the entangling
surface, and allows for some arbitrariness in defining it off the surface.  It is not clear that $Y$
can always be defined as a covariant functional of the form $Y[\delta\phi, 
\nabla_a\delta\phi,\ldots]$ without reference to additional structures, such as the normal
vectors to the entangling surface.  It would be interesting to understand better if and when 
$Y$ lifts to such a spacetime covariant form off the surface. 
\renewcommand{\baselinestretch}{1} \footnotesize}
\footnote{We thank
Aron Wall for this explanation of JKM ambiguities. \renewcommand{\baselinestretch}{1} \footnotesize} 
\beq
Y[\delta \phi] = \mu \sum_{n\geq1}\frac1n B^{(-n)} \delta C^{(n)} \,.
\eeq
The upshot of this discussion is that all terms in the entanglement entropy that are local on the
entangling surface, including all divergences, are given by a Wald entropy modified by
specific JKM terms.  
The couplings for the Wald entropy are determined by 
matching to the UV 
completion, or, in the absence of the UV description, these are simply parameters
characterizing the low energy effective theory. 
In induced gravity scenarios, the divergences are determined  by 
the matter content of the theory, and the matching
to gravitational couplings  
has 
been borne
out in explicit examples \cite{Frolov1997, Myers2013, Pourhasan2014}.

\subsection{Equilibrium condition as gravitational constraints}\label{cons}

We can now relate the variational identity (\ref{monkey}) to entanglement entropy.  
The reduced density matrix for the ball in vacuum takes the form
\beq
\rho_\Sigma = e^{- H_\text{mod}}/Z \,,
\eeq
where $H_\text{mod}$ is the modular Hamiltonian
and $Z$ is the partition function, ensuring
that $\rho_\Sigma$ is normalized.  
Since the matter is conformally invariant, the 
modular Hamiltonian takes a simple form in terms of the matter Hamiltonian $H_\zeta^m$
defined in (\ref{potati}) \cite{Hislop1982, Casini2011}
\beq \label{eqn:Hmod}
H_\text{mod} 
= \frac{2\pi}\kappa H_\zeta^m \,. 
\eeq 
Next we apply the first law of 
entanglement entropy \cite{Blanco2013a, Bhattacharya:2012mi}, 
which states that the first order perturbation to the
entanglement entropy is given by the change in modular Hamiltonian expectation value
\beq
\delta S_\text{EE} =  \delta\vev{H_\text{mod}} \, .
\eeq
Note that this equation holds for a fixed geometry and entangling surface, and hence
coincides with what was referred to as $\delta S_\text{mat}$ in section \ref{sec:intro}.   When varying the 
geometry, the divergent part of the entanglement entropy changes due to a 
change in the Wald entropy and JKM terms of the entangling surface.  
The total variation of the entanglement entropy is therefore
\beq \label{eqn:dSEEtot}
\delta S_\text{EE}=  \delta (S_\text{Wald}+S_\text{JKM})  + \delta\vev{H_\text{mod}} \, .
\eeq
At this point, we must give a prescription for defining the surface $\Sigma$ in the perturbed
geometry.  Motivated by the first law of causal diamond mechanics, we  require that $\Sigma$ 
has the same generalized volume $W'$ as in vacuum, where $W'$ differs from 
the  quantity  $W$ by a JKM term, as in equation (\ref{eqn:Wp}). This provides
a diffeomorphism-invariant criterion for defining the overall radius of the ball, since this 
radius may be adjusted until the generalized volume $W'$ is equal to its vacuum
value.  It does
not fully fix all properties of the surface, but it is enough to derive the equilibrium
condition for the entropy.  As argued in section \ref{sec:fixedflux}, the first term in 
equation (\ref{eqn:dSEEtot}) can be written instead as $\delta S_\text{Wald}\big|_W$ 
when the variation is taken holding $W'$ fixed. 
Thus, from equations (\ref{monkey}), 
(\ref{eqn:Hmod}) and (\ref{eqn:dSEEtot}), we arrive at our main result, the equilibrium condition
\beq \label{eqn:dSEEW}
\frac{\kappa}{2\pi}\delta S_{\text{EE}} \big|_{W'} = \int_\Sigma \delta C_\zeta \,,
\eeq
valid for minimally coupled, conformally invariant matter fields.

The linearized constraint equations $\delta C_\zeta=0$ may therefore be interpreted as 
an equilibrium condition on entanglement entropy for the vacuum.  
Since all first variations of the entropy vanish when the linearized gravitational constraints 
are 
satisfied, the vacuum is an extremum of entropy for 
regions with fixed generalized volume $W'$, which is necessary for it
to be an equilibrium 
state. 
Alternatively, postulating that entanglement entropy is maximal in vacuum
for all balls and in all frames would allow one to conclude that the linearized higher
derivative equations hold everywhere.

\section{Field equations from the equilibrium condition} \label{sec:equations}

The entanglement equilibrium hypothesis provides a clear connection between the 
linearized gravitational constraints and the maximality of entanglement entropy at 
fixed $W'$ in the vacuum for conformally invariant matter.  In this section, 
we will
consider whether information about the fully nonlinear 
field equations can be gleaned from the equilibrium condition. Following the approach
taken in \cite{Jacobson2015a}, we employ
a limit where the ball is taken to be much smaller than all relevant scales in the problem, but
much larger than the cutoff scale of the effective field theory, which is  set by the 
gravitational coupling constants.
By expressing the linearized equations in Riemann normal
coordinates, one can infer that the full \emph{nonlinear} field equations hold in the 
case of Einstein gravity.
As we discuss here, such a conclusion can \emph{not} be reached for higher curvature
theories.  The main issue is that higher order terms in the RNC expansion are needed to 
capture the effect of higher curvature terms in the field equations, but these contribute
at the same order as nonlinear corrections to the linearized equations.

We begin by reviewing the argument for Einstein gravity.
Near any given point, the metric looks locally flat, and has an expansion in 
terms of Riemann normal coordinates that takes the form
\beq
g_{ab}(x) = \eta_{ab}-\frac13x^c x^d R_{acbd}(0) +\mathcal{O}(x^3)\, ,
\eeq
where   $(0)$ means evaluation at the center of the ball. 
At distances small compared to the radius of curvature, the second term in this expression
is a small perturbation to the flat space metric $\eta_{ab}$.  Hence, we may apply 
the off-shell identity (\ref{eqn:dSEEW}), using the first order variation
\beq \label{eqn:dgRNC}
\delta g_{ab} = -\frac13 x^c x^d R_{acbd} (0) \,,
\eeq
and conclude that the linearized constraint $\delta C_\zeta$ 
holds for this metric perturbation.  When 
restricted to the surface $\Sigma$, 
this constraint in Einstein gravity is \cite{Seifert2007b}
\beq
C_\zeta\big|_\Sigma=-u^a \zeta^b\left(\frac1{8\pi G} G_{ab}-T_{ab}\right) \eta \,.
\eeq 
Since the background constraint is assumed to hold, the perturbed constraint is 
\beq
\delta C_\zeta\big|_\Sigma = -u^a \zeta^b\left(\frac{1}{8\pi G} \delta G_{ab}- \delta T_{ab}\right)\eta\, ,
\eeq
but in Riemann normal coordinates, we have that the linearized perturbation to the curvature is
just $\delta G_{ab} = G_{ab}(0)$, up to terms suppressed by the ball radius.  
Assuming that the ball is small enough so that the stress tensor
may be taken constant over the ball, one concludes that the vanishing constraint implies the 
nonlinear field equation  at the center of the ball\footnote{In this equation, 
$\delta T_{ab}$ should be thought of as a quantum expectation value of the 
stress tensor.  Presumably, for sub-Planckian energy densities and in the small ball limit, this first order
variation approximates the true energy density.  However, 
there exist states for which the change in stress-energy is zero at first order in perturbations
away from the vacuum, most notable for coherent states \cite{Varadarajan2016}.  
Analyzing how these states can be incorporated into the entanglement equilibrium 
story deserves further attention. \renewcommand{\baselinestretch}{1} \footnotesize }
\beq
u^a \zeta^b(G_{ab}(0) - 8\pi G \delta T_{ab}) = 0\,.
\eeq
The procedure outlined above applies at all points and all frames, allowing us to obtain the full tensorial Einstein equation.  

Since we have only been dealing with the linearized constraint, one could question
whether it gives a good approximation to the field equations at all points within 
the small ball.  This
requires estimating the size of the nonlinear corrections to this field equation.   When 
integrated over the ball, the corrections to the curvature in RNC 
are of order $\ell^2/L^2$, where 
$\ell$ is the radius of the ball and $L$ is the radius of curvature.  Since we took the ball
size to be much smaller than the radius of curvature, these terms are already suppressed
relative to the linear order terms in the field equation.  

The situation in higher derivative theories of gravity is much different.
 It is no longer the case that the linearized equations evaluated in RNC imply the full nonlinear 
field equations   in a small ball.
To see this,  consider an $L[g_{ab},R_{bcde}]$  
higher curvature theory.\footnote{Note that an analogous argument should hold for general higher derivative theories, which also involve covariant derivatives of the Riemann tensor.
\renewcommand{\baselinestretch}{1} \footnotesize} The equations of motion read
\begin{equation}\label{eomhigh}
- \frac{1}{2} g^{ab} \mathcal L  + E^{aecd} \tensor{R}{^b_{ecd}} - 2  \nabla_c \nabla_d E^{acdb}
= \frac{1}{2} T^{ab} \,.
\end{equation}
In appendix \ref{app:FLDMRNC} we show that linearizing these equations
around a Minkowski background leads to
\begin{align}\label{lineomhigh}
 \frac{\delta G^{ab}}{16\pi G} - 2  \partial_c \partial_d \delta E^{a c d b}_{\text{higher}} = \frac{1}{2} \delta T^{ab}\, ,
\end{align}
where we split $E^{abcd}=E^{abcd}_{\text{Ein}}+E^{abcd}_{\text{higher}}$ into its Einstein piece, which gives rise to the Einstein tensor, and a piece coming from higher derivative terms.
As noted before, the variation of the Einstein tensor evaluated in RNC gives the nonlinear Einstein tensor, up to corrections that are suppressed by the ratio of the ball size to the radius of curvature. 
However, in a higher curvature theory of gravity, the equations of motion 
\eqref{eomhigh} contain terms that are nonlinear in the curvature.
Linearization around a MSS background of these terms would 
produce, schematically, $\delta ( R^n ) = n \bar{R}^{n-1} \delta R$, where $\bar{R}$ denotes evaluation in the MSS background.
In Minkowski space, all such terms would vanish.
This is not true in a general MSS, but evaluating the curvature tensors in the 
background still leads to a significant loss of information about the tensor structure 
of the equation.  
We conclude that the linearized equations cannot reproduce the full nonlinear 
field equations for higher curvature gravity, and it is only the linearity of the Einstein equation
in the curvature that allows the nonlinear equations to be obtained for general relativity.

When linearizing around flat space, the higher curvature corrections to the Einstein equation  are entirely captured by the  second term in \eqref{lineomhigh}, which features  
four derivatives acting on the metric, since $E_\text{higher}^{abcd}$ is constructed from 
curvatures that already contain two derivatives of the metric.
Therefore, one is insensitive to higher curvature corrections unless at least $\mathcal{O}(x^4)$ corrections \cite{Brewin2009} are added to the Riemann normal coordinates expansion \eqref{eqn:dgRNC}
\beq\label{eqn:dg4}
\delta g^{(2)}_{ab} = x^c x^d x^e x^f\bigg(\frac2{45}R\indices{_a _c_d^g}R\indices{_b_e_f_g} -\frac1{20}\nabla_c\nabla_d R_{aebf}\bigg) \,.
\eeq
Being quadratic in the Riemann tensor, this term contributes at the same order as 
the nonlinear corrections to the linearized field equations. Hence, linearization based on 
the RNC expansion up to $x^4$ terms is not fully self-consistent.
This affirms the claim that for higher curvature theories, the nonlinear equations at a point cannot be derived by only imposing the linearized equations.


 \section{Discussion} \label{sec:conclusion}
 Maximal entanglement of the vacuum state was proposed in \cite{Jacobson2015a} as a new 
 principle in quantum gravity.  It hinges on the assumption that divergences in the 
 entanglement entropy are cut off  at short distances, so it
 is ultimately a statement about the UV complete quantum gravity theory.  However, the 
 principle can be phrased in terms of the generalized entropy, which is intrinsically UV 
 finite and well-defined within the low energy effective theory.  
 Therefore, if true, maximal vacuum entanglement provides a 
 low energy constraint on any putative
 UV completion of a gravitational effective theory.  
 
 Higher curvature terms arise generically in any such effective  field theory.  Thus, it
 is important to understand how the entanglement equilibrium argument is modified by 
 them.  
As explained in section \ref{sec:firstlaw}, the precise characterization of the entanglement equilibrium hypothesis  
relies on a classical variational identity for causal diamonds in maximally symmetric spacetimes.
This identity leads to equation (\ref{monkey}), which relates variations of the Wald entropy and 
matter energy density of the ball to the linearized constraints. The variations
are taken holding fixed a new geometric
functional $W$, defined in (\ref{eqn:W}), which we call the ``generalized volume.''

We connected this identity to entanglement equilibrium in section \ref{sec:equilibrium},
invoking the   fact that subleading entanglement entropy divergences are given by a Wald
entropy, modified by specific JKM terms, which also modify $W$ by the boundary
term (\ref{eqn:WJKM}).  
  With the additional assumption that matter is conformally invariant, we arrived at our main result \req{eqn:dSEEW}, showing that the equilibrium condition $\delta S_\text{EE}\big|_{W'}=0$ 
  applied to small balls is
  equivalent to imposing the linearized constraints $\delta C_\zeta = 0$.

In section \ref{sec:equations}, we reviewed the argument that
in the special case of Einstein gravity, 
 imposing the linearized equations within small enough balls is equivalent to 
requiring that the fully nonlinear equations hold within the ball \cite{Jacobson2015a}. Thus by considering spheres
centered at each point and in all Lorentz frames, one could 
conclude that the full Einstein equations hold everywhere.\footnote{There
is a subtlety associated with whether the solutions within each small
ball can be consistently glued together to give a solution over all of spacetime.  
One must solve for the gauge transformation relating the Riemann normal 
coordinates at different nearby points, and errors in the linearized approximation
could accumulate as one moves from point to point.  The question of whether the ball
size can be made small enough so that the total accumulated error goes to zero
deserves further attention. \renewcommand{\baselinestretch}{1} \footnotesize}  
Such an argument cannot be
made for a theory that involves higher curvature terms.  One finds that higher order terms
in the RNC expansion are needed to detect the higher 
curvature pieces of the field equations, but these terms enter at the same order as 
the nonlinear corrections to the linearized equations. This signals 
a breakdown of the perturbative
expansion unless the curvature is small.   

The fact that we obtain only linearized equations for the higher curvature theory 
is consistent with the effective field theory standpoint.  One could take 
the viewpoint that higher curvature corrections are 
 suppressed by powers of a UV scale, and the effective field theory is valid 
only when the curvature is small compared to this scale.  This suppression would suggest
that the linearized equations largely capture the effects of the higher curvature corrections
in the regime where effective field theory is reliable.

\subsection{Comparison to other ``geometry from entanglement'' approaches}
\label{sec:comparison}
Several proposals have been put forward to understand gravitational dynamics
in terms of thermodynamics and entanglement.  Here we will compare the 
entanglement equilibrium program considered in this chapter 
to two other approaches: the equation of state for
local causal horizons, and gravitational dynamics from holographic entanglement
entropy  (see \cite{Carroll2016a} for a related discussion).

\subsubsection{Causal horizon equation of state}
By assigning an entropy proportional to the area of local causal horizons, Jacobson
showed that the Einstein equation arises as an equation of state  
\cite{Jacobson1995a}.  This approach employs a physical process first law for the local 
causal horizon, defining a heat $\delta Q$ as the flux of local boost energy across the horizon.  
By assigning an entropy $S$ to the horizon proportional to its area, one finds that the 
Clausius relation $\delta Q = T\delta S$ applied to all such horizons is equivalent to the Einstein
equation.

The entanglement equilibrium approach differs in that it employs an equilibrium state first 
law [equation (\ref{firstlawhigher})], instead of  a physical process one \cite{Wald1994}.  
It therefore represents 
a different perspective that focuses on the steady-state behavior, as opposed to dynamics
involved with evolution along the causal horizon.
It is consistent therefore that we obtain constraint
equations in the entanglement equilibrium setup, since one would not expect evolution
equations to arise as an equilibrium condition.\footnote{We thank Ted Jacobson
for clarifying this point.}   That we can infer dynamical equations
from the constraints is related to the fact that the dynamics of diffeomorphism-invariant 
theories is entirely determined by the constraints evaluated in all possible Lorentz frames.  

Another difference comes from the focus on spacelike balls as opposed to local causal 
horizons.  Dealing with a compact spatial region has the advantage of providing an IR 
finite entanglement entropy, whereas the entanglement associated with local causal
horizons can depend on fields far away from the point of interest.  This allows us to give 
a clear physical interpretation for the surface entropy functional as entanglement entropy, 
whereas such an interpretation is less precise in the equation of state approaches.  

Finally, we note that both approaches attempt to obtain fully nonlinear equations by
considering ultralocal regions of spacetime.  In both cases the derivation of the field equations
for Einstein gravity is fairly robust, however higher curvature corrections present some problems.
Attempts have been made in the local causal horizon approach
that involve modifying
the entropy density functional for the horizon
\cite{Eling2006c, Elizalde2008, Chirco2011,
 Brustein2009, Parikh2016, Dey2016, Padmanabhan2009, Padmanabhan2009a,
 Guedens2011}, but they meet certain challenges.  
 These include a need for a physical interpretation of the chosen entropy density functional, 
 and dependence of the entropy on arbitrary features of the local
Killing vector in the vicinity of the horizon
 \cite{Guedens2011, Jacobson2012b}.  
While the entanglement equilibrium argument avoids these problems, it fails to get beyond
linearized higher curvature equations, even after considering the small ball limit.  The 
nonlinear equations in this case appear to involve information beyond first order perturbations,
and hence may not be accessible based purely on an equilibrium argument.

\subsubsection{ Holographic entanglement entropy }
A different approach  comes from holography and the 
Ryu-Takayanagi formula \cite{Ryu:2006bv}.  By demanding that areas of
minimal surfaces
in the bulk match the entanglement entropies of spherical regions in the boundary CFT, one
can show that the linearized gravitational equations must hold \cite{Lashkari2013, Faulkner:2013ica,
Swingle2014}.  The argument employs an equilibrium state first law for the bulk geometry,
utilizing the Killing symmetry associated with Rindler wedges in the bulk.

The holographic approach is quite similar to the entanglement equilibrium argument since
both use equilibrium state first laws.  One difference is that the holographic argument must 
utilize minimal surfaces in the bulk, which extend all the way to the boundary of AdS.  This 
precludes using a small ball limit as can be done with the entanglement 
equilibrium derivation, and is the underlying reason that entanglement equilibrium can derive fully
nonlinear field equations in the case of Einstein gravity, 
whereas the holographic approach has thus far only obtained linearized equations.  
Progress has been made to going beyond 
linear order in the holographic approach by considering higher order perturbations
in the bulk \cite{Faulkner2015, Lashkari2016, Beach2016a, Faulkner2017, Haehl2017}.
Also, by considering the equality of bulk and boundary modular flow, the linearized argument
in holography has been extended to applying in an arbitrary background \cite{Lewkowycz2018},
 which suggests 
that a fully nonlinear derivation of the dynamics from entanglement can be obtained by 
integrating the linearized result.

\subsection{Thermodynamic interpretation of the first law of causal diamond mechanics}
Apart from the entanglement equilibrium interpretation, the first law of causal diamond mechanics could also directly be interpreted  as a thermodynamic relation.   Note that the identity (\ref{barbecue}) for Einstein gravity bears a  striking resemblance  to the fundamental relation in thermodynamics
\beq \label{firstlawthermo}
dU = T dS - p dV,
\eeq
where $U(S,V)$ is the internal energy, which is a function of the   entropy $S$  and volume $V$.  
The  first law  (\ref{barbecue})  turns into the thermodynamic   relation (\ref{firstlawthermo}),   if one makes the following identifications for the temperature $T$ and pressure $p$
\beq \label{temppress}
T = \frac{\kappa \hbar}{2 \pi k_B c} \, , \quad \quad p = \frac{c^2 \kappa k}{8 \pi G} \, .
\eeq
Here we have restored fundamental constants, so that the quantities on the RHS have  the standard  units of temperature and pressure.
The expression for the temperature is the well-known Unruh    temperature  \cite{Unruh1976}. The formula for the pressure lacks a microscopic understanding at the moment, although we emphasize the expression follows from consistency of the first law.

The thermodynamic interpretation   motivates the name ``first law'' assigned to (\ref{barbecue}), and  arguably it   justifies the terminology ``generalized volume'' used for $W$   in this chapter, since it enters into the first law for higher curvature gravity (\ref{firstlawhigher}) in the place of the volume.
The only difference with the fundamental relation in thermodynamics is the minus sign in front of the energy variation. This different sign  also enters into the first law for   de Sitter horizons \cite{Gibbons1977}. In the latter case the sign appears because   empty de Sitter spacetime has maximal entropy, and adding matter only decreases the horizon entropy. Causal diamonds are rather similar in that respect.

\subsection{Generalized volume and holographic complexity} 
The emergence of  a generalized notion of volume in this analysis is interesting in
its own right.  We showed that when perturbing around a maximally symmetric background, the 
variation of the generalized volume is proportional to the variation of the gravitational part of the Hamiltonian.  The fact that the Hamiltonian could be 
written in terms of a local, geometric functional of the surface was a nontrivial consequence of 
the background geometry being maximally symmetric and $\zeta^a$ being a conformal
Killing vector whose conformal factor vanishes on $\Sigma$.  The local geometric nature
of $W$ makes it a useful, diffeomorphism invariant quantity with which to characterize the 
region under consideration, and thus should be a good state function in the thermodynamic 
description of an ensemble of quantum geometry microstates.  One might hope that such
a microscopic description would also justify the fixed-$W'$ constraint 
in the entanglement equilibrium derivation, which was only  motivated macroscopically 
by the first law of causal diamond mechanics.

Volume has  recently been identified as an important quantity in holography, where it is
conjectured to be related to complexity \cite{Susskind2014,Brown2015a}, or 
fidelity susceptibility \cite{Miyaji2015}.  The complexity$=$volume conjecture states that the complexity of some boundary state on a
time slice $\Omega$ is proportional to the  volume of the extremal codimension-one bulk hypersurface $\mathcal{B}$ which meets the asymptotic boundary on the corresponding time slice.\footnote{A similar expression has also been proposed for the complexity of subregions of the boundary time slice. In that case, $\mathcal{B}$ is the bulk hypersurface bounded by the corresponding subregion on the asymptotic boundary and
the Ryu-Takayanagi surface \cite{Ryu:2006bv} in the bulk \cite{Alishahiha2015, Ben-Ami2016},
or, more generally, the Hubeny-Rangamani-Takayanagi surface \cite{Hubeny2007} if the spacetime is time-dependent \cite{Carmi2016}.
\renewcommand{\baselinestretch}{1} \footnotesize} 

While volume is the natural functional to consider for Einstein gravity, \cite{Alishahiha2015} noted that
this should be generalized for higher curvature theories.  The functional 
proposed in that work resembles our generalized volume $W$, but suffers from 
an arbitrary dependence on the choice of foliation of the codimension-one hypersurface
on which it is evaluated.  We therefore suggest that $W$, as defined in \req{eqn:W}, may provide a suitable generalization
of volume in the context of higher curvature holographic complexity.

Observe however that our derivation of $W$ using the Iyer-Wald formalism was carried out in the particular case of spherical regions whose causal diamond is preserved by a conformal Killing vector. On more general grounds, one could speculate that
the holographic complexity functional in higher derivative gravities should involve contractions of $E^{abcd}$ with the geometric quantities characterizing $\mathcal{B}$, namely the  induced  metric $h_{ab}$ and the normal vector $u^a$.
The most general functional involving at most one factor of $E^{abcd}$ can be written as
\begin{equation}\label{wwe}
\mathcal{W}(\mathcal{B})=\int_{\mathcal{B}} \eta \left(\alpha E^{abcd}u_a h_{bc}u_d+ \beta E^{abcd} h_{ad}h_{bc}+\gamma\right)\, ,
\end{equation}
for some constants $\alpha$, $\beta$ and $\gamma$ which should be such that $\mathcal{W}(\mathcal{B})=V(\mathcal{B})$ for Einstein gravity. It would be interesting to explore the validity of this proposal in particular holographic setups, e.g., along the lines of \cite{Carmi2016}.

\subsection{Future work}
We conclude by laying out future directions for the entanglement equilibrium program.

\subsubsection{Higher order perturbations}
In this chapter we restricted attention only to first order perturbations of the entanglement
entropy and the geometry.  Working to higher order in perturbation theory could yield several
interesting results.  One such possibility would be proving that the vacuum entanglement 
entropy is maximal, as opposed to merely extremal.  The second order change in entanglement
entropy is no longer just the change in modular Hamiltonian expectation value.  The 
difference is given by the relative entropy, so a proof of maximality will likely invoke
the positivity of relative entropy.  On the geometrical side, a second order variational
identity would need to be derived, along the lines of \cite{Hollands2013}.  
One would expect that graviton contributions would appear at this order, and it 
would be interesting to examine how they play into the entanglement equilibrium story.  
Some initial steps towards such a derivation are taken in \cite{Jacobson2017b}.
Also, by considering small balls and using the higher order terms in the Riemann normal
coordinate expansion
(\ref{eqn:dg4}), in addition to   higher order perturbations,  it is possible that one could derive the fully nonlinear
field equations of any higher curvature theory.  Finally, coherent states pose a puzzle
for the entanglement equilibrium hypothesis, since they change the energy within
the ball without changing the entanglement \cite{Varadarajan2016}.  However, 
their effect on the energy density only appears at second order in perturbations, so 
carrying the entanglement equilibrium argument to higher order could shed light on 
this puzzle.


\subsubsection{Nonminimal couplings and gauge fields} \label{sec:nonmin}
 We restricted attention to minimally coupled matter throughout this chapter.  Allowing for 
 nonminimal coupling can lead to new, state-dependent divergences in the entanglement
 entropy \cite{Marolf2016}.  As before, these divergences  will be localized on 
 the entangling surface, taking the form of a Wald entropy.  It therefore
  seems plausible that an entanglement equilibrium argument will go through in this
 case, reproducing the field equations involving the nonminimally coupled field.  
 Note the state-dependent divergences could lead to variations of the couplings in the 
 higher curvature theory, which may connect to the entanglement chemistry program,
 which considers Iyer-Wald first laws involving variations of the couplings \cite{Caceres2016}.

Gauge fields introduce additional subtleties related
 to the existence of edge modes \cite{Donnelly2012, Donnelly2015E, Donnelly2016W}, and since
 these affect the renormalization of the gravitational couplings, they require special attention.  
 Gravitons are even more problematic due to difficulties in defining the entangling surface in
 a diffeomorphism-invariant manner and in finding a covariant regulator \cite{Fursaev1997,
 Cooperman:2013iqr, Solodukhin2015, Bousso2016}.  It would be 
 interesting to analyze how to handle these issues in the entanglement equilibrium
 argument.

\subsubsection{Nonspherical subregions}
 The entanglement equilibrium condition was shown to hold for spherical subregions and 
 conformally invariant matter.  One question that arises is whether an analogous 
 equilibrium statement holds for linear perturbations to the vacuum in an arbitrarily
 shaped region.  Nonspherical regions present a challenge because
there is no longer a simple relation between the modular Hamiltonian and the matter stress
tensor.  Furthermore, nonspherical regions do not admit a conformal Killing vector 
which preserves its causal development.  Since many properties of the conformal
Killing vector were used when deriving the generalized volume $W$, 
it may 
need to be modified to apply to nonspherical regions and their perturbations.

Adapting the entanglement equilibrium arguments to nonspherical regions may involve shifting 
the focus to evolution under the modular flow, as opposed to a geometrical evolution generated
by a vector field.  Modular flows are complicated in general, but one may be able to use general
properties of the flow to determine whether the Einstein equations still imply maximality
of the vacuum entanglement for the region.  Understanding the modular flow may also
shed light on the behavior of the entanglement entropy for nonconformal matter, and whether
some version of the entanglement equilibrium hypothesis continues to hold.

\subsubsection{Physical process}
As emphasized above, the first law of causal diamond mechanics is an equilibrium state construction
since it compares the entropy of  $\partial \Sigma$ on two infinitesimally related geometries
\cite{Wald1994}.  One could 
ask whether there exists a physical process version of this story, which
deals with entropy changes and energy fluxes as you evolve along the null boundary of the 
causal diamond.  For this, the notion of quantum expansion for the null surface
introduced in \cite{Bousso2016} would be a useful concept, which is defined by the derivative
of the generalized entropy along the generators of the surface.  One possible subtlety in 
formulating a physical process first law for the causal diamond is that the (classical) 
expansion 
of the null boundary is nonvanishing, so it would appear that this setup does not
correspond to a 
dynamical equilibrium configuration.  Nevertheless, it may be possible to gain useful information
about the dynamics of semiclassical gravity by considering these nonequilibrium
physical processes. 
 An alternative that avoids this issue 
is to focus on quantum extremal surfaces \cite{Engelhardt2015} whose quantum expansion 
vanishes, and therefore may lend themselves to an equilibrium physical process
first law.

\section*{Appendices}

\renewcommand\thesection{\arabic{chapter}.\Alph{section} }
\setcounter{section}{0}

\section{Conformal Killing vector in flat space}  \label{appkill}
Here we make explicit the geometric quantities introduced in section \ref{subsec:setup} in the case of a Minkowski background, whose metric we write in spherical coordinates, i.e., $ds^2 = - dt^2 + dr^2 + r^2 d \Omega_{d-2}^2$. Let $\Sigma$ be a spatial ball of radius $\ell$ in the time slice $t=0$ and with center at $r=0$. The conformal Killing vector which preserves the causal diamond  of $\Sigma$ is given by \cite{Jacobson2015a} 
\begin{equation}\label{zz}
\zeta=\left( \frac{\ell^2-r^2-t^2}{\ell^2}\right) \partial_t-\frac{2 r t}{\ell^2} \partial_r \, ,
\end{equation}
where we have chosen the normalization in a way such that $\zeta^2=-1$ at the center of the ball, which then gives the usual notion of energy for $H_\zeta^m$ (i.e. the correct units).
It is straightforward to check that 
$
\zeta(t=\pm \ell,r=0)=\zeta(t=0,r=\ell)=0\, ,
$
 i.e., the tips of the causal diamond and the maximal sphere $\partial \Sigma$ at its waist are fixed points of $\zeta$, as expected. Similarly, $\zeta$ is null on the boundary of the diamond. In particular, 
 $
 \zeta(t=\ell \pm r)=\mp 2 r(\ell \pm r)/\ell^2 \cdot (\partial_t \pm \partial_r)\, .
$
The vectors $u$ and $n$ (respectively normal to $\Sigma$ and to both $\Sigma$ and $\partial \Sigma$) read
$u=\partial_t$, $n=\partial_r$,
so that the binormal to $\partial \Sigma$ is given by 
$
n_{ab}=2\nabla_{[a} r \nabla_{b]} t \, .
$
It is also easy to check that $\lie_{\zeta}g_{ab}=2\alpha g_{ab}$ holds, where
$
\alpha\equiv \nabla_{a}\zeta^a/d=-2t/\ell^2 \,.
$
Hence, we immediately see that $\alpha=0$ on $\Sigma$, which implies that the gradient of $\alpha$ is proportional to the unit normal $u_a=-\nabla_a t$. Indeed, one finds $ \nabla_a \alpha=-2\nabla_a t/\ell^2 $, 
so in this case $N\equiv \lVert \nabla_a \alpha \rVert^{-1}=\ell^2/2$. It is also easy to show that $(\nabla_a \zeta_b)|_{\partial \Sigma}=\kappa n_{ab}$ holds, where the surface gravity reads $\kappa=2/\ell$.

As shown in \cite{Jacobson1993}, given some metric $g_{ab}$ with a conformal Killing field $\zeta^a$, it is possible to construct other metrics $\bar{g}_{ab}$ conformally related to it, for which $\zeta^a$ is a true Killing field. More explicitly, if 
$\lie_{\zeta}g_{ab}=2\alpha g_{ab}$, then $\lie_{\zeta}\bar{g}_{ab}=0$ 
as long as $g_{ab}$ and $\bar{g}_{ab}$ are related through $\bar{g}_{ab}=\Phi\, g_{ab}$, where $\Phi$ satisfies
\begin{equation}
 \lie_{\zeta} \Phi+2\alpha \Phi=0\, .
\end{equation}
For the vector \req{zz}, this equation has the general solution 
\begin{equation}\label{pii}
\Phi(r,t)=\frac{\psi(s)}{r^2}\, \quad  \text{where} \quad s\equiv \frac{\ell^2+r^2-t^2}{r} \, .
\end{equation}
Here, $\psi(s)$ can be any function. Hence, $\zeta$ in \req{zz} is a true Killing vector for all metrics conformally related to   Minkowski's  with a conformal factor given by \req{pii}. For example, setting $\psi(s)=L^2$, for some constant $L^2$, one obtains the metric of AdS$_2\times S_{d-2}$ with equal radii, namely: 
$ds^2=L^2/r^2 (-dt^2+dr^2)+L^2 d\Omega_{d-2}^2
$. Another simple case corresponds to $\psi(s)=L^2((s^2/(4L^2)-1)^{-1}$.  Through the change of variables \cite{Casini2011}: $t=L \sinh(\tau/L)/(\cosh u +\cosh (\tau/L))$, $r=L \sinh u \, /(\cosh u +\cosh (\tau/L))$, this choice leads to the $\mathbb{R}\times H^{d-1}$ metric (where $H^{d-1}$ is the hyperbolic plane): $ds^2= -d\tau^2+L^2(du^2+\sinh^2 u \, d\Omega^2_{d-2})$. 


\section{Generalized volume in higher order gravity} \label{app:W}

\noindent The generalized volume $W$ is defined in \eqref{eqn:W}.
We restate the expression here
\beq
W= \frac{1}{(d-2)E_0} \int_\Sigma \eta    \left (  E^{abcd} u_a u_d h_{bc}  - E_0  \right) \, ,
\eeq
where $E_0$ is   a theory-dependent constant defined by the tensor $E^{abcd}$
in a maximally symmetric solution to the field equations through
$E^{abcd}\overset{\text{MSS}}{=}E_0(g^{ac}g^{bd}-g^{ad}g^{bc})$. Moreover, $E^{abcd}$     is  the variation of the Lagrangian scalar $\mathcal L$ with respect to the Riemann tensor $R_{abcd}$ if we were to treat it as an independent field \cite{Iyer1994a},
\begin{align} \label{defEtensor}
E^{abcd} &= \frac{\partial   \mathcal L}{\partial R_{abcd}}  - \nabla_{a_1} \frac{\partial \mathcal L}{\partial \nabla_{a_1} R_{abcd}} + \dots   \\
&+ (-1)^m \nabla_{(a_1} \cdots \nabla_{a_m )} \frac{\partial \mathcal L}{\partial  \nabla_{(a_1} \cdots \nabla_{a_m )} R_{abcd}} \nonumber  \, ,
\end{align}
where $\mathcal L$ is then  defined  through $L = \epsilon \mathcal L$. In this section we provide explicit expressions for $W$ in $f(R)$ gravity, quadratic gravity and Gauss-Bonnet gravity. Observe that throughout this section we use the bar on $\bar{R}$ to denote evaluation on a MSS. Imposing a MSS to solve the field equations of a given higher derivative theory gives rise to a constraint between the theory couplings and the background curvature $\bar{R}$. This reads \cite{Bueno:2016ypa}
\begin{equation}\label{emb}
E_0=\frac{d}{4\bar{R}} \mathcal L(\bar{R}) \, ,
\end{equation} 
where $ \mathcal L(\bar{R})$ denotes the Lagrangian scalar evaluated on the background.

\paragraph{$f(R)$ gravity.} 
A simple higher curvature gravity is obtained by replacing $R$ in the Einstein-Hilbert action by a function of $R$
\beq
L_{f(R)} = \frac{1}{16\pi G}  \epsilon f(R) \, .
\eeq
To obtain the generalized volume we need
\beq
E^{abcd}_{f(R)} = \frac{f'(R)}{32 \pi G} \left( g^{ac} g^{bd}- g^{ad} g^{bc} \right) \, , \quad E_0 = \frac{f'(\bar{R})}{32 \pi G} \, .
\eeq
The generalized volume then reads
\beq
W_{f(R)} = 
\frac{1}{d-2} \int_{\Sigma} \eta \left [ (d-1) \frac{f'(R)}{f'(\bar{R})} - 1\right] \, .
\eeq
\paragraph{Quadratic gravity.}
A general quadratic   theory of gravity is given by the Lagrangian
\begin{align}\notag
  L_\text{quad}  = & \,   \epsilon \bigg [ \frac{1}{16 \pi G} \big( R -2 \Lambda \big)   + \alpha_1  R^2 + \alpha_2  R_{ab} R^{ab} \\
& + \alpha_3   R_{abcd}R^{abcd}  \bigg ] \, .
\end{align}
Taking the derivative of the Lagrangian with respect to the Riemann tensor leaves us with
\begin{align}\notag
E^{abcd}_\text{quad} = & \, \left ( \frac{1}{32 \pi G} +  \alpha_1 R \right) 2 g^{a[c} g^{d]b} \\
& + \alpha_2 \left (   R^{a[c} g^{d]b} + R^{b[d} g^{c]a} \right) + 2 \alpha_3R^{abcd} \, ,
\end{align}
and using \req{msb} one finds
\beq
E_0 = \frac{1}{32 \pi G} + \left( \alpha_1 + \frac{\alpha_2}{d} +  \frac{2\alpha_3}{d(d-1)} \right) \bar{R}\, .
\eeq
The generalized volume for quadratic gravity thus reads
\begin{align}\notag
&W_\text{quad}
=  \frac{1}{(d-2)E_0} \int_\Sigma \eta \bigg [ (d-1)\left(\frac{1}{32 \pi G}+\alpha_1 R \right) - E_0 \\
&+ \frac{1}{2} \alpha_2 \left (R - R^{ab} u_a u_b (d-2)\right) - 2 \alpha_3 R^{ab} u_a u_b \bigg ].
\end{align}
An interesting instance of quadratic gravity is Gauss-Bonnet theory, which is obtained by restricting to $\alpha_1=-\frac{1}{4}\alpha_2=\alpha_3=\alpha$.
The generalized volume then reduces to
\begin{align} \notag
W_{\text{GB}}  = & \, \frac{1}{(d-2)E_0}\int_\Sigma \eta \left[ \frac{1}{32 \pi G}(d-1) - E_0 \right. \\ \label{eq:WGB}
&\left.+ (d-3) \alpha \Big( R + 2 R^{ab} u_a u_b \Big) \right] \, ,
\end{align}
with $E_0=1/(32 \pi G)+\alpha \bar{R} (d-2)(d-3)/(d(d-1))$.
Since the extrinsic curvature of $\Sigma$ vanishes in the background, 
the structure $R+2R^{ab}u_a u_b$ is equal to the intrinsic Ricci scalar of $\Sigma$,
in the background and at first order in perturbations.


\section{Linearized equations of motion for higher curvature gravity using RNC} \label{app:FLDMRNC}
 
The variational identity (\ref{eqn:offshelllocalgeo}) states that the vanishing of the linearized constraint equations $\delta C_\zeta$ is equivalent to a relation between the variation of the Wald entropy, generalized volume, and matter energy density.
In \cite{Jacobson2015a}, Jacobson used this relation to extract  the Einstein equations,
 making use of Riemann normal coordinates. 
Here we perform a similar calculation for the higher curvature generalization of the first law of causal diamond mechanics which will produce the linearized equations of motion.
In this appendix we will restrict to theories whose Lagrangian depends on the metric and the Riemann tensor, $L[g_{ab},R_{abcd}]$, and to linearization around flat space.

The equations of motion for such a general higher curvature theory  read
\begin{equation}\label{eomhighapp}
\begin{split}
  - \frac{1}{2} g^{ab} \mathcal L  + E^{aecd} \tensor{R}{^b_{ecd}} - 2  \nabla_c \nabla_d E^{acdb}    = \frac{1}{2} T^{ab}.
\end{split}
\end{equation}
Linearizing the equations of motion around flat space leads to
\begin{align}\label{lineomhighapp}
 &  - \frac{1}{32\pi G} \eta^{ab} \delta R  + E^{aecd}_\text{Ein} \delta \tensor{R}{^b_{ecd}} - 2  \partial_c \partial_d \delta E^{a c d b}_{\text{higher}} \nonumber\\
&\quad= \frac{\delta G^{ab}}{16\pi G} - 2  \partial_c \partial_d \delta E^{a c d b}_{\text{higher}} = \frac{1}{2} \delta T^{ab}\, ,
\end{align}
where we split $E^{abcd}=E^{abcd}_{\text{Ein}}+E^{abcd}_{\text{higher}}$ into an  Einstein piece, which goes into the Einstein tensor, and a piece coming from higher derivative terms.
We used the fact that many of the expressions in \eqref{lineomhighapp} significantly simplify when evaluated in the Minkowski background because the curvatures vanish.
For example, one might have expected additional terms proportional to the variation of the Christoffel symbols coming from $\delta(\nabla_c\nabla_d E^{acdb})$.
To see why these terms are absent, it is convenient to split this expression into its Einstein part and a part coming from higher derivative terms.
The Einstein piece does not contribute since $E^{acdb}_{\text{Ein}}$ is only a function of the metric and therefore its covariant derivative vanishes.
The higher derivative piece will give $\partial_c \partial_d \delta E^{a c d b}_\text{higher}$ as well as terms such as $\delta \Gamma^{c}_{ce} \nabla_d E^{eadb}_{\text{higher}}$ and $\Gamma^{c}_{ce} \nabla_d \delta E^{eadb}_{\text{higher}}$.
However, the latter two terms are zero because both the Christoffel symbols
 and $E^{eadb}_{\text{higher}}$ vanish when evaluated in the Minkowski background with
 the standard coordinates.

We now want to evaluate each term in \eqref{newvarid} using Riemann normal coordinates.
Taking the stress tensor $T^{ab}$ to be  constant for small enough balls, the variation of \eqref{potati} reduces to
\begin{equation}
\delta H^{m}_{\zeta} = \frac{\Omega_{d-2}\ell^{d}}{d^2-1} \kappa u_a u_b \delta T^{ab} + \mathcal{O}\left(\ell^{d+2}\right)\, ,
\end{equation}
where $\Omega_{d-2}$ denotes the area of the $(d-2)$-sphere, $\ell$ is the radius of our geodesic ball and $u_a$ is the future pointing unit normal.
As was found in \cite{Jacobson2015a}, the Einstein piece of the symplectic form will combine with the area term of the entropy to produce the Einstein tensor.
Therefore, we focus on the higher curvature part of $\delta H_\zeta^g$.
Combining \eqref{eqn:hamilton} and \eqref{eqn:symplform}, we find 

\begin{align}\label{eq:symplformhigh}
\delta H^g_{\zeta,\text{higher}} 
&= - \frac{4\kappa}{\ell} \int d\Omega \int dr r^{d-2} u_a u_d \eta_{bc } \Big(\delta  E^{abcd}_{\text{higher}}(0)
+ \partial_i \delta  E^{abcd}_{\text{higher}}(0) r n^i  
\nonumber  \\ 
& 
\qquad \qquad\qquad \qquad + \frac{1}{2}\partial_i \partial_j \delta  E^{abcd}_{\text{higher}}(0) r^2 n^i n^j + \mathcal{O}\left(r^3\right)\Big) \nonumber\\
&= - 4\kappa \Omega_{d-2} \ell^{d-2} u_a u_d \eta_{bc } \bigg(\frac{\delta  E^{abcd}_{\text{higher}}(0)}{(d-1)}
 + \frac{\ell^{2}  \delta^{ij} \partial_i \partial_j \delta  E^{abcd}_{\text{higher}}(0)}{2(d^2-1)} \bigg) + \mathcal{O}\left(\ell^{d+2}\right)\, .
\end{align}
Here, $n^i$ is the normal vector to $\partial \Sigma$ and the indices $a,b$ run over space-time directions, while the indices $i,j$ run only over spatial directions, and $\partial_i$ is the derivative operator compatible with the flat background metric on $\Sigma$.
In the first line, we simply use the formula for the Taylor expansion of a quantity $f$
in the coordinate
system compatible with $\partial_i$, 
\begin{equation}\label{Taylor}
f(x)=f(0)+\partial_a f(0) x^a + \frac{1}{2} \partial_a \partial_b f(0) x^a x^b + \mathcal{O}\left(x^3\right) \, ,
\end{equation}
where $(0)$ denotes that a term is evaluated at $r=0$.
Since we evaluate our expressions on a constant timeslice at $t=0$, we have $x^t=0$ and $x^i=r \, n^i$, where $r$ is a radial coordinate inside the geodesic ball and the index $i$ runs only over the spatial coordinates.
To evaluate the spherical integral, it is useful to note that spherical integrals over odd powers of $n^i$ vanish and furthermore
\begin{align}
\int d\Omega \, n^i n^j &= \frac{\Omega_{d-2}}{d-1}\delta^{ij} \, , \\
\int d\Omega \, n^i n^j n^k n^l &= \frac{\Omega_{d-2}}{d^2-1}\left(\delta^{ij}\delta^{kl}+\delta^{ik}\delta^{jl}+\delta^{il}\delta^{jk}\right)  \;.
\end{align}
Next, we evaluate $\delta S_\text{higher}$, the variation of the higher curvature part of the Wald entropy given in \eqref{eqn:SWald}, in a similar manner.
\begin{align}
\delta S_{\text{higher}} = 8 \pi \Omega_{d-2}\ell^{d-2} u_a u_d
& \left(\frac{\eta_{bc}\delta E^{abcd}_{\text{higher}}(0)}{(d-1)} 
 + \frac{\ell^2\left[\eta_{bc}\delta^{ij} \partial_i \partial_j \delta E^{abcd}_{\text{higher}}(0) 
 	+ 2 \partial_b \partial_c \delta E^{abcd}_{\text{higher}}(0) \right] }{2(d^2-1)}\right) \nonumber \\
	& + \mathcal{O}\left(\ell^{d+2}\right) \; ,
\end{align}
We are now ready to evaluate the first law of causal diamond mechanics \eqref{newvarid}.
Interestingly, the leading order pieces of the Hamiltonian and Wald entropy exactly cancel against each other.
Note that these two terms would have otherwise dominated over the Einstein piece.
Furthermore, the second term in the symplectic form and Wald entropy also cancel, leaving only a single term from the higher curvature part of the identity.
Including the Einstein piece, we find the first law  for higher curvature gravity reads in Riemann normal coordinates 
\begin{equation}\label{eomRNC}
\begin{split}
- \frac{\kappa \Omega_{d-2}\ell^{d} }{d^2-1} u_a u_d \Big ( \frac{\delta G^{ad}(0)}{8 \pi G} - 4 \partial_b \partial_c \delta E^{abcd}_{\text{higher}}(0) - \delta T^{ad} \Big) 
 +\mathcal{O}\left(\ell^{d+2}\right) =0 \, ,
\end{split}
\end{equation}
proving equivalence to the linearized equations (\ref{lineomhighapp}).



\renewcommand\thesection{\thechapter.\arabic{section}}


\chapter{Local phase space and edge modes for diffeomorphism-invariant theories}
This chapter is based on my paper ``Local phase space and edge modes for 
diffeomorphism-invariant theories," published in the Journal of High Energy Physics in 
2018 \cite{Speranza2018}. 
\label{ch:lps}
\section{Introduction}
In gravitational theories, the problem of defining local subregions and observables is complicated
by diffeomorphism invariance.  Because it is a gauge symmetry, diffeomorphism invariance
leads to constraints that must be satisfied by initial data for the field equations.  These constraints
relate the values of fields 
in one subregion of a Cauchy slice to their values elsewhere, 
so that the fields 
cannot be interpreted as observables localized to a particular region.   
While this is true in any gauge theory, a further challenge for diffeomorphism-invariant theories
is that specifying a particular subregion is nontrivial, since diffeomorphisms can change the 
subregion's coordinate position.  

A related issue in quantum gravitational theories is the problem of defining entanglement entropy
for a subregion.  The usual definition of entanglement entropy assumes a factorization of the 
Hilbert space $\Hi = \Hi_A\otimes \Hi_{\bar{A}}$ into tensor factors $\Hi_A$ and $\Hi_{\bar{A}}$
associated with a subregion $A$ and its complement $\bar{A}$. However, all physical states
in a gauge theory are required to be annihilated by the constraints, and the nonlocal relations
the constraints impose on the physical Hilbert space prevents 
such a factorization from occurring.\footnote{Strictly speaking, the factorization of a Hilbert space 
of any continuum field theory is  formal, 
and only makes sense after regulating, e.g.\ with a lattice.
The nonfactorization due to gauge constraints is more fundamental, and persists even
in the regulated theory. \renewcommand{\baselinestretch}{1} \footnotesize}
One way of handling this nonfactorization is to define the entropy in terms of the algebra of 
observables for the local subregion \cite{Casini2014a}.  This necessitates a choice of 
center for the algebra, which roughly corresponds to
 Wilson lines that are cut by the entangling surface.   
This procedure is further complicated in gravitational theories, since the local subregion 
and its algebra of observables must be defined in a diffeomorphism-invariant manner.  
Thus, the issues of local observables
and entanglement in gravitational theories are intertwined.  

Despite these challenges, there are indications that a well-defined notion of local observables
and entanglement should exist in gravitational theories.  Holography provides a compelling  example, 
where the entanglement of bulk regions bounded by an extremal surface may be expressed 
in terms of entanglement in the CFT via the Ryu-Takayanagi formula and its quantum corrections
\cite{Ryu:2006bv, Faulkner2013a}.
Such regions are defined relationally relative to a fixed region
on the boundary, and hence give a diffeomorphism-invariant characterization of the local subregion.
Work regarding bulk reconstruction suggests that the algebra of 
observables for this subregion is 
fully expressible in terms of the subregion algebra of the CFT 
\cite{Dong2016, Jafferis2015, Czech2012, Cotler2017, Harlow2016, Almheiri2015}.

As discussed in section
\ref{sec:SgenSEE}, 
entanglement entropy provides a natural explanation for the proportionality between
black hole entropy and horizon area 
\cite{Sorkin:2014kta, Bombelli1986, Srednicki1993a, Frolov1993}, 
while finessing the issue of entanglement
divergences through renormalization of the gravitational couplings 
\cite{Susskind1994, Jacobson1994a, Cooperman:2013iqr}.  
However, in the case of gauge theories, the matching between entanglement 
entropy divergences and the 
renormalization of gravitational couplings is subtle.  
The entropy computed using conical methods \cite{Callan1994a}
contains contact
terms \cite{Kabat1995, Fursaev1997, Solodukhin2015}, 
which are related to the presence of edge modes on the entangling surface.  
These arise as a consequence of the nonfactorization of the Hilbert space due to 
the gauge constraints.  
Only when  the 
entanglement from these edge modes 
is properly handled does the black hole entropy 
have a statistical interpretation in terms of a von Neumann entropy \cite{Donnelly2012,
Donnelly2015E, Donnelly2016W}.  

Recently, Donnelly and Freidel presented a continuum description of the edge modes that arise 
both in Yang-Mills  theory and general relativity \cite{Donnelly2016F}.  Using covariant 
phase space techniques \cite{Witten1986, Crnkovic1987, Crnkovic:1987tz, Ashtekar1991}, 
they construct
a symplectic potential and symplectic form associated with a local subregion.  
These are expressed as local integrals of the fields and their variations over a Cauchy surface
$\Sigma$.  However, one finds that they are not fully gauge-invariant:
gauge transformations that are 
nonvanishing at the boundary $\partial\Sigma$ change the symplectic form by 
boundary terms.  Invariance is restored by introducing 
new fields  in a neighborhood
of the boundary, whose change under gauge transformations cancels the boundary term from the 
original symplectic form.  These new edge modes thus realize the idea that boundaries 
break gauge invariance, and cause some would-be gauge modes to become  
degrees of freedom associated with the subregion \cite{Carlip1994, Carlip1997}.

The analysis of diffeomorphism-invariant theories in \cite{Donnelly2016F} was restricted to 
general relativity with vanishing cosmological constant.  However, the 
construction can be generalized to arbitrary diffeomorphism-invariant theories, and it is the 
purpose of the present chapter to show how this is done.  The  symplectic
potential for the edge modes can be expressed in terms of the Noether charge and the 
on-shell Lagrangian
of the theory, and the symplectic form derived from it has contributions from the edge modes 
only
at the boundary.  These edge modes come equipped with set of symmetry transformations,
and the symmetry algebra is represented on the phase space as a Poisson bracket algebra.  
The generators of the surface symmetries are given by the Noether charges associated with the 
transformations.  We find that for generic diffeomorphism-invariant theories, the transformations
that preserve the entangling surface generate the algebra 
$\text{Diff}(\partial\Sigma)\ltimes\left(SL(2,\mathbb{R})
\ltimes \mathbb{R}^{2\cdot(d-2)}\right)^{\partial\Sigma}$.  In certain cases, including general 
relativity, the algebra is reduced to $\text{Diff}(\partial\Sigma)\ltimes 
SL(2,\mathbb{R})^{\partial\Sigma}$, consistent with the results of \cite{Donnelly2016F}.
Furthermore, for any other theory, there always exists a modification of the symplectic
structure in the form of a Noether charge ambiguity \cite{Jacobson1994b} that reduces the algebra
down to $\text{Diff}(\partial\Sigma)\ltimes 
SL(2,\mathbb{R})^{\partial\Sigma}$.  We also discuss what happens when the algebra
is enlarged to include surface translations, the transformations that do not map $\partial\Sigma$
to itself.  In order for these transformations to be Hamiltonian, the dynamical fields generically have 
to satisfy boundary conditions at $\partial\Sigma$.  Assuming the appropriate boundary
conditions can be found, the full surface symmetry algebra is a central extension of either
$\text{Diff}(\partial\Sigma)\ltimes\left( SL(2,\mathbb{R})\ltimes\mathbb{R}^2 \right)^{\partial\Sigma}$,
or a larger, simple Lie algebra.
The appearance of central charges in these algebras is familiar from similar constructions
involving edge modes at asymptotic infinity or black hole horizons 
\cite{Brown1986a, Carlip1997, Carlip2011}.

The construction of the extended phase space for arbitrary diffeomorphism-invariant theories
is useful for a number of reasons.  For one, higher curvature corrections
to the Einstein-Hilbert action generically appear  due to quantum gravitational effects.  
It is useful to have a formalism that can compute the corrections to the edge mode entanglement
coming from these higher curvature terms.  Additionally, there are several 
diffeomorphism-invariant theories that are simpler than general relativity in four dimensions, 
such as 2 dimensional dilaton gravity or 3 dimensional gravity in Anti-de-Sitter space.  
These could be useful testing grounds in which to understand the edge mode 
entanglement entropy, before trying to tackle the problem in four or higher dimensions.  
Finally, the general construction clarifies the relation of the extended phase space to the 
Wald formalism \cite{Wald1993a, Iyer1994a}, a connection that was also noted in
\cite{Geiller2017}.  

This chapter begins with a review of the covariant canonical formalism in section \ref{sec:cps}.
Care is taken to describe vectors and differential forms on this infinite-dimensional space, 
and also to understand the effect of diffeomorphisms of  the spacetime manifold on 
the covariant phase space.  Section \ref{sec:edge} discusses the $X$ fields that appear
in the extended phase space, which give rise to the edge modes.  Following this,
the construction of the extended phase space is given in section \ref{sec:eps}, which
describes how the edge mode fields contribute to the extended symplectic form.  Ambiguities
in the construction are characterized in section \ref{sec:jkm}, and the surface symmetry 
algebra is identified in section \ref{sec:ssa}.  Section \ref{sec:disc} gives a summary of results 
and ideas for 
future work.

\section{Covariant canonical formalism} \label{sec:cps}
The covariant canonical formalism \cite{Witten1986, Crnkovic1987, Crnkovic:1987tz, Ashtekar1991}
provides a Hamiltonian description of a field theory's degrees of freedom
while maintaining spacetime covariance.  This is achieved by working with the space $\SSS$ of 
solutions to the field equations.  
As long as the field equations admit a well-posed initial 
value formulation, each solution is in one-to-one correspondence with its initial data
on some Cauchy slice.  
$\SSS$ 
may therefore
be used to construct a phase space
 that is equivalent to  Hamiltonian formalisms coordinatized by initial positions and momenta.
Since a solution need not refer to a choice of initial Cauchy
slice and decomposition into spatial and time coordinates, spacetime covariance remains 
manifest in a phase space constructed from $\SSS$.  
The specification of a Cauchy surface and time variable can be viewed as a choice 
of coordinates on $\SSS$, with each solution being identified by its initial data.  

An important subtlety in this construction occurs for field equations with gauge symmetry.  The 
space $\SSS$ involves all  solutions to the field equations, so, in particular, treats two 
solutions that differ only by a gauge transformation as distinct.\footnote{Identifying solutions
with initial data is still possible if one  supplements the original field equations with
suitable gauge-fixing conditions.  One could therefore consider $\SSS$ as being coordinatized
by initial data along with a choice of gauge.  \renewcommand{\baselinestretch}{1} \footnotesize}  
In this case, $\SSS$ is too 
large to be the correct phase space for the theory, since gauge-related solutions should represent
physically equivalent configurations.  Instead, the true phase space $\PP$ should be obtained 
by quotienting $\SSS$ by the action of the gauge group.  It is useful to view $\SSS$ as 
 a fiber bundle, with
each fiber consisting of all solutions related to each other by a gauge transformation, in which
case $\PP$ is simply the base space of this fiber bundle.  
As discussed in section \ref{sec:eps}, the Lagrangian for the theory imbues
$\SSS$ with the structure of a presymplectic manifold, equipped with a degenerate presymplectic
form.  This degeneracy is necessary in order for it to project to a well-defined symplectic form
on $\PP$.    The remainder of this section is devoted
to describing the geometry of the space $\SSS$, while the requirements for various functions and 
forms (including the presymplectic form) 
to descend to well-defined objects on $\PP$ are discussed in section \ref{sec:edge}.

Working directly
with $\SSS$ allows   coordinate-free techniques to be applied  to both the 
spacetime manifold and the solution space itself.  In particular, the exterior calculus on the 
$\SSS$ gives a powerful language for describing the phase space symplectic geometry.  
We will follow the treatment of the exterior calculus given in 
\cite{Donnelly2016F},\footnote{For an extended review of this formalism, see
\cite{Compere2007} and references therein. \renewcommand{\baselinestretch}{1} \footnotesize
} where it was 
used to provide an extremely efficient way of identifying edge modes for a local subregion
in a gauge theory.   
This section provides a review of 
the formalism,
on which the remainder of this chapter heavily relies.  

The theories under consideration consist of dynamical fields, including the metric $g_{ab}$ and 
any matter fields, propagating on a spacetime manifold $M$.  
These fields satisfy diffeomorphism-invariant equations of motion, and the phase space is 
constructed from the infinite-dimensional space of solutions to these equations, $\SSS$.  
Despite being infinite-dimensional, many concepts from finite-dimensional differential 
geometry, such as vector fields, one-forms, and Lie derivatives, 
extend straightforwardly to $\SSS$, assuming it satisfies some technical
requirements such as being a Banach manifold \cite{Lee1990a, Lang1985}.
One begins by understanding the functions on $\SSS$, 
a wide
class of which is provided by the dynamical fields themselves.  Given 
a spacetime point $x\in M$ and a field $\phi$, the function $\phi^x$ associates to each 
solution the value of $\phi(x)$  in that solution.  
More generally, functionals of the dynamical fields, such
as integrals over regions of spacetime, also define functions on $\SSS$ by simply evaluating 
the functional in a given solution.  
We will often denote $\phi^x$ simply by $\phi$, with the 
dependence on the spacetime point $x$ implicit. 

A vector at a point of $\SSS$ describes an infinitesimal displacement away from a particular
solution, and hence corresponds to a solution of the linearized field equations.  Specifying 
a linearized solution about each full solution then defines a vector field $V$ 
on all of $\SSS$.  
The vector field acts on $\SSS$-functions  as a directional derivative, and in particular its
action on the functions $\phi^x$ is to give a new function $\Ph_V^x\equiv V[\phi^x]$, 
which, given a solution, evaluates the linearization $\Ph$ of the field $\phi$ 
at the point $x$.  This also allows us to define the exterior derivative of the 
functions $\phi^x$, denoted $\delta \phi^x$.  When contracted with the vector field
$V$, the one-form $\delta\phi^x$ simply returns the scalar function $\Ph_V^x$.  The 
one-forms $\del\phi^x$ form an overcomplete basis, so that arbitrary one-forms may be 
expressed as sums (or integrals over the spacetime point $x$) of $\del\phi^x$.  
This basis is overcomplete because the functions $\phi^x$ at different
points $x$ are related through the equations of motion, so that the forms $\del\phi^x$ are 
related as well.

Forms of higher degree can be constructed from the $\del\phi^x$ one-forms by 
taking exterior products.  The exterior product of a $p$-form $\alpha$ and a $q$-form $\beta$
is simply written $\alpha\beta$, and satisfies $\alpha\beta = (-1)^{pq} \beta\alpha$.  Since we only
ever deal with exterior products of forms defined on $\SSS$ instead of more general tensor
products, no ambiguity arises by omitting the $\wedge$ symbol, which we instead reserve 
for spacetime 
exterior products.  The action of the exterior derivative on arbitrary forms is fixed as usual by 
its action on scalar functions, along with the requirements of linearity, nilpotency $\delta^2=0$, 
and that it acts as an antiderivation,
\beq
\delta(\alpha\beta) = (\delta\alpha)\beta + (-1)^p \alpha \delta\beta.
\eeq
The exterior derivative $\delta$ always increases the degree of the form by one.  On the other 
hand, each vector field
$V$ defines an antiderivation $I_V$ that reduces the degree by one through contraction.  
$I_V$ can be completely characterized by its action on one-forms $I_V \del\phi^x=\Phi_V^x$,
along with the antiderivation property, linearity, nilpotency $I_\Ph^2
=0$, and requiring that it annihilate scalars.  
Just as in finite dimensions, the action of the $\SSS$ Lie derivative, denoted $L_V$, 
is related to $\del$ and $I_V$ via Cartan's magic formula \cite{Lang1985}
\beq \label{eqn:cartanmagic}
L_V = I_V \delta + \delta I_V.
\eeq
$L_V$ is a derivation, $L_V (\alpha\beta) = (L_V\alpha)\beta + \alpha L_V \beta$, that 
preserves the degree of the form.

We next discuss the consequences of working with  diffeomorphism invariant theories.  
A diffeomorphism $Y$ is a smooth, invertible map, $Y:M\rightarrow M$, sending the 
spacetime manifold $M$ to itself.  The diffeomorphism  induces a map of
tensors at $Y(x)$ to tensors at $x$ through the pullback $Y^*$ \cite{Wald1984}. 
Diffeomorphism invariance is simply the statement that if a configuration of tensor fields
$\phi$ satisfy the equations of motion, then so do the pulled back fields $Y^*\phi$. 
Now consider a one-parameter family of diffeomorphisms $Y_\lambda$, with $Y_0$ the identity.  
This yields a family of fields $Y_\lambda^*\phi$ that all satisfy the equations of motion.  The first
order change induced by $Y^*_\lambda$ defines the spacetime Lie derivative $\lie_\xi$ 
with respect to $\xi^a$, the tangent vector to the flow of $Y_\lambda$.
Consequently, $\lie_\xi \phi$ must be a solution to the linearized field 
equations, and the infinitesimal diffeomorphism generated by $\xi^a$ defines a vector field on
$\SSS$, which we  denote $\hat{\xi}$, 
whose action on $\delta\phi$ is 
\beq
I_{\hat\xi} \delta \phi\equiv \lie_\xi \phi.
\eeq

The diffeomorphisms we have considered so far have been taken to act the same on all solutions.
A useful generalization of this are the solution-dependent diffeomorphisms, defined through a 
function,
$\mathscr{Y}:\SSS\rightarrow \text{Diff}(M)$, valued in the diffeomorphism
group of the manifold, $\text{Diff}(M)$.  
Letting $Y$ denote the image of this function, 
we would like
to understand how the Lie derivative $L_V$ and exterior derivative $\del$ on $\SSS$ 
combine with the action of the pullback $Y^*$.  
In the
case $\mathscr{Y}$ is constant on $\SSS$, the Lie derivative simply commutes with $Y^*$, and so
$L_V Y^*\al = Y^* L_V \al$, where $\al$ is any form constructed from fields
and their variations at a single spacetime point.  When $Y$ is not constant, 
$V$ generates one-parameter families of diffeomorphisms $Y_\lambda$ and 
forms $\alpha_\lambda$ along the flow in $\SSS$.  At a given solution $s_0$, define a 
solution-independent diffeomorphism  
$Y_0 \equiv \mathscr{Y}(s_0)$ by the value of $\mathscr{Y}$ at $s_0$.
Then $Y^*_\lambda \alpha_\lambda$ and $Y^*_0\alpha_\lambda$ are  related
to each other at all values of $\lambda$ by a diffeomorphism, $Y_\lambda^* (Y_0^{-1})^*$.  
The first order change in these quantities at $\lambda=0$ is given by 
$L_V$, and since the two quantities differ at first order by an infinitesimal diffeomorphism, we find 
\beq \label{eqn:LphY*a}
L_V Y^*\al = L_V Y_0^*\al + Y^* \lie_{\dar(Y;V)} \al = Y^*(L_V \al + \lie_{\dar(Y;V)} \al).
\eeq
It is argued in appendix \ref{app:ids}, identity \ref{id:LVY*a},
that the vector $\dar^a(Y;V)$ depends linearly on 
$V$, and hence defines a one-form on $\SSS$, denoted $\dar_Y^a$.\footnote{In
\cite{Donnelly2016F}, $\dar_Y^a$ was denoted $\del_Y^a$.  We choose a different notation
to emphasize
that $\dar_Y^a$ is not an exact form, and to avoid confusion with the exterior derivative $\del$.
\renewcommand{\baselinestretch}{1} \footnotesize}
This yields the pullback formula for $L_V$,
\beq
L_V Y^*\al = Y^*(L_V\al + \lie_{I_V \dy} \alpha).
\eeq
Applying (\ref{eqn:cartanmagic}) to this equation, one can derive the pullback
formula for exterior derivatives  from \cite{Donnelly2016F} (see \ref{id:dY*a} for details),
\beq \label{eqn:dY*a}
\delta Y^*\alpha = Y^*(\delta\al+\lie_{\dy}\al).
\eeq

A number of properties of the variational vector field $\dar_Y^a$ follow from the 
formulas above.  First, note $\dar_Y^a$ is not an exact form on $\SSS$; rather, 
its exterior derivative
can be deduced from (\ref{eqn:dY*a}),
\beq
0 = \delta\del Y^*\al = Y^*(\del \lie_\dy\al +\lie_\dy \del\al +\lie_\dy \lie_\dy \al) = Y^*(\lie_{\del
(\dy)} \al + \lie_\dy \lie_\dy\al),
\eeq
and applying \ref{id:liedarliedar}, we conclude
\beq \label{eqn:ddx}
\delta(\dy)^a = -\frac12[\dy,\dy]^a.
\eeq
Another useful formula relates $\dar_Y^a$ to the vector $\dar_{Y^{-1}}^a$ associated with the 
inverse of $Y$.  Using that $Y^*$ and $(Y^{-1})^*$ are inverses of each other, we find
\beq
\delta \al = \delta Y^* (Y^{-1})^*\al = Y^*[\delta(Y^{-1})^*\al + \lie_{\dy} (Y^{-1})^* \al]
= \del\al +\lie_{\del_{Y^{-1}}}\al + \lie_{Y^*\dy} \al,
\eeq
where the last equality involves the identity \ref{id:liexiYi}. This implies
\beq \label{eqn:dy-1}
\dar_{Y^{-1}}^a = -Y^*\dar_Y^a.
\eeq
Additional identities are derived in appendix \ref{app:ids}.

Finally, as a spacetime vector field, $\dar_Y^a$ also defines a vector-valued one-form 
$\hdy$ on $\SSS$, which acts as $I_{\hdy} \delta\phi = \lie_\dy \phi$.  The contraction
$I_{\hdy}$ defines a derivation that preserves the degree of the form, 
in contrast to $I_{\hat\xi}$, 
which is an antiderivation that reduces the degree.  Similarly, $\delta(\dy)^a$
defines a vector-valued two-form on $\SSS$, and produces an antiderivation $I_{\delta(\dy)\,\hat{}}$
that increments the degree.

\section{Edge mode fields} \label{sec:edge}
Edge modes appear when a gauge symmetry is broken due to the presence of a 
boundary $\partial\Sigma$ of a Cauchy surface $\Sigma$.  
The classical phase space  or quantum mechanical Hilbert space associated with 
$\Sigma$ transforms nontrivially under gauge transformations that act at the boundary.  
This can be understood from the perspective of Wilson loops that are cut by the 
boundary.   A closed Wilson loop is gauge-invariant, but the cut Wilson loop
becomes a Wilson line in $\Sigma$, whose endpoints transform in some representation
of the gauge group.  To account for these cut-Wilson-loop degrees of freedom, one can 
introduce fictitious charged fields at $\partial \Sigma$, which can be attached to the 
ends of the Wilson lines to produce a gauge-invariant object.   These new fields are the 
edge modes of the local subregion.  They account for the possibility of charge 
density existing outside of $\Sigma$, which would affect the fields in $\Sigma$ due to 
Gauss law constraints.  The contribution of the edge modes to the entanglement can 
therefore be interpreted as parameterizing  ignorance of such localized charge densities
away from $\Sigma$.

A similar picture arises in the classical phase space of a diffeomorphism-invariant theory.  
The edge modes appear when attempting to construct a symplectic
structure  associated with $\Sigma$ for the solution space $\SSS$.  
Starting with the Lagrangian of the theory, one can construct from its variations a 
symplectic current $\omega$, a spacetime $(d-1)$-form whose integral over a spatial 
subregion $\Sigma$ provides a candidate presymplectic form. 
However, this form fails to be 
diffeomorphism invariant for two reasons.  First, a diffeomorphism moves points on the mainfold
around, and hence changes the shape and coordinate location of the 
surface.  
Second, since solutions related to each other by a diffeomorphism  represent
the same physical 
configuration,  the true phase space $\PP$ is obtained by projecting all solutions in 
a  gauge orbit in $\SSS$ down to a single representative.  In order for the symplectic form
to be compatible with this projection, the infinitesimal diffeomorphisms must be  degenerate 
directions of the presymplectic form \cite{Lee1990a}.\footnote{Indeed, only 
functions on $\SSS$ that are constant along the gauge orbits descend to well-defined functions
on $\PP$.  Similarly, the only forms that survive the projection must be both constant along
gauge orbits and annihilate vectors tangent to the gauge orbits.  In particular, the functions 
$\phi^x$ constructed from the dynamical fields do not survive the projection, while 
diffeomorphism-invariant functionals of $\phi^x$ do survive.  Note that this is one reason for 
working with $\SSS$: it is technically simpler to derive relations involving the local field functions
$\phi^x$ in $\SSS$ than always working with diffeomorphism-invariant objects in $\PP$.  Most of
the relations in this chapter are derived in $\SSS$, and then are argued to hold in $\PP$ if they
involve diffeomorphism-invariant functionals and are properly degenerate. \renewcommand{\baselinestretch}{1} \footnotesize
}  
This is equivalent to saying that the Hamiltonian generating the diffeomorphism may be chosen
to vanish. 
While the symplectic form obtained by integrating $\omega$ over a surface 
is degenerate for diffeomorphisms that vanish sufficiently quickly at its boundary, those that do not
produce boundary terms that spoil degeneracy. 

The problem of non-invariance due to diffeomorphisms that move the surface is solved by 
defining the surface's location in a diffeomorphism-invariant manner.  There are a variety of 
ways that this can be done.  One example comes from the Ryu-Takayanagi prescription in
holography, where the bulk entangling surface $\partial\Sigma$ is defined as the extremal surface
that asymptotes to a given subregion on the boundary of AdS \cite{Ryu:2006bv}.  Another set of 
techniques are the relational constructions of \cite{Giddings2006}, where one set of fields can
be used to define a coordinate system, and subregions can be defined relationally to these
coordinate fields.  An important point about the edge modes is that they are necessary even
after dealing with this first source of non-invariance: the presymplectic form may still not be 
appropriately degenerate even after specifying the subregion invariantly.  The remainder of this
chapter will primarily be focused on how this second issue is resolved, although 
the extended phase space provides a formal solution to the first issue as well.  

As demonstrated in \cite{Donnelly2016F}, both problems can be handled by introducing a collection
of additional fields $X$ whose contribution to the symplectic form restores diffeomorphism 
invariance.  These fields are the edge modes of the extended phase space.  This section
 is devoted to describing these fields and their 
 transformation properties under diffeomorphisms; 
the precise way in which they contribute to the symplectic form is discussed in section
\ref{sec:eps}.  

The fields $X$ can be defined through a $\text{Diff($M$)}$-valued function $\mathscr{X}:\SSS
\rightarrow \text{Diff}(M)$.  In a given solution $s$, $X$ is identified with 
the diffeomorphism in the image of the map, 
$X=\mathscr{X}(s)$.  One way to interpret $X$ is as defining
a map from (an open subset of) $\mathbb{R}^d$ into the spacetime manifold $M$,
and hence can be thought of as a choice of coordinate system covering the local
subregion $\Sigma$.\footnote{We assume
for simplicity that the subregion of interest can be covered by a single coordinate
system.  For topologically nontrivial subregions, the fields may consist of a collection of maps
$X_i$, one for each coordinate patch needed to cover the region.
\renewcommand{\baselinestretch}{1} \footnotesize}  The problem of 
defining the subregion $\Sigma$ is  solved by declaring it to be the image under the $X$ 
map of some fiducial subregion $\sigma$ in $\mathbb{R}^d$.
A full solution
to the field equations now consists of specifying the map $X$ as well as the value of the dynamical
fields $\phi(x)$ at each point in spacetime.  
The transformation law for $X$ 
under a diffeomorphism $Y:M\rightarrow M$ is given by the pullback along
$Y^{-1}$,  $\bar{X} = Y^{-1} \circ X$.   

Since $X$ defines a diffeomorphism from $\mathbb{R}^d$ to $M$, it can be used to pull back 
tensor fields  on $M$ to $\mathbb{R}^d$.  We can argue as before that the Lie derivative $L_V$
and exterior derivative $\delta$ satisfy pullback forumlas analogous to equations (\ref{eqn:LphY*a})
and (\ref{eqn:dY*a}),
\begin{align}
L_V X^*\al &= X^*(L_V \al+\lie_{I_V \dar_X} \al) \label{eqn:LphX*a} \\
\delta X^*\al &= X^*(\del \al + \lie_{\dar_X} \al),  \label{eqn:dX*a}
\end{align}
which serve as defining relations for the variational spacetime vector $\dar_X^a$.  
The result of contracting $\dar_X^a$
with a vector field $\hat\xi$ corresponding to a spacetime 
diffeomorphism
 can be deduced by first noting that the pulled back fields $X^*\phi$ are invariant under
 diffeomorphisms, since 
\beq \label{eqn:barX*}
\bar{X}^* Y^*\phi = X^*(Y^{-1})^* Y^*\phi = X^*\phi.  
\eeq
In particular,
the $\SSS$ Lie derivative $L_{\hat\xi}$ must annihilate $X^* \phi$ for any $\xi$, so from 
(\ref{eqn:LphX*a}),
\beq \label{eqn:LxiX*phi}
0=L_{\hat\xi} X^*\phi = X^*(L_{\hat\xi} \phi +\lie_{I_{\hat{\xi}}\dar_X}\phi) = X^*(\lie_\xi \phi+ 
\lie_{I_{\hat\xi} \dar_X} \phi),
\eeq
and hence
\beq \label{eqn:Ixidar}
I_{\hat\xi} \dar_X^a = -\xi^a.
\eeq

We can also derive the transformation law for $\dar_X^a$ under a diffeomorphism from the 
pullback formulas (\ref{eqn:dY*a}) and (\ref{eqn:dX*a}).  On the one hand we have 
\beq
\delta \bar{X}^*\al = \bar{X}^*(\del \al + \lie_{\dar_{\bar{X}}} \al),
\eeq
while on the other hand this can also be computed as 
\beq
\del\bar{X}^*\al = \del X^* (Y^{-1})^* \al = X^*[\del (Y^{-1})^* \al + \lie_{\dar_X} (Y^{-1})^* \al]
= \bar{X}^*(\del \al  + \lie_{\dar_{Y^{-1}}} \al + \lie_{Y^*\dar_X} \al)
\eeq
where the last equality employed identity \ref{id:liexiYi}.  Comparing these expressions
and applying the formula (\ref{eqn:dy-1}) for $\dar_{Y^{-1}}^a$ gives the transformation law
\beq\label{eqn:dbarXa}
\dar_{\bar{X}}^a = Y^*(\dar_X^a-\dar_Y^a).
\eeq

The $X$ fields lead to an easy prescription for forming diffeomorphism-invariant quantities: simply
work with the pulled back fields $X^*\phi$.  These are diffeomorphism-invariant due to equation
(\ref{eqn:barX*}), and consequently the variation $\delta X^*\phi$ is as well.  We can explicitly 
confirm that $\del X^*\phi$ are annihilated by infinitesimal diffeomorphisms $\hat\xi$:
\beq
I_{\hat\xi}\del X^*\phi = I_{\hat\xi}X^*(\del\phi + \lie_{\dar_X}\phi) = X^*(\lie_\xi\phi - \lie_\xi\phi)=0.
\eeq
Note that these relations ensure that $X^*\phi$ and $\delta X^*\phi$ descend to functions on
 the reduced phase space $\PP$, after quotienting $\SSS$ by the degenerate directions of the 
presymplectic form.  
Another combination of one-forms that appears frequently is $\alpha + I_{\hat\dar_X}\alpha$,
and it is easily checked that $I_{\hat\xi}$ annihilates this sum.  Finally, we note that when
no confusion will arise, we will simply denote $\dar_X^a$ by $\dar^a$ to avoid excessive
clutter.  When referring to other diffeomorphisms besides $X$, we will explicitly include the 
subscript, as in $\dar_Y^a$.

\section{Extended phase space} \label{sec:eps}
We now turn to the problem of defining a gauge-invariant symplectic form to associate with 
the local subregion.  In this chapter, the precise meaning of a local subregion is 
the domain of dependence of some spacelike hypersurface $\Sigma$,\footnote{The requirement
that $\Sigma$ be spacelike is necessary in order to interpret the symplectic form constructed
on it as  characterizing a subset of the 
theory's degrees of freedom.  While the construction  would seem to also apply to timelike
hypersurfaces, such a hypersurface has an empty domain of dependence, and so  there 
is no sense in which it  determines the dynamics in some open subset of the manifold.
\renewcommand{\baselinestretch}{1} \footnotesize}
which serves as a Cauchy surface for the subregion.  We further require that $\Sigma$ 
have a boundary $\partial\Sigma$, so that it may be thought of as a subspace of a larger Cauchy 
surface for the full spacetime.  
The standard procedure of \cite{Lee1990a, Wald1993a, Iyer1994a} for constructing 
a symplectic form for a diffeomorphism-invariant field theory
begins with a Lagrangian $L[\phi]$, 
a spacetime $d$-form constructed covariantly from the dynamical fields $\phi$.  Its variation 
takes the form
\beq \label{eqn:dLCH4}
\delta L = E\cdot\delta\phi + d\theta,
\eeq
where $E=0$ are the dynamical field equations, and the exact form $d\theta$, where 
$d$ denotes the spacetime exterior derivative, defines the 
symplectic potential current $(d-1)$-form $\theta\equiv\theta[\phi; \delta\phi]$, which is a one-form
on solution space $\mathcal{S}$.  The $\SSS$-exterior derivative of $\theta$ defines the symplectic
current $(d-1)$-form, $\omega=\delta\theta$, whose integral over $\Sigma$ normally
defines the presymplectic form $\Om_0$ for the phase space.  As a consequence of 
diffeomorphism-invariance, $\Om_0$ contains degenerate directions: it annihilates 
any infinitesimal diffeomorphism generated by vector field $\xi^a$
that vanishes sufficiently quickly near the boundary.  This is succinctly expressed 
for such a vector field by 
$I_{\hat\xi}\Om_0=0$.  The true phase space $\PP$  
is obtained by quotienting out these degenerate directions by mapping all 
diffeomorphism-equivalent solutions to a single point in $\PP$.  $\Omega_0$ then defines a 
nondegenerate symplectic form on $\PP$ through the process of phase space reduction
\cite{Lee1990a}.  

This procedure is deficient for a local subregion  because $\Om_0$
fails to be degenerate for diffeomorphisms that act near the Cauchy surface's 
boundary $\partial\Sigma$.  
If the boundary were at asymptotic infinity, such diffeomorphisms could be disallowed by
imposing boundary conditions on the fields, 
or could otherwise be regarded as true
time evolution with respect to the fixed asymptotic structure, 
in which case degeneracy would 
not be expected \cite{Ashtekar1991}.  For a local subregion, however, neither option is acceptable.   
Imposing a boundary condition on the fields at $\partial\Sigma$ has a nontrivial effect on the 
dynamics \cite{Andrade2015b, Andrade2015a, Andrade2016}, whereas we are interested in a  phase space that locally reproduces the same dynamics
as the theory defined on the full spacetime manifold $M$.  
Furthermore, the diffeomorphisms acting
at $\partial\Sigma$ cannot be regarded as true time evolution generated by nonvanishing 
Hamiltonians, because these diffeomorphisms  are degenerate directions of a presymplectic 
form for the entire manifold $M$.  

Donnelly and Freidel \cite{Donnelly2016F}
proposed a resolution to this issue by extending
the local phase space to 
include the $X$
fields described in section  \ref{sec:edge}.  The minimal prescription for introducing
them into the theory is to simply replace the Lagrangian with its pullback $X^*L$.  Since the 
Lagrangian is a covariant functional of the fields, $X^*L[\phi]=L[X^*\phi]$, so that the 
pulled back Lagrangian
depends only on the redefined fields $X^*\phi$, and is otherwise independent of $X$.  The
variation of this Lagrangian gives
\beq \label{eqn:dLX*phi}
\delta L[X^*\phi] = E[X^*\phi] \cdot \delta X^*\phi + d\theta[X^*\phi; \delta X^*\phi].
\eeq
Thus the redefined fields satisfy the same equations of motion $E[X^*\phi]=0$ as the original
fields, and, due to diffeomorphism invariance, this implies that the original $\phi$ fields must 
satisfy the equations as well.  Additionally,  the Lagrangian had no further dependence
on $X$, which means the $X$ fields do not satisfy any field equations.  
 If $X$ is understood as defining a coordinate system
for the local subregion, the dynamics of the extended $(\phi,X)$ system is simply given by the 
original field equations, expressed in an arbitrary coordinate system determined by $X$.

The symplectic potential current is read off from (\ref{eqn:dLX*phi}),
\beq\label{eqn:th'}
\theta' = \theta[X^*\phi; \delta X^*\phi] = \theta[X^*\phi; X^*(\delta\phi + \lie_{\dx}\phi)]
= X^*(\theta+I_{\hdx}\theta).
\eeq
This object is manifestly invariant with respect to solution-dependent diffeomorphisms, since 
both $X^*\phi$ and $\delta X^*\phi$ are.  In particular, $\theta'$ annihilates any 
infinitesimal diffeomorphism $I_{\hat\xi}$, as a consequence of the fact that $I_{\hat\xi}
\del X^*\phi = 0$ (see equation \ref{eqn:LxiX*phi}).  An equivalent 
expression for $\theta'$ can be obtained by introducing the Noether current 
for a vector field $\xi^a$,
\beq
J_\xi = I_{\hat\xi} \theta-i_\xi L,
\eeq
where $i_\xi$ denotes contraction with the spacetime vector $\xi^a$.
Due to diffeomorphism invariance, 
$J_\xi$ is an exact form when the equations of motion hold \cite{Wald1993a, Iyer1994a}, 
and may be 
written
\beq
J_\xi = dQ_\xi + C_\xi,
\eeq 
where $Q_\xi$ is the Noether charge and 
$C_\xi=0$ are combinations of the field equations that comprise the constraints for the 
theory \cite{Iyer1995b}.  Then $\theta'$ in (\ref{eqn:th'}) may be expressed
on-shell
\beq\label{eqn:th'2}
\theta' = X^*(\theta+i_{\dx} L + d Q_{\dx}).
\eeq

As an aside, note that we can vary the Lagrangian with respect to $(\phi, X)$ instead of 
the redefined fields $(X^*\phi, X)$, and equivalent dynamics arise.  This variation
produces  
\beq \label{eqn:dX*L}
\delta X^*L[\phi] = X^*(\delta L+\lie_{\dx}L) = X^*(E\cdot\del\phi) +d X^*(\theta+i_{\dx}L),
\eeq
where Cartan's magic formula $\lie_\dx = i_\dx d + d i_\dx$ was used, along with the fact that $d$
commutes with pullbacks.
Again, $\phi$ satisfies the same field equation $E[\phi]=0$, and $X$ is subjected to no
dynamical equations.  This variation suggests a  potential current
$\theta''=X^*(\theta
+i_{\dx}L)$, which differs from (\ref{eqn:th'2}) by the exact form $dX^*Q_\dx$.  This 
difference is simply an ambiguity in the definition of the  potential current, since shifting 
it by an exact form does not affect equation (\ref{eqn:dLCH4}) \cite{Jacobson1994b, Iyer1994a}.
  However,
$\theta''$ does not annihilate infinitesimal diffeomorphisms $I_{\hat\xi}$, making $\theta'$ the 
preferred choice.  The degeneracy requirement for the symplectic potential current therefore
gives a prescription to partially fix its ambiguities \cite{Geiller2017}, although additional
ambiguities remain, and are discussed in section \ref{sec:jkm}.

The symplectic potential $\Theta$ 
is now constructed by integrating $\theta'$ over $\Sigma$.  
Since $\theta'$ is defined as a pullback by $X^*$, its integral must be over the pre-image $\sigma$, 
for which  $X(\sigma)=\Sigma$.  This gives
\begin{align}
\Theta &= \int_\sigma \theta[X^*\phi;\delta X^*\phi] \label{eqn:Ths} \\
&= \int_\Sigma(\theta+i_{\dx}L) + \int_{\partial\Sigma} Q_{\dx}\label{eqn:ThS}.
\end{align}
The second line uses the alternative expression (\ref{eqn:th'2}) for 
$\theta'$, and is written as an integral 
of fields defined on the
original Cauchy surface $\Sigma$, without pulling back by $X$.   This makes use of the general
formula 
$\int_\sigma X^*\alpha = \int_{X(\sigma)} \alpha$, and also applies  
Stoke's theorem $\int_\Sigma d\alpha
= \int_{\partial\Sigma}\alpha$ to write the Noether charge as a boundary integral.
Equation (\ref{eqn:ThS}) differs from the symplectic potential for the nonextended phase space,
$\Theta_0 = \int_\Sigma\theta$, by both a boundary term depending on the Noether charge, as 
well as a bulk term coming from the on-shell value of the Lagrangian.  For vacuum
general relativity with no cosmological constant, this extra bulk contribution vanishes, being
proportional to the Ricci scalar \cite{Donnelly2016F}.  
However, when matter is present or the cosmological constant
is nonzero, this extra bulk contribution to $\Theta$ can survive.  As we  discuss below,
this bulk term imbues the symplectic form on the reduced phase space $\PP$ 
with nontrivial cohomology. 

Taking an exterior derivative of $\Th$ yields the symplectic form, $\Omega=\delta\Theta$.  The expression
(\ref{eqn:Ths}) leads straightforwardly to
\beq\label{eqn:Oms}
\Om = \int_\sigma \omega[X^*\phi; \delta X^*\phi, \delta X^*\phi], 
\eeq
where we recall the definition of the symplectic current $\omega=\delta\theta$.  This expression
for $\Omega$ makes it clear that it is invariant with respect to all diffeomorphisms, and  
that infinitesimal diffeomorphisms are degenerate directions, again because $I_{\hat\xi}\delta
X^*\phi = 0$.  
The symplectic form can also be expressed as an integral over $\Sigma$ and its boundary
using the original fields $\phi$, by computing the exterior derivative of (\ref{eqn:ThS}).   
Noting that the integrands implicitly involve a pullback by $X^*$, we find
\beq
\Om = \int_\Sigma(\omega +\lie_\dx \theta+\del i_\dx L+\lie_\dx i_\dx L) 
+\int_{\partial\Sigma} (\delta Q_{\dx} + \lie_\dx Q_\dx)
\eeq
The first term is the symplectic form for 
the nonextended theory, $\Omega_0 = \int_\Sigma \omega$.
The remaining three terms in the bulk $\Sigma$ integral simplify to an exact form on-shell 
 $d(i_\dx \theta +\frac12 i_\dx i_\dx L)$ (see identity \ref{id:liedxth}),  so the final expression
 is
 \beq\label{eqn:OmS}
 \Om = \int_\Sigma\omega + \int_{\partial\Sigma} \left[\delta Q_{\dx}+\lie_\dx Q_\dx + i_\dx\theta
 +\frac12i_\dx i_\dx L\right].
 \eeq 
This expression is related to one obtained in a similar context in \cite{Azeyanagi2009}.

Hence, we arrive at the important result that the symplectic form differs from $\Omega_0$
by terms localized on the boundary $\partial\Sigma$ involving $\dx^a$.  This immediately
implies that $\Omega$ has degenerate directions: any phase space vector field $V$ that 
vanishes on $\delta \phi$ and whose contraction with $\dx^a$ vanishes sufficiently
quickly near $\partial \Sigma$ will annihilate $\Omega$.  In fact, only the values of $\dx^a$
and  $\nabla_b\dx^a$ 
 at $\partial\Sigma$ contribute to (\ref{eqn:OmS}); all other freedom in $\dx^a$ is pure gauge.
To see why these are the only relevant pieces of $\dx^a$ for the symplectic form, 
we can use
the 
explicit expression for the Noether charge given in \cite{Iyer1994a}.  Up to ambiguities which
are discussed in section \ref{sec:jkm}, the Noether charge is given by
\beq\label{eqn:Qxi}
Q_\xi = -\ep_{ab}E\indices{^a^b^c_d}\nabla_{c}\xi^{d} + W_c\xi^c,
\eeq
where $\ep_{ab}$ is the spacetime volume form with all but the first two indices suppressed,
$E^{abcd}=\frac{\delta\mathcal{L}}{\delta R_{abcd}}$ is the variational derivative of the Lagrangian
scalar $\mathcal{L}=-(*L)$ 
with respect to the Riemann tensor, and inherits the index symmetries of the Riemann
tensor, and $W_c[\phi]$ is a tensor with $(d-2)$ covariant, antisymmetric indices suppressed, 
constructed locally from the dynamical fields; its precise form
is not needed in this work.

The last two terms in (\ref{eqn:OmS}) depend only on the value of $\dx^a$ on $\partial \Sigma$,
while the terms involving $Q_\dx$ can depend on 
derivatives of $\dx^a$.  From  (\ref{eqn:Qxi}), $Q_\dx$ 
involves one  derivative of $\dx^a$, and 
(\ref{eqn:OmS}) has terms involving the derivative of $Q_\dx$, so  that up to two 
derivatives of $\dx^a$ could contribute to the symplectic form.  
To see how these derivatives appear, we decompose $\delta Q_\dx$ as
\beq \label{eqn:dQdX}
\del Q_\dx =  Q_{\del (\dx)} + \qo_\dx,
\eeq
where $\qo_\xi=\qo_\xi[\phi;\del\phi]$\footnote{$\qo$ is the archaic Greek letter ``qoppa.''
\renewcommand{\baselinestretch}{1} \footnotesize } 
is a variational one-form depending on a vector $\xi$ (which can be a differential form
on $\SSS$), given by
\beq\label{eqn:qoxi}
\qo_\xi = -\del(\ep_{ab} E\indices{^a^b^c_d})\nabla_c \xi^d-\ep_{ab}E\indices{^a^b^c_d} 
\del\Gamma^d_{ce}
\xi^e + \del W_c \xi^c,
\eeq
and $\del \Gamma^d_{ce}$ is the variation of the Christoffel symbol,
\beq
\del\Gamma^d_{ce} = \frac12g^{df}(\nabla_c\del g_{fe}+\nabla_e\del g_{fc}-\nabla_f \del g_{ce}).
\eeq
This decomposition is useful because $\qo_\dx$ contains only first derivatives of $\dx^a$, while
$Q_{\del\dx} = -\frac12 Q_{[\dx,\dx]}$ involves second derivatives through the derivative of the 
vector field Lie bracket.  

In appendix \ref{app:edge}, 
it is argued that 
the second derivatives of $\dx^a$ in $Q_{\del(\dx)}+\lie_\dx Q_\dx$ cancel out, 
 so
that the boundary contribution in (\ref{eqn:OmS}) 
depends on only $\dx^a$ and $\nabla_b \dx^a$ at $\partial\Sigma$. 
This means that $\Om$ has a large number of degenerate directions, corresponding to all
values of $\dx^a$ on $\Sigma$ that are not fixed by the values of $\dx^a$ and $\nabla_b\dx^a$
at the boundary.  The true phase space $\PP$ is then obtained by quotienting out these 
pure gauge degrees of freedom.  
In doing so, $\Omega$ descends 
to a nondegenerate, closed two-form on the quotient space \cite{Lee1990a}.
However, the symplectic potential 
$\Theta$ does not survive this projection.   It depends nontrivially on the value of $\dx^a$ 
everywhere on $\Sigma$ through the term involving the Lagrangian in (\ref{eqn:ThS}), 
which causes
it to become a multivalued form on the quotient space.  
One way to see its multivaluedness is to note that $i_\dx L$ is a top rank 
form on $\Sigma$, so, by the Poincar\'e
lemma applied to $\Sigma$, it can be expressed as the exterior derivative of a $(d-2)$-form,
\beq \label{eqn:dhX}
i_\dx L\big|_\Sigma = d h_X i_\dx L.
\eeq 
Here, $h_X$ is the homotopy operator that inverts the exterior derivative $d$ on closed forms on
$\Sigma$ \cite{Edelen2005}.  
As the notation suggests, it depends explicitly on the value of the $X$ fields throughout
$\Sigma$, which we recall can be thought of as defining a coordinate system for the subregion.
Since $h_X i_\dx L$ is a 
 spacetime $(d-2)$-form 
 and an $\SSS$ one-form, evaluated 
at $\partial\Sigma$ it may be expressed in terms of $\dx^a$ and $\del\phi$ at $\partial\Sigma$,
which provide a basis for local variational forms.  Hence,
\beq \label{eqn:inthX}
\int_\Sigma i_\dx L = \int_{\partial\Sigma} h_X i_\dx L,
\eeq
and we see that this latter expression depends on $\dx^a$ at $\partial\Sigma$, so therefore
will project to the quotient space.  However, $h_X$ will be a different operator depending on the 
values of the  
$X$ fields on $\Sigma$, and hence this boundary integral will give a different form on the 
reduced phase space for different bulk values of $X$.  This shows that the Lagrangian 
term in $\Theta$
projects  to a multivalued form on the quotient space. 

The failure of $\Theta$ to be single-valued implies that the reduced phase space $\PP$ 
has nontrivial
cohomology.  In particular, the projected 
symplectic form $\Omega$ is not exact, despite being closed.  For a given choice of the value of $\Theta$, the equation 
$\Om = \del\Theta$ still holds locally near a given solution 
in the reduced phase space, but there can be global 
obstructions since $\Theta$ may not return to the same value after tracing out a closed loop
in the solution space.  It would be interesting to investigate the consequences of this nontrivial
topology of the reduced phase space, and in particular whether it has any relation to the 
appearance of central charges in the surface symmetry algebra.  

Finally, note that for vacuum general relativity with no cosmological constant, 
the Lagrangian vanishes on shell, being proportional to the Ricci scalar.  In this special 
case, $\Theta$ is not multivalued and descends to a well-defined one-form on the 
reduced phase space, suggesting that the phase space topology simplifies. However, the 
inclusion of a cosmological constant or the presence of matter anywhere in the local subregion
leads back to the generic case in which $\Theta$ is multivalued.

\section{JKM ambiguities} \label{sec:jkm}
The constructions of the symplectic potential current $\theta$ and Noether charge $Q_\xi$ 
are subject to a number of ambiguities identified by Jacobson, Kang and Myers (JKM) 
\cite{Jacobson1994b, Iyer1994a}.  
These ambiguities correspond to the ability to add an exact form to 
the Lagrangian $L$, the potential current $\theta$, or the Noether charge $Q_\xi$ without 
affecting the dynamics or the defining properties of these forms.  Normally it is required that 
the ambiguous terms be locally constructed from the dynamical fields in a spacetime-covariant
manner.  In the extended phase space, however, there is additional freedom provided by the $X$
fields as well as the surfaces $\Sigma$ and $\partial\Sigma$ to construct forms that would 
otherwise fail to be  covariant.  The freedom provided by the $X$ fields is 
considerable, given that they can be used to construct homotopy operators as in (\ref{eqn:dhX})
and (\ref{eqn:inthX}) that mix the local dynamical fields $\phi$ at different spacetime points.  
For this reason, we refrain from using the $X$ fields in such an explicit manner to construct
ambiguity terms.  However, we allow for ambiguity terms that are constructed using the 
structures provided by $\Sigma$ and $\partial \Sigma$, such as their induced metrics and extrinsic
curvatures.  This allows for a wider class of Noether charges, including those that appear in 
holographic entropy functionals and the second law of black hole mechanics 
for higher curvature  theories \cite{Dong2014, Camps2013,
Miao2014, Wall2015}.

A simple example of which types of objects are permitted in constructing the ambiguity terms is 
provided by the unit normal $u_a$  to $\Sigma$ versus the lapse function $N$.  
Interpreting $X^\mu$ as a coordinate system for the local subregion, we can take $\Sigma$ to 
lie at $X^0=0$.  Then the lapse and unit normal are related by
\beq
u_a = - N \nabla_a X^0.
\eeq
The form $\nabla_a X^0$ depends explicitly on the $X$ field, and hence is not allowed in our
constructions.  However, the unit normal $u_a$ can be constructed using only the surface $\Sigma$
and the metric, and hence is independent of the $X$ fields.  This then implies that $N$ also depends
on the $X$ fields, and so the lapse function cannot explicitly be used in constructing ambiguity terms.

\subsection{$L$ ambiguity}
The first ambiguity corresponds to adding an exact form $d\alpha$ to the Lagrangian.   This 
does not affect the equations of motion; however, its variation now contributes to $\theta$.  
The following changes occur from adding this term to the Lagrangian:
\begin{subequations}
\begin{align}
L&\rightarrow L + d\alpha \\
\theta&\rightarrow \theta+\del\alpha\\
J_\xi&\rightarrow J_\xi + d i_\xi \alpha \\
Q_\xi&\rightarrow Q_\xi + i_\xi \alpha. \label{eqn:QLambig}
\end{align} 
\end{subequations}
Note that since $\theta$ changes by an $\SSS$-exact form, the symplectic current $\omega$
is unaffected.  Incorporating these changes into the definition of the symplectic potential 
 (\ref{eqn:ThS}) changes $\Theta$ by
 \beq
 \Th \rightarrow \Th + \int_\Sigma(\del\alpha+i_\dx d\alpha)+\int_{\partial\Sigma} i_\dx \alpha
 = \Th + \del\int_\Sigma\alpha.
 \eeq
 We point out that the new term annihilates infinitesimal
 diffeomorphisms $I_{\hat \xi}$, so that $\Th$ remains fully diffeomorphism-invariant.
 Since $\Theta$ changes by an $\SSS$-exact form, the symplectic form $\Omega =\del\Th$ 
receives no change from this type of ambiguity, which can also be checked by tracking
the changes of all quantities in (\ref{eqn:OmS}).  
Given that only $\Omega$, and not $\Theta$, is needed in the construction of the phase space,
this ambiguity in $L$ has no effect on the phase space.  However, it
has some relevance to the surface symmetry algebra discussed in section \ref{sec:ssa}.  
The generators of this algebra are given by the Noether charge, and for surface symmetries that 
move $\partial \Sigma$ (the ``surface translations''), this ambiguity would appear to 
have an effect.  However, as discussed in subsection \ref{sec:trans}, once the appropriate
boundary terms are included in the generators, 
the result is independent of this ambiguity.  The 
 form of the generator does motivate a natural prescription for fixing the ambiguity 
 such that the Lagrangian has 
a well-defined variational principle, so that it is completely stationary on-shell, as opposed to 
being stationary up to boundary contributions.

\subsection{$\theta$ ambiguity} \label{sec:thambig}
The second ambiguity comes from the freedom to add an exact form $d\beta$ to $\theta$,
since doing so does not affect its defining equation (\ref{eqn:dLCH4}).  Here, $\beta\equiv\beta[\phi;
\del\phi]$ is a spacetime $(d-2)$-form and a one-form on $\SSS$.  
The changes that arise from this addition are
\begin{subequations}
\begin{align}
\theta&\rightarrow \theta + d\beta \label{eqn:betaCH4}\\
\omega&\rightarrow \omega + d \del \beta\\
J_\xi &\rightarrow J_\xi + d I_{\hat\xi}\beta \\
Q_\xi & \rightarrow Q_\xi + I_{\hat\xi}\beta. \label{eqn:Q+Ixib}
\end{align}
\end{subequations}
Under these transformations, the symplectic potential (\ref{eqn:ThS}) changes to
\beq
\Th\rightarrow \Th +\int_{\partial\Sigma} (\beta + I_{\hdx}\beta).
\eeq

Hence, the symplectic potential is modified by an arbitrary boundary term $\beta$, 
accompanied by $I_{\hdx}\beta$ that ensures that $\Th$ retains degenerate directions
along linearized diffeomorphisms.  Unlike the $L$ ambiguity, this modification is not $\SSS$-exact,
and changes the boundary terms in the symplectic form,
\beq
\Om\rightarrow\Om +\int_{\partial\Sigma} (\del\beta+\del I_{\hdx}\beta+\lie_{\dx} \beta
+\lie_\dx I_{\hdx}\beta).
\eeq
Because $\beta$ can in principle involve arbitrarily many derivatives of $\del\phi$, its presence
can cause $\Omega$ to depend on second or higher derivatives of $\dx^a$ on the boundary.  
This affects which parts of $\dx^a$ correspond to degenerate directions, and will lead to 
different numbers of boundary degrees of freedom in the reduced phase space.  As discussed
in section \ref{sec:ssa}, this ambiguity can also be used to reduce the surface symmetry 
algebra to a subalgebra.  

Give that $\beta$  contributes to $\Th$ and $\Om$ only at the boundary, it can involve tensors 
associated with the surface $\partial\Sigma$ that do not 
correspond
to spacetime-covariant tensors, such as the extrinsic curvature.  This allows the Dong entropy
\cite{Dong2014, Camps2013, Miao2014}, 
which differs from the Wald entropy \cite{Wald1993a, Iyer1994a} by extrinsic curvature
terms, to 
be viewed as a Noether charge with a specific choice of ambiguity terms.  This is the
point of view 
advocated for in \cite{Wall2015}, where the ambiguity was resolved by requiring that the 
entropy functional derived from the resultant Noether charge satisfy a linearized second law.  
In general, fixing the ambiguity requires some additional input, motivated by the particular
application at hand.

\subsection{$Q_\xi$ ambiguity}
The final ambiguity is the ability to shift $Q_\xi$ by a closed form $\gamma$, with $d\gamma=0$.
Since $Q_\xi$ depends linearly on $\xi^a$ and its derivatives, 
$\gamma$ should be chosen 
to also satisfy this requirement.  If $\gamma$ is identically closed for all $\xi^a$, it
then follows that it must be exact, $\gamma = d\nu$ \cite{Wald1990a}.  
Its integral over the closed surface
$\partial\Sigma$ then vanishes, so that it has no effect on $\Theta$ or $\Omega$.

\section{Surface symmetry algebra}\label{sec:ssa}
The extended phase space constructed in section \ref{sec:eps} contains new edge mode
fields $\dx^a$  on the boundary of the Cauchy surface for the local subregion, 
whose presence is required 
in order to
have a gauge-invariant symplectic form.  Associated with the edge modes are a new
class of transformations that leave the symplectic form and the equations of motion 
invariant.  These new transformations
comprise the surface symmetry algebra.  This algebra 
plays an important role  in the quantum theory 
when describing the edge mode contribution to the entanglement entropy, thus it is necessary
to identify the algebra and its canonical generators.  

As discussed in \cite{Donnelly2016F}, the surface symmetries coincide with diffeomorphisms 
in the preimage space, $Z: \mathbb{R}^d\rightarrow \mathbb{R}^d$, where $\mathbb{R}^d
\supset X^{-1}(M)$. 
These leave the spacetime fields $\phi$ unchanged, but transform
the $X$ fields by $X\rightarrow X\circ Z$.  This also transforms the pulled back fields 
$X^*\phi\rightarrow Z^* X^*\phi$, and due to the diffeomorphism invariance of the field
equations, the pulled back fields still define solutions.  These transformations therefore
comprise a set of symmetries for the dynamics in the local subregion.  
Infinitesimally, these transformations are generated
by vector fields $w^a$ on $\mathbb{R}^d$.  Analogous to 
vector fields defined on $M$, $w^a$ defines
a vector $\hat{w}$ on $\SSS$, whose action on the pulled back fields $X^*\phi$ is given by the 
Lie derivative,
\beq
L_{\hat{w}} X^*\phi = \lie_w X^*\phi = X^*\lie_{(X^{-1})^*w}\phi,
\eeq
while its action on $\phi$ is trivial, $L_{\hat w}\phi = 0$.  On the other hand, we may apply the 
pullback formula (\ref{eqn:LphX*a}) to this equation to derive
\begin{align}
X^*\lie_W\phi = X^* I_{\hat w} \lie_\dx \phi,
\end{align}
where $W^a = ({X^{-1}})^*w^a$.  The contractions of the vector $\hat{w}$ with the basic $\SSS$ 
one-forms are therefore
\beq
I_{\hat w} \dar^a = W^a,\qquad I_{\hat w}\del\phi=0.
\eeq
We also will assume that $w^a$ is independent of the solution, so that $\del w^a=0$.  
Writing this as $0=\del X^* W^a$, and applying the pullback formula (\ref{eqn:LphX*a}), one 
finds
\beq\label{eqn:dWa}
\del W^a = -\lie_{\dx} W^a.
\eeq 

In order for the transformation to be a symmetry of the phase space, it must generate a 
Hamiltonian flow.  This means that $I_{\hat w} \Omega$ is exact, and determines the Hamiltonian
$H_{\hat w}$
for the flow via $\del H_{\hat w} = - I_{\hat w}\Om$.  The contraction with the symplectic form
can be computed straightforwardly from (\ref{eqn:OmS}) by first using the decomposition 
(\ref{eqn:dQdX}) for $\del Q_\dx$.  Then
\begin{align}
I_{\hat w}\Om &= \int_{\partial\Sigma}(-\qo_W-Q_{[W,\dx]} -\lie_\dx Q_W +\lie_W Q_\dx+i_W\theta
+i_W i_\dx L) \\
&=-\del \int_{\partial\Sigma}Q_W + \int_{\partial\Sigma} i_W(\theta+I_{\hdx}\theta). 
\label{eqn:diQW}
\end{align}
The first three terms of the first line combine into the first term in the second line, using  formula
(\ref{eqn:dWa}) for $\del W^a$, formula (\ref{eqn:dQdX}) for $\del Q_W$, and recalling that 
the integral involves an implicit pullback by $X^*$, so that $\del \int_{\partial\Sigma} Q_W
 = \int_{\partial \Sigma} (\del Q_W +\lie_\dx Q_W)$.  
 
It is immediately apparent that if the second integral in (\ref{eqn:diQW}) vanishes, the flow 
is Hamiltonian.  This  occurs if $W^a$ is tangent to $\partial \Sigma$ or vanishing at 
$\partial\Sigma$, and hence defines a mapping of the surface into itself.  If $W^a$ is tangential,
it generates a diffeomorphism $\partial\Sigma$, while vector fields that vanish on $\partial \Sigma$
generate transformations of the normal bundle to the surface while holding all points on the 
surface fixed.  These transformations were respectively called surface diffeomorphisms and 
surface boosts in \cite{Donnelly2016F}.  The remaining transformations consist of the 
surface translations, where 
$W^a$ has components normal to the surface, and the second integral in (\ref{eqn:diQW})
does not vanish.  In general, this term does not give a Hamiltonian flow, except when the 
fields satisfy certain boundary conditions.  We will briefly discuss the surface translations
in subsection \ref{sec:trans}, where we show that they can give rise to central charges in the 
surface symmetry algebra.

Returning to the surface-preserving transformations, we find that the Hamiltonian is 
given by the Noether charge integrated over the boundary,
\beq
H_{\hat w} = \int_{\partial \Sigma} Q_W.
\eeq
The surface symmetry algebra is generated through the Poisson bracket of the Hamiltonians
for all possible surface-preserving vectors.  The Poisson bracket is given by
\beq
\{ H_{\hat w}, H_{\hat v}\} = I_{\hat w} I_{\hat v}\Omega = -I_{\hat w} \del\int_{\partial \Sigma} Q_V
=-I_{\hat w}\int_{\partial \Sigma}(\qo_V+Q_{\del V} + \lie_\dx Q_V) = \int_{\partial\Sigma}Q_{[W,V]},
\eeq
where the last equality uses equation (\ref{eqn:dWa}) applied to $\del V^a$ and that 
$\int_{\partial\Sigma}\lie_WQ_V = \int_{\partial\Sigma} i_W dQ_V$
vanishes when integrated over the surface since $W^a$ is parallel to $\partial\Sigma$.
This shows that the algebra generated by the Poisson bracket is compatible with the 
Lie algebra of surface preserving vector fields,
\beq
\{H_{\hat w}, H_{\hat v}\} = H_{{[w,v]}\,\hat{}},
\eeq
without the appearance of any central charges, i.e.\ the map $w^a\mapsto H_{\hat w}$ 
is a Lie algebra homomorphism.  
Note that the algebra of surface-preserving vector fields is much larger than the 
surface symmetry algebra.  This is because the  generators of surface symmetries
depend only on the values of the vector field and its derivative at $\partial \Sigma$.  Vector fields
that die off sufficiently quickly near $\partial\Sigma$ correspond to vanishing Hamiltonians.  
The transformations they induce on $\SSS$ are pure gauge, and they drop out after passing 
to the reduced phase space.  

To identify the surface symmetry algebra, it is useful to first describe the larger algebra of 
surface-preserving diffeomorphisms, which contains the surface symmetries as a subalgebra. 
It takes the form of a semidirect product, $\text{Diff}(\partial\Sigma)\ltimes\ddiff$
where 
$\text{Diff}(\partial \Sigma)$ is the diffeomorphism group of $\partial\Sigma$, and $\ddiff$ is 
the normal subgroup of diffeomorphisms that fix all points on $\partial\Sigma$.\footnote{``$\text{Dir}$'' stands for ``Dirichlet,'' since these are the diffeomorphisms that would be 
consistent with fixed, Dirichlet boundary conditions at $\partial\Sigma$. 
\renewcommand{\baselinestretch}{1} \footnotesize }  $\ddiff$ is 
generated by vector fields $W^a$ that vanish on $\partial\Sigma$, and it is a normal subgroup 
because the vanishing property is preserved
under commutation with all surface-preserving vector fields:
\beq
[W,V]^a\big|_{\partial\Sigma} = (W^b\partial_bV^a-V^b\partial_bW^a)\big|_{\partial\Sigma}=0,
\eeq
where the first term vanishes since $W^b$ vanishes at $\partial\Sigma$, and the second term
vanishes because $V^b$ is parallel to $\partial\Sigma$, and $W^a$ is zero everywhere along 
the surface.  A general surface preserving vector field can then be expressed as 
\beq
W^a = W^a_\parallel+ W^a_0,
\eeq
where $W^a_0$ vanishes on $\partial\Sigma$ and $W^a_\parallel$ is tangent to $\partial \Sigma$.  
Note that this decomposition is not canonical; away from $\partial\Sigma$ there is some freedom
in specifying which components of the vector field correspond to the tangential direction.  
However, given any such
choice, it is clear that 
if $W^a_\parallel$ is nonvanishing at $\partial\Sigma$, then it will be nonzero in a 
neighborhood of $\partial\Sigma$, and hence the parallel vector fields  
act nontrivially on the $V_0^a$ component of 
 other vector fields.  Finally, the commutator of two purely parallel vector fields $[W_\parallel, 
 V_\parallel]$ will remain purely parallel, since they are tangent to an integral submanifold.  
The map $W^a\mapsto W^a_\parallel$ is therefore a homomorphism from the surface-preserving
diffeomorphisms onto $\text{Diff}(\partial\Sigma)$, with kernel  $\ddiff$.  This
establishes that the group of surface-preserving diffeomorphisms is $\text{Diff}(\partial\Sigma)
\ltimes \ddiff$.

The surface symmetry algebra is represented as a subalgebra of $\text{Diff}(\partial\Sigma)
\ltimes \ddiff$.  The Hamiltonian for a surface-preserving vector field is determined by the 
Noether charge $Q_W$, which depends only on the value of $W^a$ and its first derivative at 
$\partial\Sigma$.  Hamiltonians for vector fields that are nonvanishing at $\partial\Sigma$ provide
a faithful representation of the $\text{Diff}(\partial\Sigma)$ algebra; however, the vanishing
vector fields only represent a subalgebra of $\ddiff$. To determine it, note that only the first
derivative of $W^a$ contributes to the Noether charge, and its tangential derivative vanishes.  
Letting $x^i$, $i=0,1$, 
represent coordinates in the normal directions that vanish on $\partial\Sigma$, 
the components of the vector field may be expressed $W^\mu = x^i W\indices{_i^\mu}
+\mathcal{O}(x^2)$, $\mu=0,\ldots,d-1$, 
and the $\mathcal{O}(x^2)$ 
terms are determined by the second derivatives, which do not contribute
to the Noether charge.  Then
the  commutator of two vectors is
\beq
[W,V]^\mu = x^i(W\indices{_i^j}V\indices{_j^\mu}-V\indices{_i^j}W\indices{_j^\mu})+
\mathcal{O}(x^2),
\eeq
which is seen to be determined by the matrix commutator of $W\indices{_i^\mu}$ and 
$V\indices{_j^\nu}$, by allowing the $i,j$ indices to run over $0,\ldots, d-1$, setting all entries
with $i,j>1$ to zero.  

This algebra gives a copy of $SL(2,\mathbb{R})\ltimes \mathbb{R}^{2\cdot(d-2)}$ for each
point on $\partial \Sigma$.  The abelian normal subgroup $\mathbb{R}^{2\cdot(d-2)}$ is 
generated by vectors for which the $\mu$ index in $W\indices{_i^\mu}$ is tangential, i.e.\
$W\indices{_i^j}\equiv W\indices{_i^\mu}\nabla_\mu x^j=0$.
These vectors represent shearing transformations of the normal bundle: they 
generate flows that vanish on $\partial\Sigma$, and are parallel to $\partial \Sigma$ 
away from the surface.   By specifying a normal direction,
one obtains a homomorphism sending $W\indices{_i^\mu}$ 
to its purely normal part, $W\indices{_i^j}$.  The fact that only the traceless part of $\nabla_aW^b$
contributes to the Noether charge, which follows from the antisymmetry of $E^{abcd}$ from 
equation (\ref{eqn:Qxi}) in $c$ and $d$, translates to the requirement that 
$W\indices{_i^j}$ be traceless when $W^a$ vanishes on $\partial\Sigma$.  This means that 
the $2\times 2$ matrices $W\indices{_i^j}$ generate an $SL(2,\mathbb{R})$ algebra.  The 
generators $V\indices{_i^\mu}$  of $\mathbb{R}^{2\cdot(d-2)}$ transform as a collection of 
$(d-2)$ vectors 
under the $SL(2,\mathbb{R})$ algebra, verifying the semidirect product structure 
$SL(2,\mathbb{R})\ltimes \mathbb{R}^{2\cdot(d-2)}$ for the vector fields vanishing at 
$\partial\Sigma$.  Under diffeomorphisms of $\partial\Sigma$, $V\indices{_i^\mu}$ transforms
as a pair of vectors; hence,
the full surface symmetry algebra is  $\text{Diff}(\partial\Sigma)
\ltimes\left(SL(2,\mathbb{R})\ltimes \mathbb{R}^{2\cdot(d-2)}\right)^{\partial\Sigma}$.

The extra factor of $\mathbb{R}^{2\cdot(d-2)}$ is a novel feature of this analysis, appearing 
for generic higher curvature theories, but not for general relativity \cite{Donnelly2016F}.  
Its presence or absence 
is explained by the particular structure of $E^{abcd}$, the variation of the Lagrangian scalar with
respect to $R_{abcd}$.  When $E^{abcd}$ is determined by its trace, i.e., equal to 
$\frac{E}{d(d-1)} (g^{ac}g^{bd}-g^{ad}g^{bc})$ with $E$ a scalar, the $\mathbb{R}^{2\cdot(d-2)}$ 
transformations are pure gauge.  The Noether charge for a vector field vanishing
at the surface evaluates to\footnote{The binormal is defined to be $n_{ab} = 2u_{[a}n_{b]}$ where
$u_a$ is the timelike unit normal and $n_a$ is the inward-pointing
spacelike unit normal.  The spacetime
volume form at $\partial \Sigma$ is then $\ep_{ab}\big|_{\partial\Sigma} = -n_{ab}\wedge\mu$.
\renewcommand{\baselinestretch}{1} \footnotesize} 
\beq
Q_W\big|_{\partial\Sigma} =\mu\, n_{ab}E\indices{^a^b^c_d}\nabla_cW^d
= \mu \frac{E}{d(d-1)} n\indices{^c_d}\nabla_c W^d,
\eeq
where $\mu$ is the volume form on $\partial\Sigma$ and $n_{ab}$ is the binormal;
 $n\indices{^c_d}$ projects out the tangential component in $\nabla_cW^d$, leaving only
the $SL(2,\mathbb{R})$ transformations as physical symmetries.  A particular class of theories
in which this occurs are $f(R)$ theories (which include general relativity), 
where the Lagrangian is a function of the Ricci scalar,
and $E^{abcd} = \frac12f'(R)(g^{ac}g^{bd}-g^{ad}g^{bc})$.  In more general theories, however,
$n_{ab}E\indices{^a^b^c_d}$ will have a tangential component on the $d$ index, and the algebra
enlarges to include the $\mathbb{R}^{2\cdot(d-2)}$ tranformations.

Curiously, there always exists a choice of ambiguity terms, discussed in subsection 
\ref{sec:thambig},
that eliminates the $\mathbb{R}^{2\cdot(d-2)}$ symmetries.  Namely, 
the symplectic potential current $\theta$ can be modified 
as in equation (\ref{eqn:betaCH4}), with $\beta$ 
chosen to be 
\beq \label{eqn:bmod}
\beta = \ep_{ab}E^{abed}s\indices{_e^c} \del g_{cd},
\eeq
and $s\indices{_e^c} = -u_e u^c + n_e n^c$ is the projector onto the normal bundle
of $\partial\Sigma$.  Note that the explicit use of  normal vectors to $\partial \Sigma$
makes this $\beta$ not spacetime-covariant.
This is nevertheless in line with the broader set of allowed ambiguity terms discussed above.  
From equation (\ref{eqn:Q+Ixib}), this term changes 
the Noether charge of a vector vanishing at $\partial\Sigma$ to 
\beq
Q_W\big|_{\partial\Sigma}=\mu \,n_{ab}\left(E\indices{^a^b^c_d}-E\indices{^a^b^e_d}s\indices{_e^c}
-E\indices{^a^b^e^c}s\indices{_e_d}  \right)\nabla_c W^d.
\eeq
The additional terms involving $s\indices{_e^c}$ drop out when contracted with the normal 
component on the $d$ index of $\nabla_c W^d$; however, on the tangential component the
additional terms
cancel against the first term.  This choice of ambiguity thus reduces the 
surface symmetry algebra to coincide with the algebra for general relativity, 
$\text{Diff}(\partial\Sigma)\ltimes SL(2,\mathbb{R})^{\partial\Sigma}$.

Whether or not to use this choice of $\beta$ depends on the application at hand, and it is 
unclear at the moment how exactly $\beta$ should be fixed when trying to characterize the 
edge mode contribution to the entanglement entropy of a subregion.  The above choice is 
natural in the sense that it gives the same surface symmetry algebra for any 
diffeomorphism-invariant theory.  This would mean that the surface symmetry algebra is 
determined by the gauge group of the theory, while the Hamiltonians for the symmetry generators 
change depending on the specific dynamical theory under consideration.  Note also that there
are additional ambiguity terms that could be added, some of which enlarge the symmetry algebra
by introducing dependence on higher derivatives of the vector field.  Determining how to fix 
the ambiguity remains an important open problem for the extended phase space program.

\subsection{Surface translations} \label{sec:trans}
While the surface-preserving transformations are present for generic surfaces, in situations 
where the fields satisfy certain boundary conditions at $\partial\Sigma$, the surface-symmetry
algebra can enhance to include surface translations.  These are generated by vector fields
that contain a normal component to $\partial\Sigma$ on the surface.  For such a vector field,
the second integral in (\ref{eqn:diQW}) does not vanish, so for this transformation to be 
Hamiltonian, this integral must be an exact $\SSS$ form.  To understand when this can
occur, it is useful to first rewrite the integral in terms of pulled back fields on $\partial\sigma$,
the preimage of $\partial\Sigma$ under the $X$ map:
\beq
\int_{\partial\Sigma} i_W(\theta+I_{\hdx}\theta) = 
\int_{\partial\sigma}X^* i_W\theta[\phi;\del\phi +\lie_\dx\phi] 
=\int_{\partial\sigma} 
i_w\theta[X^*\phi; \del X^*\phi].
\eeq
Since $\delta w^a=0$, it is clear from this last expression that the flow will be Hamiltonian 
only if at the boundary, $\theta$ is exact when contracted with $w^a$,
\beq \label{eqn:iwthds}
i_w\theta[X^*\phi;\del X^*\phi]\big|_{\partial\sigma} = i_w\delta X^*B,
\eeq
where $B[\phi]$ is some functional of the fields, possibly involving structures
defined only at $\partial\Sigma$ such as the extrinsic curvature.  When this condition is 
satisfied, the second integral in (\ref{eqn:diQW}) simply becomes $\del\int_{\partial\Sigma} i_W B$,
and so the full Hamiltonian for an arbitrary vector field $w^a$ is
\beq \label{eqn:Hhw}
H_{\hat w}=\int_{\partial\Sigma} \left(Q_W-i_WB\right).
\eeq

Next we compute the algebra of the surface symmetry generators under the Poisson bracket.  
It is worth noting first that by contracting equation (\ref{eqn:iwthds}) with $I_{\hat v}$, 
we find that the $B$ functional satisfies
\beq
i_W \lie_V B\big|_{\partial\Sigma} = i_W I_{\hat V}\theta = i_W(dQ_V+i_VL).
\eeq
With this, the Poisson bracket is given by
\begin{align}
\{H_{\hat w}, H_{\hat v} \} &= -I_{\hat w} \del\int_{\partial\Sigma} \left(Q_V-i_V B\right)   \nonumber \\
&= \int_{\partial\Sigma} \left(-I_{\hat w}\del Q_V - \lie_W Q_V+I_{\hat w} i_{\del V} B + \lie_W
i_V B  \right) \nonumber \\
&=\int_{\partial\Sigma}\left( Q_{[W,V]}-i_{[W,V]}B\right) + \int_{\partial\Sigma}i_W\left(-dQ_V
+\lie_V B - i_V dB\right) \nonumber \\
&= H_{[w,v]\,\hat{}} + \int_{\partial\Sigma} i_W i_V(L-dB). \label{eqn:Hwvh}
\end{align}
Hence, the commutator algebra of the vector fields $w^a$ is represented by the algebra
provided by the Poisson bracket, except when both vector fields have normal components 
at the surface, in which case the second term in (\ref{eqn:Hwvh}) gives a modification.
In fact, the quantities
\beq\label{eqn:Kwhvh}
K[\hat w, \hat v]\equiv \int_{\partial\Sigma} i_W i_V (L-dB)
\eeq
provide a central extension of the algebra, which is verified by showing that 
they are locally constant on the phase space, and hence commute with all generators.  
The exterior derivative is
\begin{align}
\del K[\hat w, \hat v] &= \int_{\partial\Sigma} \big[\del i_W i_V(L-dB) + \lie_\dx i_W i_V (L-dB)\big] 
\nonumber\\
&= \int_{\partial\Sigma} i_W i_V (\del L-d\del B).
\end{align}
On shell, we have $\del L =  d\theta$, and from (\ref{eqn:iwthds}) we can argue
that the replacement $ i_W i_Vd\del B\rightarrow i_W i_Vd\theta$ is valid at $\partial\Sigma$.  
Hence, the above variation vanishes, and $K[\hat w, \hat v]$ indeed defines a central extension
of the algebra.  

The modification that $B$ makes to the symmetry generators takes the same form as a Noether
charge ambiguity arising from changing the Lagrangian $L\rightarrow L+d\alpha$, with $\alpha=-B$.
Using the 
modified Lagrangian
$L-dB$, the potential current changes to $\theta-\del B$.  The boundary condition
(\ref{eqn:iwthds}) then implies that the terms involving $\theta$ in (\ref{eqn:diQW}) vanish.  
The symmetry generators are simply given by the integrated Noether charge, which is modified to
modified to $Q_W\rightarrow Q_W-i_W B$ by the ambiguity.  Hence, the generators $H_{\hat w}$ 
are the same as in (\ref{eqn:Hhw}), and their Poisson brackets  still involve the central 
charges $K[\hat w, \hat v]$. Finally, note that the constancy of the central charges 
requires the variation of the modified Lagrangian $L-dB$ be zero
when evaluated on 
$\partial\Sigma$.  Requiring that variations of the Lagrangian have no boundary term on
shell generally determines the boundary conditions for the theory.  The same is true
here: a choice of $B$ satisfying (\ref{eqn:iwthds}) can generally only be found if the fields 
obey certain boundary conditions, and different boundary conditions lead to different choices
for $B$.  

The surface translations 
can be parameterized
by normal vector fields $W^i$ defined on $\partial\Sigma$.  Assuming 
$\partial_i W^j=0$ in some coordinate system, where $i,j$ are normal indices, 
we can work out their commutation relations
with generators of the rest of the algebra:
\begin{align}
[W^i, V^j] &=0 \label{eqn:WV}\\
[W^i, x^j V\indices{_j^k}] &= W^i V\indices{_i^k} \\
[W^i, V^A] &= -V^A\partial_A W^i\\
[W^i, x^j V\indices{_j^A}] &= W^i V\indices{_i^A}- x^j V\indices{_j^A}\partial_A W^i, 
\label{eqn:WxV}
\end{align}
where $A$ denotes a tangential index.  The first relation shows that the new generators commute
among themselves (although the corresponding Poisson bracket is equal to the central charge
$K[\hat w, \hat v]$), while the second and third show that $W^i$ transforms as a vector under
$SL(2,\mathbb{R})$ and as a scalar under $\text{Diff}(\partial \Sigma)$.  If the Noether charge
ambiguity is chosen as in equation (\ref{eqn:bmod}) 
so that the normal shearing generators $x^j V\indices{_j^A}$  
drop out of the algebra, the resulting surface symmetry algebra is $\text{Diff}(\partial\Sigma)\ltimes
\left(SL(2,\mathbb{R})\ltimes \mathbb{R}^2\right)^{\partial\Sigma}$.
However, if the normal shearing transformations are retained, equation (\ref{eqn:WxV}) shows
that the surface translations are no longer a normal subgroup, since the commutator gives 
rise to generators of $\text{Diff}(\partial\Sigma)$ and $SL(2,\mathbb{R})^{\partial\Sigma}$.  
In this case, the full surface symmetry would be algebra is simple.  However, if one checks
the Jacobi identity between two normal shears and a surface translation, one sees
that it is violated.  Hence, the normal shears are not compatible with including the surface
translations. 

The above analysis was carried out assuming that all normal vectors generate a surface 
symmetry.  In practice, equation (\ref{eqn:iwthds}) may only be obeyed for some 
specifically chosen normal vectors \cite{Brown1986a}.  The resulting 
algebra will then be a subalgebra of the generic case considered in this section.

\section{Discussion} \label{sec:disc}

Building on the results of \cite{Donnelly2016F}, this chapter has described a general procedure 
for constructing the extended phase space in a 
diffeo\-morphism-invariant theory for a local subregion.  
The integral of the symplectic current for the unextended theory fails to be degenerate for 
diffeomorphisms that act at the boundary, and this necessitates the introduction of new fields,
$X$, to ensure degeneracy.  These fields can be thought of as defining a coordinate system
for the local subregion, and the extended solution space consists of fields satisfying 
the equations of motion in all possible coordinate systems parameterized by $X$.  
While the $X$ fields do not satisfy dynamical equations themselves, it was shown in 
section \ref{sec:eps} 
 that their variations contribute to the symplectic form through the boundary
integral  in equation (\ref{eqn:OmS}).  

There are a few novel features of the extended phase space for arbitrary diffeomorphism-invariant
theories that do not arise in vacuum
general relativity with zero cosmological constant.  First, in any
theory whose Lagrangian does not vanish on-shell, the symplectic potential $\Theta$ is not
a single-valued one form on the reduced phase space $\PP$.  This is due to the bulk integral
of the Lagrangian that appears in  equation (\ref{eqn:ThS}), along with the fact that variations 
for which $\dx^a$ has support only away from the boundary $\partial\Sigma$ are 
degenerate directions of the extended symplectic form, (\ref{eqn:OmS}).  
Because of this, $\Om$ fails to be exact, despite satisfying $\del \Om=0$.  Investigating the 
consequences of this nontrivial cohomology for $\PP$ remains an interesting topic for 
future work.  

Another new result comes from the form of the surface symmetry algebra.  
As in general relativity, any phase space transformation generated by $\hat w$ 
for which $W^a\equiv I_{\hat w} \dx^a$ is tangential  at $\partial \Sigma$
is Hamiltonian. These generate the group $\text{Diff}(\partial\Sigma)\ltimes \ddiff$ of 
surface-preserving diffeomorphisms, but only a subgroup is represented on the phase space.  
This subgroup was found in section \ref{sec:ssa} 
to be $\text{Diff}(\partial\Sigma)\ltimes\left(SL(2,\mathbb{R})\ltimes
\mathbb{R}^{2\cdot(d-2)} \right)^{\partial\Sigma}$, which is larger than the surface symmetry
group $\text{Diff}(\partial\Sigma)\ltimes SL(2,\mathbb{R})^{\partial\Sigma}$ found in
\cite{Donnelly2016F} for general relativity.  The additional abelian factor $\mathbb{R}^{2\cdot(d-2)}$
arises generically; however, 
it is not present in 
$f(R)$ theories, in which the tensor $E^{abcd}$ is constructed solely
from the metric and scalars.  We also noted that for any theory, there exists a choice
(\ref{eqn:bmod})
of ambiguity terms that can be added to $\theta$, with the effect of eliminating the 
$\mathbb{R}^{2\cdot(d-2)}$ factor of the surface symmetry algebra.  

The inclusion of  surface 
translations into the surface symmetry algebra was discussed in section \ref{sec:trans}.  
This requires the existence of a $(d-1)$-form $B$ satisfying the relation
(\ref{eqn:iwthds}) for at least some vector fields that are normal to the boundary.  
If such a form can be found, the surface translations are generated by the Hamiltonians
(\ref{eqn:Hhw}).
Interestingly, the Poisson brackets of these Hamiltonians acquire central charges
given by (\ref{eqn:Kwhvh}), which depend on the on-shell value of the modified Lagrangian
$L-dB$ at $\partial\Sigma$.  
Such central charges are a common occurrence in surface symmetry algebras that include
surface translations \cite{Brown1986a, Carlip1999, Silva2002, Barnich2002, 
Compere2007,Carlip2011, Freidel2017}.
In general, the existence of $B$ requires that the fields satisfy boundary conditions at 
$\partial\Sigma$.  An important topic for future work would be to classify which boundary
conditions the fields must satisfy in order for $B$ to exist.  For example, with 
Dirichlet boundary conditions where
the field values are specified at $\partial\Sigma$, $B$ is given by the Gibbons-Hawking
boundary term, constructed from the trace of the extrinsic curvature in the normal
direction \cite{Gibbons1977}.   However, 
such boundary conditions are quite restrictive on the dynamics.  For a local subsystem in which
$\partial\Sigma$ simply represents a partition of a spatial slice, one would not expect Dirichlet
conditions to be compatible with all solutions of the theory.  An alternative approach would be
to impose conditions that specify the location of the surface in a diffeomorphism-invariant manner,
without placing any restriction on the dynamics.  One example is requiring that the surface
extremize its area or some other entropy functional, as is common in holographic entropy 
calculations \cite{Ryu:2006bv, Hubeny2007, Dong2014, Camps2013, Miao2014, Dong2017}.  
Since extremal surfaces exist in generic solutions, these boundary conditions put no
dynamical restrictions on the theory, but rather restrict where the surface $\partial\Sigma$ lies.  

The effects of JKM ambiguity terms in the extended phase space 
construction were discussed in section \ref{sec:jkm}.  It was noted that the $B$ form that appears
when analyzing the surface translations could be interpreted as a Lagrangian ambiguity,
 $L\rightarrow L-dB$.  Note that this type of ambiguity does not affect the symplectic 
form (\ref{eqn:OmS}), and, as a consequence, the generators of the surface 
symmetries do not depend on this replacement.  In fact, the generators (\ref{eqn:Hhw}) are 
invariant with respect to additional changes to the Lagrangian $L\rightarrow L+d\alpha$,
since such a change shifts the Noether charge $Q_W\rightarrow Q_W + i_W\alpha$, but
also induces the change $B\rightarrow B+\alpha$.  An ambiguity that does affect the phase
space is the shift freedom in the symplectic potential current, $\theta\rightarrow\theta+d\beta$.
We noted that certain choices of $\beta$ can change the number of edge mode degrees of 
freedom, and also can affect the surface symmetry algebra.  
In the future, we would like to understand how this ambiguity should be fixed.  One idea 
would be to use the ambiguity to ensure some $B$ can be found satisfying equation 
(\ref{eqn:iwthds}).  In this case, the ambiguity is fixed as an integrability condition
for $\theta$.  Such an approach seems related to the ideas of \cite{Wall2015} in which
the ambiguity was chosen to give an entropy functional satisfying a linearized second law.  
Another approach discussed in \cite{Miao2014, Miao2015b, Camps2016, Dong2017} fixes
the ambiguity through the choice of metric splittings that arise when performing the
replica trick in the computation of holographic entanglement entropy.

As discussed in section \ref{sec:edgemodes}, one of the main motivations for 
constructing the extended phase space is to understand entanglement entropy
in diffeomorphism-invariant theories \cite{Donnelly2016F}.  
The Hilbert space for such a theory does not factorize
across an entangling surface due to the constraints.  However, one can instead construct an 
extended Hilbert space for a local subregion  as a quantization of the 
extended phase space constructed above.  This extended Hilbert space will contain edge mode
degrees of freedom that transform in representations of the surface symmetry algebra.  
A similar extended Hilbert space can be constructed for the complementary region 
with Cauchy surface $\bar{\Sigma}$,
whose edge modes and surface symmetries will match those associated with $\Sigma$.  
The physical Hilbert space for $\Sigma\cup \bar\Sigma$ is given by the so-called entangling
product of the two extended Hilbert spaces, which is  the tensor product modded out by the 
action of the surface symmetry algebra.  One then finds that the density matrix associated with
$\Sigma$ splits into a sum over superselection sectors, labelled by the representations of the 
surface symmetry group.  

This block diagonal form of the density matrix leads to a 
von Neumann entropy that is the sum of three types of terms,
\beq
S = \sum_i \left( p_i S_i  - p_i \log p_i + p_i \log \dim R_i\right),
\eeq
where the sum is over the representations $R_i$ of the surface symmetry group, $p_i$ give the 
probability of being in a given representation, and $S_i$ is the von Neumann entropy within
each superselection sector.  The first term represents the average entropy of the interior
degrees of freedom, while the second term is a classical Shannon entropy coming from 
uncertainty in the surface symmetry representation corresponding to the state.  The last 
term arises from entanglement between the edge modes themselves, and is only present for a 
nonabelian surface symmetry algebra \cite{Donnelly2012a, Donnelly2014a}.
The dimension of the representation has some expression in terms of the Casimirs of the 
group, and hence this term will take the form of an expectation value of local operators
at the entangling surface.  It is conjectured that this term provides a statistical
interpretation for the Wald-like contributions in the generalized entropy, $S_\text{gen}
= S_\text{Wald-like} + S_\text{out}$ \cite{Donnelly2016F}.  Put another way, given a UV 
completion for the quantum gravitational theory, the edge modes keep track of the
 entanglement between the UV modes that are in a fixed state, corresponding
to the low energy ``code subspace'' \cite{Lin2017, Harlow2016}.  

On reason for considering the extended phase space in the context of entanglement
entropy comes from issues of divergences in entanglement entropy.  These divergences
arise generically in quantum field theories, and a regulation prescription is needed 
in order to get a finite result.  A common regulator for Yang-Mills theories is a lattice 
\cite{Casini2014a, Donnelly2012a, Donnelly2014a},
which preserves the gauge invariance of the theory.  Unfortunately, a lattice 
breaks diffeomorphism invariance, which can be problematic when 
using it as a regulator for gravitational
theories (see \cite{Hamber2009} for a review of the lattice approach to quantum
gravity).  The extended phase space provides a continuum description of the edge modes that
respects diffeomorphism invariance.  As such, it should be amenable to finding a regulation
prescription that does not spoil the gauge invariance of the gravitational theory.  
Note that edge modes have been successfully quantized using the extended phase
space for abelian Chern-Simons theories \cite{Fliss2017a, Wong2017a}, and it has 
also been applied to string field theory \cite{Balasubramanian2018}.  Finding a way to 
quantize the edge modes and compute their entanglement in a gravitational theory
is an important next step in this program.

There are a number of directions for future work on the extended phase space itself, outside 
of its application to entanglement entropy.  One topic of interest is to clarify the fiber bundle
geometry of the solution space $\SSS$, which arises due to diffeomorphism invariance.  
A fiber in this space consists of all solutions that are
related by diffeomorphism, and the $\dx^a$ fields define a flat connection
on the bundle.  Flatness in this case is equivalent to the equation $\del(\dx^a)+\frac12[\dx,\dx]^a
=0$ for the variation of $\dx^a$.  
See \cite{Gomes2017, Gomes2018} for related discussion of this fiber bundle description
of $\SSS$.  
Another technical question that arises is whether $\SSS$ truly carries a smooth manifold 
structure.  One obstruction to smoothness would be if the equations of motion are not well-posed
in some coordinate system.  In this case, the solutions do not depend smoothly on the initial
conditions on the Cauchy slice $\Sigma$, calling into question the smooth manifold structure
of $\SSS$.  If $X$ is used to define the coordinate system, this would mean that for
some values of $X$ the solution space is not smooth.  A possible way around this is to always work 
in a coordinate system in which the field equations are well-posed, and the gauge transformation
to this
coordinate system would impose dynamical equations on the $X$ fields.  Another obstruction 
to smoothness comes from issues related to ergodicity and chaos in totally constrained systems
\cite{Dittrich2015}.  It would be interesting to understand if these issues are problematic 
for the phase space construction 
given here, and whether the $X$ fields ameliorate any of these problems.

Another interesting application would be to formulate the first law of black hole mechanics 
and various related ideas in terms of the extended phase space.  This could be particularly 
interesting in clarifying certain gauge dependence that appears when looking at  
second order perturbative identities, such as described in \cite{Hollands2013}.  The edge modes
should characterize all possible gauge choices, and they may inform some of the relations 
found in \cite{Lashkari2016, Beach2016a, Faulkner2017} when considering 
different gauges besides the Gaussian null coordinates 
used in \cite{Hollands2013}.  They could also be useful in understanding
quasilocal gravitational energy, and in particular how to define the gravitational energy inside a 
small ball.  This can generally be determined by integrating a pseudotensor over the ball, but 
there is no preferred choice for a gravitational pseudotensor, so this procedure is ambiguous.  
It would be interesting if a preferred choice presented itself by considering second order variations
of the first law of causal diamonds \cite{Jacobson2015a, Bueno2017}, 
using the extended phase space.
Some ideas in this
direction are considered in \cite{Jacobson2017b}, but it is difficult to find a quasilocal
gravitational energy that satisfies the desirable property of being proportional to the Bel-Robinson
energy density in the small ball limit \cite{Szabados2009, Senovilla2000a}.

Finally, it would be very useful to recast the extended phase space construction in 
vielbein variables.  
Some progress on the vielbein formulation was reported in \cite{Geiller2017}.  
Since vielbeins have an additional internal gauge symmetry associated with local 
Lorentz invariance,  care must be taken when applying  covariant
canonical constructions \cite{Jacobson2015b, Prabhu2017}.  It would be particularly interesting
to analyze the surface symmetry algebra that arises in this case, which could differ
from the algebra derived using metric variables because the gauge group is different.  
Comparing the algebras and edge modes in both cases would weigh on the question of 
how physically relevant
and universal their contribution to entanglement entropy is.  
This problem was recently addressed for three dimensional gravity in \cite{Geiller2018},
which interestingly found that the collection of edge modes obtained was the 
same as when using metric variables.  This suggests that edge modes degrees of freedom 
arise independent of the choice of field variables, hinting at their fundamental importance to the 
underlying theory.

\renewcommand\thesection{\thechapter.\Alph{section}}
\setcounter{section}{0}

\section{List of identities} \label{app:ids}
This appendix gives a collection of identities for the exterior calculus on solution space $\SSS$
along with their proofs.  
\setenumerate[1]{label=\thesection.\arabic*}

\begin{enumerate}
\item $L_V = I_V \delta + \delta I_V$ \label{id:LV}
\begin{proof}
This follows from standard treatments of the exterior calculus \cite{Lang1985}.
\end{proof}

\item $L_V I_U = I_{[V, U]} + I_U L_V$ \label{id:LVIU}
\begin{proof}
This is simply the derivation property of the Lie derivative applied to all tensor fields on 
$\SSS$.   $I_U\alpha$ is a contraction of the vector $U$ with the one-form $\alpha$, so 
the Lie derivative first acts on $U$ to give the vector field commutator $L_V U = [V,U]$,
and then acts on $\alpha$, with the contraction $I_U$ now being applied to 
$L_V\alpha$.  Hence,
on an arbitrary form, $L_V I_U \alpha = I_{[V,U]}\alpha + I_U L_V\alpha$.
\end{proof}

\item $L_V Y^*\alpha = Y^*(L_V\alpha + \lie_{(I_V\dy)}\al)$ \label{id:LVY*a}
\begin{proof}
The discussion of section \ref{sec:cps} derived  equation
(\ref{eqn:LphY*a}), so all that remains is to show that $\dar^a(Y;V)$ is linear in the vector $V$.
This can be demonstrated inductively on the degree of $\alpha$.    For scalars, it is enough 
to show it holds on the functions $\phi^x$.  Applying \ref{id:LV}, we have on the one hand
\beq
L_V Y^*\phi = I_V \del Y^*\phi,
\eeq
while on the other hand,
\beq
L_V Y^*\phi = Y^*\left(L_V\phi + \lie_{\dar(Y;V)} \phi \right) = I_V Y^*\del\phi + Y^*\lie_{\dar(Y;V)}
\phi
\eeq
since $I_V$ commutes with $Y^*$.  Equating these expressions, we find
\beq \label{eqn:Y*liedar}
Y^* \lie_{\dar(Y;V)} \phi = I_V\left(\del Y^*\phi - Y^*\del\phi  \right).
\eeq
Since the right hand side of this expression is linear in $V$, $\chi(Y;V)$ must be as well.  

Now suppose \ref{id:LVY*a} holds for all forms of degree $n-1$, and take $\alpha$ to be 
degree $n$.  Then for an arbitrary vector $U$, $I_U Y^*\alpha$ is degree $n-1$, so 
\beq
L_V I_U Y^*\alpha = Y^*\left(L_V I_U\alpha + \lie_{(I_V\dar_Y)} I_U\alpha\right) = I_{[V,U]}Y^*\alpha
+ I_UY^*\left(L_V\al + \lie_{(I_V\dar_Y)}\alpha \right),
\eeq
where identity \ref{id:LVIU} was applied along with the fact that $I_U$ commutes with $\lie_\xi$. 
On the other hand,
\begin{align}
L_VI_U Y^*\al = I_{[V,U]} Y^*\alpha + I_U L_V Y^*\alpha = I_{[V,U]} Y^*\alpha + I_U
Y^*\left(L_V\al + I_{\bar\dar(Y;V)}\al \right).
\end{align}
Since $U$ was arbitrary, equating these expressions shows that 
$\bar\dar^a(Y;V)=I_V\chi_Y^a$, showing that the formula holds for 
forms of degree $n$.
\end{proof}

\item $I_V \lie_{\dar_Y} = \lie_{(I_V\dar_Y)} - \lie_{\dy} I_V$ \label{id:IVliedy}
\begin{proof}
This is essentially the antiderivation property applied to $\lie_{\dy}$.  The spacetime Lie 
derivative $\lie_\dy$ acting on a tensor can be written in terms of $\dar_Y^a$ and its derivatives 
contracted with the tensor, where all instances of $\dar_Y^a$ appear to the left.  It is straightforward
to see that when $I_V$ contracts with $\dar_Y^a$ in this expression, the terms will combine into
$\lie_{(I_V\dy)}$, and since $I_V$ does not change the spacetime
tensor structure of the object it contracts, the remaining terms will combine into $-\lie_\dy I_V$,
with the minus coming from the antiderivation property of $I_V$. 
\end{proof}

\item $\delta Y^* \alpha = Y^*(\delta\alpha +\lie_\dy\alpha)$ \label{id:dY*a}
\begin{proof}
This may also be demonstrated inductively on the degree of $\alpha$.  For scalars, we 
simply note that equation (\ref{eqn:Y*liedar}) is valid for arbitrary vectors $V$, and since 
$\dar^a(Y;V) = I_V \dar_Y^a$, we derive $\del Y^*\phi = Y^*(\del\phi +\lie_{\dy}\phi)$.
Assume now \ref{id:dY*a} holds for all $(n-1)$-forms, and take $\alpha$ an $n$-form and 
$V$ an arbitrary vector.  Then
\begin{align}
I_V \del Y^*\al &= L_V Y^*\alpha - \del I_V Y^*\al \nonumber \\
&= Y^*\left(L_V\al + \lie_{(I_V \dy)}\al - \del I_V \al - \lie_{\dy} I_V\al\right) \nonumber \\
&= I_V Y^*(\del\al + \lie_{\dy}\al )
\end{align}
The first equality applies \ref{id:LV}, the second uses \ref{id:LVY*a} and the fact that $I_V Y^*\al$
is an $(n-1)$-form, and the last equality follows from \ref{id:LV} and \ref{id:IVliedy}.  Since
$V$ is arbitrary, this completes the proof.
\end{proof}

\item $\frac12[\dy,\dy]^a = \dar_Y^b\nabla_b\dar_Y^a$ \label{id:dardar}
\begin{proof}
This is a consequence of the formula for the commutator of two vectors, $[\xi,\zeta] = 
\xi^b\nabla_b \zeta^a - \zeta^b\nabla_b \xi^a$, along with the fact that since $\dar^a$ is an $\SSS$
one-form, it anticommutes with itself.  Alternatively, the formula may be checked by 
contracting with arbitrary vectors $V$ and $U$.  Letting $I_V\dar_Y^a = - \xi^a$ and $I_U \dar_Y^a
=-\zeta^a$, we have
\beq
I_VI_U\frac12[\dy, \dy]^a = I_V[\dy,\zeta]^a = [\zeta,\xi]^a = 
\zeta^b\nabla_b\xi^a-\xi^b\nabla_b\zeta^a = I_V I_U \dar_Y^b\nabla_b\dar_Y^a.
\eeq
\end{proof}

\item $\lie_\dy \lie_\dy = \lie_{\frac12[\dy,\dy]}$ \label{id:liedarliedar}
\begin{proof}
For ordinary spacetime vectors $\xi^a$ and $\zeta^a$, the Lie derivative satisfies \cite{Edelen2005}
\beq
\lie_\xi \lie_\zeta = \lie_{[\xi,\zeta]} + \lie_\zeta \lie_\xi.
\eeq
Since $\dar_Y^a$ are anticommuting, this formula is modified to 
\beq
\lie_\dy \lie_\dy = \lie_{[\dy,\dy]} - \lie_\dy \lie_\dy,
\eeq
from which the identity follows.  Note that \ref{id:dardar} provides a formula for $[\dy,\dy]^a$.
\end{proof}

\item $\lie_\xi (Y^{-1})^* = (Y^{-1})^*\lie_{Y^*\xi}$ \label{id:liexiYi}
\begin{proof}
This identity is a standard property of the Lie derivative, see e.g.\ \cite{Kolar1993}.
\end{proof}

\item $\lie_\dx i_\dx = \frac12(i_{[\dx,\dx]} + d i_\dx i_\dx - i_\dx i_\dx d)$ \label{id:liedxidx}
\begin{proof}
The identity for ordinary spacetime vectors $\xi^a$ and $\zeta^b$ \cite{Edelen2005}
\beq
\lie_\xi i_\zeta = i_{[\xi,\zeta]} + i_\zeta \lie_\xi
\eeq
along with the fact that $\dx^a$ are anticommuting gives
\begin{align}
\lie_\dx i_\dx &= i_{[\dx,\dx]} - i_\dx \lie_\dx \nonumber \\
&= i_{[\dx,\dx]} - i_\dx di_\dx - i_\dx i_\dx d \nonumber\\
&= i_{[\dx,\dx]} - \lie_\dx i_\dx + di_\dx i_\dx - i_\dx i_\dx d, 
\end{align}
and moving $-\lie_\dx i_\dx$ to the left hand side proves the identity.
\end{proof}

\item $\lie_\dx \theta + \del i_\dx L + \lie_\dx i_\dx L = d\left(i_\dx\theta +\frac12 i_\dx i_\dx L\right)$
\label{id:liedxth}
\begin{proof}
The first term in this expression is $\lie_\dx\theta = d i_\dx \theta + i_\dx d\theta$, which gives 
one of the terms on the right hand side of the identity, along with $i_\dx d\theta$.  Next we have
\beq
\del i_\dx L = i_{\del \dx} L - i_\dx \del L = -\frac12 i_{[\dx,\dx]}L - i_\dx d\theta,
\eeq
where we applied equation (\ref{eqn:ddx}) for $\del \dx^a$, and used that $\del L = d\theta$
on shell.  The $-i_\dx d\theta$ term cancels against the similar term appearing in $\lie_\dx \theta$,
so that the remaining pieces are 
\beq
-\frac12 i_{[\dx,\dx]}L + \lie_\dx i_\dx L = \frac12 d i_\dx i_\dx L,
\eeq
which follows from identity \ref{id:liedxidx} and $dL=0$.  Hence, the terms on the left of the 
\ref{id:liedxth} combine into the exact form $d(i_\dx\theta +\frac12 i_\dx i_\dx L)$.
\end{proof}

\item $[V, \hat\xi] = (I_V\delta \xi^a)\,\hat{}$ \label{id:Vhxi}
\begin{proof}
Here we can use that on local $\SSS$-scalars, $L_{\hat\xi}\phi = \lie_\xi\phi$.  Then
\begin{align}
L_{[V,\hat\xi]}\phi = L_VL_{\hat\xi} \phi - L_{\hat\xi} L_V \phi  = L_V \lie_\xi\phi - \lie_\xi I_V\del\phi
=\lie_{(I_V\del \xi)\,\hat{}}\,\phi = L_{(I_V\del\xi)\,\hat{}} \,\phi,
\end{align}
hence, $[V,\hat\xi] = (I_V\del\xi^a)\,\hat{}$.
\end{proof}

\item $L_{\hat\xi} = \lie_\xi +I_{\del\xi\,\hat{}}$ \label{id:Lhxi}
\begin{proof}
This formula is meant to apply to local functionals of the fields defined at a single spacetime point.
Since $I_{\del\xi\,\hat{}}$ annihilates scalars, it clearly is true for that case.  Then assume the 
formula has been shown for all $(n-1)$-forms, and take $\alpha$ to be an $n$-form.  For an 
arbitrary vector $V$, since
$I_V\alpha$ is an $(n-1)$-form, we have 
\begin{align}
I_V L_{\hat\xi}\al &= L_{\hat\xi}I_V\al - I_{[\hat\xi,V]}\al = \lie_\xi I_V\al + I_{\del\xi\,\hat{}}I_V\al 
-I_{[\hat\xi,V]}\al \nonumber \\
& = I_V(\lie_\xi \al + I_{\del\xi\,\hat{}}\,\al) - I_{(I_V\del\xi)\,\hat{}}\,\al - I_{[\hat\xi,V]} \al,
\end{align}
and the last two terms in this expression cancel due to identity \ref{id:Vhxi}. Since $V$ was 
arbitrary, we conclude that the identity holds for all $n$ forms, and by induction for all 
$\SSS$ differential forms.
\end{proof}

\item $L_\hdx = I_\hdx \delta - \delta I_\hdx$ \label{id:Lhdxdef}
\begin{proof}
This is essentially a definition of what is meant by $L_\hdx$.  The left hand side is the graded
commutator of the derivation $I_\hdx$ and the antiderivation $\del$, which defines the 
the antiderivation $L_\hdx$ \cite{Kolar1993}.
\end{proof}

\item $[V,\hdx] = (\del I_V \dx^a)\,\hat{} - [I_V\dx,\dx]\,\hat{}$ \label{id:Vhdx}
\begin{proof}
This follows from the defining relation of the bracket \cite{Kolar1993},
\beq
L_V L_\hdx - L_\hdx L_V = L_{[V,\hdx]}.
\eeq
Applied to $\phi$ and defining $\nu^a = -I_V\dx^a$, this gives
\begin{align}
L_{[V,\hdx]}\phi &= (L_V L_{\hdx}-L_{\hdx} L_V)\phi \nonumber \\
&=I_V\del \lie_\dx \phi -\del\lie_\nu\phi - \lie_\dx I_V\del\phi \nonumber \\
&= I_V(\lie_{\del\dx} \phi - \lie_{\dx} \del \phi) - \lie_{\del\nu}\phi 
- \lie_{\nu}\del \phi+ I_V\lie_\dx \del\phi \nonumber \\
&= (\lie_{[\nu,\chi]} - \lie_{\del\nu})\phi \nonumber \\
&= (L_{[\nu,\chi]\,\hat{}} - L_{\del\nu\,\hat{}} )\phi,
\end{align}
To get to the third line, the expression (\ref{eqn:ddx}) for 
$\del\dx^a$ was used.  We then conclude $[V,\hdx] = [\nu,\dx]\,\hat{} - \del\nu\,\hat{}$, proving
the identity.
\end{proof}

\item $L_\hdx = \lie_\dx - I_{\del\dx\,\hat{}}$ \label{id:Ldx}
\begin{proof}
The formalism of graded commutators developed in \cite{Kolar1993} is a useful tool in 
proving this identity.  Given two graded derivations $D_1$ and $D_2$, their graded
commutator $D_1 D_2 - (-1)^{k_1 k_2} D_2 D_1$ is another 
graded derivation, where $k_i$ are the degrees of the 
respective derivations, i.e.\ the amount the derivation increases or decreases the degree of 
the form on which it acts.  Hence, since $I_V$ and $L_{\hdx}$ are  derivations of degrees $-1$
and
$1$, they satisfy
\beq \label{eqn:IVLhdx}
I_V L_{\hdx} + L_{\hdx} I_V = -L_{\hat{\nu}} + I_{[\hdx,V]},
\eeq
where $-\nu^a = I_V \dx^a$.  Similarly, we have 
\beq\label{eqn:IVhddx}
I_V I_{\del \dx\;\hat{}} + I_{\del\dx\;\hat{}} I_V = I_{(I_V\del\dx)\,\hat{}} = I_{[\nu,\dx]\,\hat{}},
\eeq
where equation (\ref{eqn:ddx}) was used in the last equality. 

We then prove the identity through induction on the degree of the form on which it acts.  
It is true for scalars because $I_{\del\dx} \phi=0$.  Then suppose it is true for all $(n-1)$-forms,
and take $\alpha$ to be an $n$-form.  For an arbitrary vector $V$ we have
\begin{align}
I_VL_\dx\al &= I_{[\hdx,V]}\al - L_{\hat\nu}\al - L_\dx I_V\al\nonumber \\
&= I_{\del\nu\,\hat{}\,} \al - I_{[\nu,\dx]\,\hat\,}\al - \lie_\nu\al - I_{\del\nu\,\hat{}\,}\al
 - \lie_\dx I_V\al + I_{\del \dx} I_V \al \nonumber \\
 &= I_V(\lie_\dx \al - I_{\del\dx\,\hat{}\,}\al).
\end{align}
The first line employs equation (\ref{eqn:IVLhdx}), the second line uses identities \ref{id:Vhdx}
and \ref{id:Lhxi} as well as the fact that $I_V\al$ is an $(n-1)$-form, and the third line
employs equation (\ref{eqn:IVhddx}). Since $V$ is arbitrary, we conclude the identity holds
for all $n$-forms, which completes the proof.
\end{proof}

\end{enumerate}

\section{Edge mode derivatives in the symplectic form} \label{app:edge}
In this appendix, we derive the result advertised in section 
\ref{sec:eps}, that the 
symplectic form (\ref{eqn:OmS}) does not depend on second or higher derivatives of 
$\dx^a$.  Derivatives of $\dx^a$
appear in $\Om$ through the terms $\del Q_{\dx} + \lie_{\dx}Q_\dx$.  The Lie derivative
term may be expressed
\begin{align}
\lie_{\dx}Q_{\dx} &= L_{\hdx} Q_{\dx} + I_{{\del(\dx)} \,\hat{}}\,Q_{\dx} \nonumber  \\
&= I_\hdx \del Q_\dx-\del I_\hdx Q_\dx-Q_{\del(\dx)}  \nonumber \\
&= I_\hdx \qo_\dx +I_\hdx Q_{\del(\dx)}  +\del Q_\dx - Q_{\del(\dx)} \nonumber \\
&= \qo_\dx + I_{\hdx}\qo_\dx +Q_{[\dx,\dx]}.
\end{align}
These steps invoke the identities \ref{id:Ldx}, \ref{id:Lhdxdef} and equations (\ref{eqn:ddx}) 
and (\ref{eqn:Ixidar}), as well as the defining relation
(\ref{eqn:dQdX}) for  
$\qo_\dx$. Adding $\del Q_\dx = \qo_\dx -\frac12 Q_{[\dx,\dx]}$ to this yields
\beq\label{eqn:dQdx}
\del Q_{\dx} +\lie_\dx Q_\dx = 2 \qo_\dx+I_{\dx} \qo_\dx + \frac12Q_{[\dx,\dx]}.
\eeq
From the derivation property of  $I_\hdx$ acting on $\SSS$-forms and the identity 
$I_\hdx \dx^a = -\dx^a$, it follows that $\qo_\dx+I_\dx \qo_\dx=  \qo[\phi;\lie_\dx\phi]_\dx$,
so that (\ref{eqn:dQdx}) can equivalently be expressed
\beq \label{eqn:dQlieQ2}
\del Q_\dx +\lie_\dx Q_\dx = \qo[\phi;\del\phi]_\dx  + \qo[\phi;\lie_\dx \phi]_\dx + \frac12 Q_{[\dx,\dx]}.
\eeq

This expression is now amenable to determining  how the derivatives of $\dx^a$ appear.
Both $\qo[\phi;\lie_\dx\phi]_\dx$ and $Q_{[\dx,\dx]}$ contain second derivatives.  The relevant 
term in $\qo[\phi;\lie_\dx\phi]_\dx$ 
comes from the variation of the Christoffel symbol in (\ref{eqn:qoxi}),
which gives
\begin{align}
&\,-\ep_{ab}E^{abcd}\left(\nabla_c \nabla_{(d}\dx_{e)}+\nabla_e\nabla_{(d}\dx_{c)}
-\nabla_d\nabla_{(c}
\dx_{e)}\right) \dx^e \nonumber \\
=&\,-\frac12\ep_{ab}E^{abcd}\left(  \nabla_{c}\nabla_{e}\dx_d - \nabla_d\nabla_e\dx_c\right)
\dx^e + \text{n.d.} \nonumber \\
=&\, -\ep_{ab}E\indices{^a^b^c_d} \left(\nabla_{(c}\nabla_{e)} \dx^d\right) \dx^e + \text{n.d.}, 
\label{eqn:epEdddX}
\end{align}
where ``$\text{n.d.}$" represents terms with no derivatives acting on $\dx^a$.
This derivation invokes the 
antisymmetry of $E^{abcd}$ on $c$ and $d$, and collects all terms involving antisymmetrized
derivatives of $\dx_d$ into the $\text{n.d.}$ piece, since these can be replaced by a Riemann
tensor contracted with an undifferentiated $\dx_d$. 

Second derivatives of $\dx^d$ also appear in $\frac12Q_{[\dx,\dx]}$ through the 
$E\indices{^a^b^c_d}$ term in the equation (\ref{eqn:Qxi}) for the Noether charge.  This term 
evaluates to 
\begin{align}
&\, -\frac12\ep_{ab}E\indices{^a^b^c_d}\nabla_c[\dx,\dx]^d \nonumber  \\
&=\,-\ep_{ab}E\indices{^a^b^c_d} \nabla_c(\dx^e\nabla_e\dx^d) \nonumber \\
&=\, -\ep_{ab}E\indices{^a^b^c_d}\left(\dx^e\nabla_{(c}\nabla_{e)}\dx^d + \nabla_c\dx^e\nabla_e
\dx^d  \right) +\text{n.d.},
\end{align}
which uses identity \ref{id:dardar}.  When added to (\ref{eqn:epEdddX}), the second derivative
terms cancel since $\dx^e$ is an $\SSS$ one form, so 
$(\nabla_{(c}\nabla_{e)}\dx^d)\dx^e = -\dx^e\nabla_{(c}\nabla_{e)}\dx^d$.  This 
shows that (\ref{eqn:dQlieQ2}) does not depend on second derivatives of $\dx^d$.

\renewcommand\thesection{\thechapter.\arabic{section}}


\renewcommand{\baselinestretch}{1}
\small\normalsize


\newpage
\Urlmuskip=0mu plus 1mu\relax
\bibliographystyle{JHEPthesis}
\bibliography{bibliographies/intro_refs,bibliographies/refs-enthigher,bibliographies/cps,bibliographies/diamondEE} 

\end{document}